\documentclass[twoside]{article}

\usepackage[accepted]{aistats2021}
\usepackage{natbib}
\setcitestyle{authoryear}

\usepackage{booktabs}       
\usepackage{amsfonts}       
\usepackage{nicefrac}       
\usepackage{microtype}      
\usepackage{xcolor}

\usepackage{subfigure}
\usepackage{graphicx}
\usepackage{amsmath}
\usepackage{amssymb}
\usepackage{dsfont}

\usepackage{amsthm}

\newtheorem{my_definition}{Definition}

\newtheorem{my_claim}{Claim}
\newtheorem{my_lemma}{Lemma}
\newtheorem{my_theorem}{Theorem}

\newtheorem{my_assumption}{Assumption}

\begin{document}


\twocolumn[

\aistatstitle{Causal Inference under Networked Interference and Intervention Policy Enhancement}

\aistatsauthor{ Yunpu Ma \And Volker Tresp }

\aistatsaddress{ Ludwig Maximilian University of Munich \\ cognitive.yunpu@gmail.com
 \And  Ludwig Maximilian University of Munich $\&$ \\ Siemens CT \\ volker.tresp@siemens.com } ]

\begin{abstract}
Estimating individual treatment effects from data of randomized experiments is a critical task in causal inference. The Stable Unit Treatment Value Assumption (SUTVA) is usually made in causal inference. However, interference can introduce bias when the assigned treatment on one unit affects the potential outcomes of the neighboring units. This interference phenomenon is known as spillover effect in economics or peer effect in social science. Usually, in randomized experiments, or observational studies with interconnected units, one can only observe treatment responses under interference. Hence, the issue of how to estimate the superimposed causal effect and recover the individual treatment effect in the presence of interference becomes a challenging task. In this work, we study causal effect estimation under general network interference using Graph Neural Networks, which are powerful tools for capturing node and link dependencies in graphs. After deriving causal effect estimators, we further study intervention policy improvement on the graph under capacity constraint. We give policy regret bounds under network interference and treatment capacity constraint. 
\end{abstract}


\section{Introduction}

In causal inference one commonly makes the consistency and the interference-free assumptions, i.e., the Stable Unit Treatment Value Assumption (SUTVA)~\citep{rubin1980randomization}, under which the individual treatment response is consistently defined and unaffected by variations in other individuals. However, this assumption is problematic under a social network setting since peers are not independent; ``no man is an island,'' as written by the poet John Donne.

Interference occurs when the treatment response of an individual is influenced through the exposure to its social contacts' treatments or affected by its social neighbors' outcomes through peer effects~\citep{bowersFP2013,toulis2013estimation}. For instance, the treatment effect of an individual under a vaccination against an infectious disease might influence the health conditions of its surrounding individuals; or a personalized online advertisement might affect other individuals' purchase of the advertised item through opinion propagation in social networks. Separating individual treatment effect and peer effect in causal inference becomes a difficult problem under interference since, in randomized experiments or observational studies, one can only observe the superposition of both effects. In this work we study the issue of how to estimate causal responses and make optimal policies on the network.

One of the main objectives of treatment effect estimation is to derive optimal treatment decision rules for individuals according to their characteristics. Population-averaged utility functions have been studied in~\citep{manski2009identification,athey2017efficient,kallus2018balanced,kallus2018confounding}. In those publications, a policy learner can adapt and improve its decision rules through the utility function. However, interactions among units are always ignored. On the other hand, a policy learner usually faces a capacity or budget constraint, as studied in~\citep{kitagawa2017should}. In this work, we develop a new type of utility function defined on interconnected units and investigate provable policy improvement with budget constraints.

\subsection{Related Work}

Causal inference with interference was studied in~\citep{hudgens2008toward,tchetgen2012causal,liu2014large}. However, the assumption of group-level interference, having partial interference within the groups and independence across different groups, is often invalid. Hence, several works focus on unit-level causal effects under cross-unit interference and arbitrary treatment assignments, such as~ \citep{aronow2017estimating,forastiere2016identification,ogburn2017causal,viviano2019policy}. Other approaches for estimating causal effects on networks use graphical models, which are studied in~\citep{arbour2016inferring,tchetgen2017auto}.

\subsection{Notations and Previous Approaches}

Let $\mathcal{G}=(\mathcal{N}, \mathcal{E}, A)$ denote a directed graph with a node set $\mathcal{N}$ of size $n$, an edge set $\mathcal{E}$, and an adjacency matrix $A \in \{0, 1\}^{n \times n}$. For a node, or unit, $i \in \mathcal{N} $, let $\mathcal{N}_i$ indicate the set of neighboring nodes with $A_{ij} = 1$ excluding the node $i$ itself, and let $\mathbf{X}_i$ denote a vector of covariates for node $i$ which is defined in some space $ \boldsymbol{ \chi }$. Let's first focus on the Neyman–Rubin causal inference model~\citep{rubin1974estimating,splawa1990application}. Let $T_i$ be a binary variable with $T_i=1$ indicating that node $i$ is in the treatment group, and $T_i=0$ if $i$ is in the control group. Moreover, let $Y_i$ be the outcome variable with $Y_i(T_i=1)$ indicating the potential outcome of $i$ under treatment $T_i=1$ and $Y_i (T_i=0)$ the potential outcome under control $T_i=0$. Moreover, we use $T_{\mathcal{N}_i}$ and $Y_{\mathcal{N}_i}$ to represent the treatment assignments and potential outcomes of neighboring nodes $\mathcal{N}_i$, and $ \mathbf{T} $ the entire treatment assignments vector.

In the SUTVA assumption, the individual treatment effect on node $i$ is defined as the difference between outcomes under treatment and under control, i.e., $ \tau ( \mathbf{X}_i ) := \mathbb{E} [ Y_i (T_i = 1) - Y_i ( T_i = 0) | \mathbf{X}_i ] $. To estimate treatment effects under network interference, an exposure variable $G$ is proposed in~\citep{toulis2013estimation,bowersFP2013,aronow2017estimating}. The exposure variable $G_i$ is a summary function of neighboring treatments $T_{\mathcal{N}_i}$. 


Under the assumption that the outcome only depends on the individual treatment and neighborhood treatments, \citep{forastiere2016identification} defines an individual treatment effect under the exposure $ G_i = g $ as
\begin{align}
  \tau ( \mathbf{X}_i, G_i = g ) & := \mathbb{E} [ Y_i (T_i= 1, G_i = g ) \nonumber \\
  & \ - Y_i (T_i = 0, G_i = g) | \mathbf{X}_i ]. 
  \label{eq:ind_treatment_defined}
\end{align}
Moreover, the spillover effect under the treatment $T_i = t$ and the exposure $G_i = g$ is defined as $ \delta (\mathbf{X}_i, T_i = t, G_i = g ) := \mathbb{E} [ Y_i (T_i = t, G_i = g) - Y_i (T_i = t, G_i = 0) | \mathbf{X}_i ] $. Treatment and spillover effects are then estimated using generalized propensity score (GPS) weighted estimators.

In general, the outcome model can be more complicated, depending on network topology and covariates of neighboring units. \citep{ogburn2017causal} investigates more general causal structural equations under dimension-reducing assumption, and the potential outcome reads $ Y_{i, t} := f_Y ( \mathbf{X}_i, s_X ( \{ \mathbf{X}_j | j \in \mathcal{N}_i \} ), T_i , s_T ( \{ T_j | j \in \mathcal{N}_i \} ) ) $, where $ s_X $ and $ s_T $ are summary functions of neighborhood covariates and treatment, e.g., they could be the summation or average of neighboring treatment assignments and covariates, respectively. In this work, we incorporate Graph Neural Network (GNN)-based causal estimators with appropriate covariates and treatment aggregation functions as inputs. GNNs can learn and aggregate feature information from distant neighbors, which makes it a right candidate for capturing the spillover effect given by the neighboring units.


\textbf{Contributions }  This work has the following major contributions. First, we propose GNN-based causal estimators for causal effect prediction and to recover the direct treatment effect under interference (Section~\ref{sec:GNN_estimator}). Second, we define a novel utility function for policy optimization on a network and derive a graph-dependent policy regret bound (Section~\ref{sec:policy}). Third, we provide policy regret bounds for GNN-based causal estimators (Section~\ref{sec:policy} and Appendix I and J).  Last, we conduct extensive experiments to verify the superiority of GNN-based causal estimators and show that the accuracy of a causal estimator is crucial for finding the optimal policy (Section~\ref{sec:exp}).

\section{GNN-based Causal Estimators} 
\label{sec:GNN_estimator}

In this section, we introduce our Graph Neural Network-based causal effect estimators under general network interference.

\subsection{Structural Equation Model}

Given the graph  $ \mathcal{G} $, the covariates of all units in the graph $ \mathbf{X} $, and the entire treatment assignments vector $ \mathbf{T}$, the structural equation model describing the considered data generation process is given as follows
\begin{align}
  T_i & = f_{T} ( \mathbf{X}_i ) \nonumber \\
  Y_i & = f_{Y} ( T_i, \mathbf{X}, \mathbf{T},  \mathcal{G} ) + \epsilon_{ Y_i },
  \label{eq:structural} 
\end{align}
for units $i = 1, \dots, n$. This structural equation model encodes both the observational studies and the randomized experiments setting. In observational studies, e.g., on the Amazon dataset (see Section~\ref{sec:datasets}), the treatment $\mathbf{T}_i$ depends on the covariate $ \mathbf{X}_i $ and the unknown specification of $f_{T}$, or even on the neighboring units under network interference. In the setting of the randomized experiment, e.g., experiments on Wave1 and Pokec datasets, the treatment assignment function is specified as $ f_{T} = \mathrm{Bern} (p) $, where $ p $ represents predefined treatment probability. Function $f_Y$ characterizes the causal response, which, in addition to $ \mathbf{X}_i $ and $ \mathbf{T}_i $, depends on the graph and neighboring covariates and treatment assignments.  If only influences from first-order neighbors are considered, the response generation can be specified as $ Y_i = f_{Y} ( T_i, \mathbf{X}_{ \mathcal{N}_i }, \mathbf{T}_{ \mathcal{N}_i },  \mathcal{G} ) + \epsilon_{ Y_i } $. When the graph structure is given and fixed, we leave out $ \mathcal{G} $ in the notation.

\subsection{Distribution Discrepancy Penalty}


Even without network interference, a covariate shift problem of counterfactual inference is commonly observed, namely the factual distribution $ \Pr (\mathbf{X}, T)$ differs from the counterfactual distribution $ \Pr (\mathbf{X}, 1 - T)$. To avoid biased inference, \citep{johansson2016learning,shalit2017estimating} propose a balancing counterfactual inference using domain-adapted representation learning. Covariate vectors are first mapped to a feature space via a feature map $ \Phi $. In the feature space, treated and control populations are balanced by penalizing the distribution discrepancy between $ \Pr ( \Phi( \mathbf{X} ) | T=0 ) $ and $ \Pr ( \Phi( \mathbf{X} ) | T=1 ) $ using the \emph{Integral Probability Metric}. This approach is equivalent to finding a feature space such that the treatment assignment $T$ and representation $ \Phi ( \mathbf{X} ) $ become approximately disentangled, namely $ \Pr ( \Phi ( \mathbf{X} ), T ) \approx \Pr ( \Phi( \mathbf{X} ) ) \Pr  ( T ) $. We use the Hilbert-Schmidt Independence Criterion (HSIC) as the dependence test in the feature space, whose form is provided in Appendix A. We observe that incorporating the feature map and the representation balancing penalty is important to tackle the imbalanced assignments in observational studies, e.g., on the Amazon dataset (see Section~\ref{sec:datasets}).

\subsection{Graph Neural Networks}

Different GNNs are employed and compared in our model, and we briefly provide a review.

\textbf{Graph Convolutional Network (GCN)}~\citep{kipf2016semi} The graph convolutional layer in GCN is one special realization of GNNs, which is defined as $ \mathbf{X}^{(l+1)} = \sigma \left( \hat{ \mathbf{D} }^{-1/2 } \hat{ \mathbf{A} } \hat{ \mathbf{D} }^{-1/2} \mathbf{X}^{(l)} \mathbf{W}^{(l)} \right) $, where $ \mathbf{X}^{(l+1)} $ is the hidden output from the $l$-th layer with $ \mathbf{X}^{(0)} $ being the input features matrix, and $\sigma$ is the activation function, e.g., ReLU. The modified adjacency $ \hat{ \mathbf{A} } $ with inserted self-connections is defined as $ \hat{ \mathbf{A} } := \mathbf{A} + \mathbf{I} $, and $ \hat{ \mathbf{D} }  $ denotes the node degree matrix of $ \hat{ \mathbf{A} } $.

\textbf{GraphSAGE} GraphSAGE~\citep{hamilton2017inductive} is an inductive framework for calculating node embeddings and aggregating neighbor information. The mean aggregation operator of the GraphSAGE in this work reads $ \mathbf{X}_i^{ (l+1) } = \mathrm{norm} \left( \mathrm{mean}_{ j \in \mathcal{N}_i \cup \{ i \} } \mathbf{X}_j^{ (l) }  \mathbf{W}^{ (l) } \right) $, with $ \mathrm{norm}$ being the normalization operator. Traditional GCN algorithms perform spectral convolution via eigen-decomposition of the full graph Laplacian. In contrast, GraphSAGE computes a localized convolution by aggregating the neighborhood around a node, which resembles the simulation protocol of linear treatment response with spillover effect for semi-synthetic experiments (see Section~\ref{sec:datasets}). Due to the resemblance, a better causal estimator is expected when using GraphSAGE as the aggregation function (see the beginning of Appendix H for more heuristic motivations.).

\textbf{$1$-GNN} $1$-GNN~\citep{morris2018weisfeiler} is a variation of GraphSAGE, which performs separate transformations of  node features and aggregated neighborhood features. Since the features of the considered unit and its neighbors contribute differently to the superimposed outcome, it is expected that the $1$-GNN is more expressive than GraphSAGE. The convolutional operator of $1$-GNN has the form $ \mathbf{X}_i^{ (l+1) } = \sigma \left(  \mathbf{X}_i^{ (l) } \mathbf{W}_1^{ (l) } + \mathrm{mean}_{ j \in \mathcal{N}_i }  \mathbf{X}_j \mathbf{W}_2^{ (l) } \right) $.

\begin{figure}[ht]
\centering
  \includegraphics[width=0.85\linewidth]{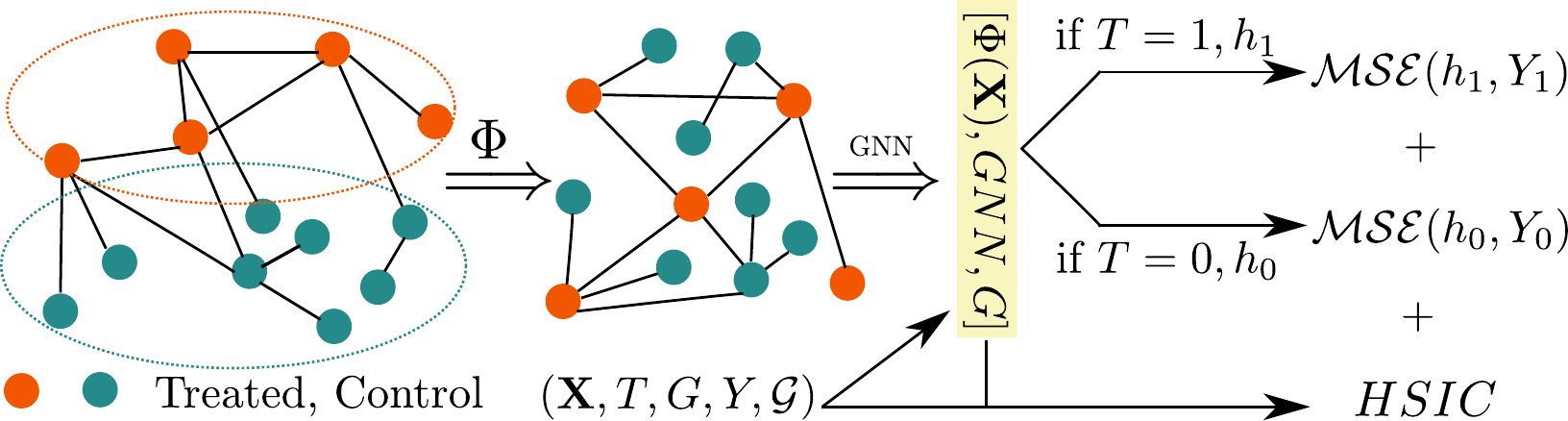}
  \caption{Treated and control populations have different distributions in the covariate vectors space. Through a map $\Phi$ and distribution discrepancy term HSIC, features and treatment assignments become disentangled in the feature space. On top of $ \Phi $, we apply GNNs, where $\Phi$ and GNNs have $2$ or $3$ hidden layers, depending on the dataset. After applying GNNs, for each node $i$, the concatenation $ [ \Phi( \mathbf{X}_i ) , \textit{GNN}( \Phi ( \mathbf{X} ) , \mathbf{T} )_i, G_i ]$ is fed into outcome prediction network $h_1$ or $h_0$ depending on the treatment assignment. The loss function combines outcome prediction error and distribution discrepancy in the feature space.}
  \label{fig:model}
\end{figure}

\subsection{GNN-based Causal Estimators}
\label{sec:gnn_estimator_bound}

We use the random variable $ G_i := \frac{ \sum_{j \in \mathcal{N}_i } T_j }{ | \mathcal{N}_i | } $ that indicates the level of exposure to the treated neighbors as the treatment summary function, and the output of GNNs as the covariate aggregation function. The concatenation $[ \Phi ( \mathbf{X}_i ), \textit{GNN}( \Phi( \mathbf{X} ) , \mathbf{T} )_i, G_i ]$ of node $i$ is then fed into the outcome prediction network $h_1$ or $h_0$, depending on $T_i$, where $ h_1 $ and $ h_0 $ are neural networks with a scalar output. Note that $ \textit{GNN}( \Phi( \mathbf{X} ) , \mathbf{T} )_i $ indicates that the treatment vector $ \mathbf{T} $ is also a GNNs' input. During the implementation, the treatment assignment vector masks the covariates, and GNN models use the masked covariates $ T_i \mathbf{X}_i $, for $i = 1, \dots, n $, as inputs. In summary, given $ ( \Phi (X)_i, T_i, G_i, Y_i) $ and graph $ \mathcal{G} $, the loss function for GNN-based estimators is defined as \\
$ \mathcal{L}_{ \mathrm{ est } } := \mathcal{MSE} \left( h_{T_i} ( [ \Phi ( \mathbf{X}_i ), GNN ( \Phi ( \mathbf{X} ) , \mathbf{T} )_i, G_i ] ), Y_i  \right) + \kappa \hat{HSIC}_{ \mathcal{K}_{\sigma} }$, where $ \kappa $ and $ \sigma $ are tunable hyperparameters. Our model is illustrated in Fig.~\ref{fig:model}. During the implementation, we incorporate two types of empirical representation balancing: balancing the outputs of representation network $ \Phi $ to tackle imbalanced assignments, denoted as $ \hat{HSIC}^{ \Phi } $, and balancing the outputs of the GNN representations to tackle imbalanced spillover exposure, denoted as $ \hat{HSIC}^{\textit{GNN}} $.  


It is necessary to emphasize that only the causal responses of a fractional of units in the graph $\mathcal{G}$ can be observed by the models. GNN-based causal estimators use this part of causal responses, the network structure $ \mathcal{G} $, and covariates $ \mathbf{X} $ as input, and predict the superimposed causal effects of the remaining units. Note that for GNN-based models, the identifiability of causal response is guaranteed under reasonable assumptions similar to those given in Section 3.2 of~\citep{ogburn2017causal}. The proof is relegated to Appendix B.

We briefly discuss an error bound that can be derived for GNN-based causal estimators to estimate the superimposed causal effects. Especially, in the following claim, we provide evidence about the dependency of the error bound on the maximal node degree through network interference.
\begin{my_claim}
GNN-based causal estimators restricted to a particular class for predicting the superimposed causal effects have an error bound $ \mathcal{O} ( \sqrt{ \frac{ D_{max}^3 \ln D_{ max } }{ n } } ) $, where $ D_{ max } := 1 + d_{ max } + d_{ max }^2 $, and $ d_{ max } $ is the maximal node degree in the graph.
  \label{claim:estimator}
\end{my_claim}
The above claim indicates that an accurate and consistent causal estimator is difficult to achieve with large network interference. In the worst case is that the $ \mathcal{O} ( \sqrt{1 / n} ) $ convergence rate, or sample dependency, becomes unreachable when the maximal node degree increases with the number of nodes of the network, namely $ d_{ \max } (n) $. The exact convergence rate of causal estimators is difficult to derive since it depends on the topology of the network, and it beyond the theoretical scope of this work. The derivation of Claim~\ref{claim:estimator} is relegated to Appendix H.

Notice that the outcome prediction networks $h_0$ and $h_1$ (see Fig.~\ref{fig:model}) are trained to estimate the superposition of individual treatment effect and spillover effect. Still, after fitting the observed outcomes, we expect to extract the non-interfered individual treatment effect from the causal estimators by assuming that the considered unit is isolated. An individual treatment effect estimator can be defined similarly to Eq.~\ref{eq:ind_treatment_defined}.  To be more specific, the individual treatment effect of unit $i$ is expected to be extracted from GNN-based estimators by setting its exposure to $ G_i = 0 $ and its neighbors' covariates to $\mathbf{0}$, namely
\begin{equation}
  \hat{ \tau } ( \mathbf{X}_i ) = h_1 ( [ \Phi ( \mathbf{X}_i ), \mathbf{0}, 0 ] ) - h_0 ( [ \Phi ( \mathbf{ X }_i ), \mathbf{0}, 0 ] ). 
  \label{eq:tau_estimation}
\end{equation}


\section{Intervention Policy on Graph}
\label{sec:policy}

After obtaining the treatment effect estimator, we develop an algorithm for learning intervention assignments to maximize the utility on the entire graph; the learned rule for assignment is called a policy. As suggested in~\citep{athey2017efficient}, without interference a utility function is defined as $  A( \pi ) = \mathbb{E} [ ( 2 \pi ( \mathbf{X}_i ) - 1 ) ( Y_i( T_i=1 ) - Y_i( T_i=0 ) ) ] = \mathbb{E} [ ( 2 \pi ( \mathbf{ X }_i ) - 1 ) \tau ( \mathbf{X}_i ) ] $.  An optimal policy $ \hat{ \pi }_n $ is obtained by maximizing the $n$-sample empirical utility function $ \hat{A}_n^{ \tau } ( \pi ) := \frac{1}{n} \sum_{ i=1 }^n ( 2 \pi ( \mathbf{ X }_i )  -1 ) \hat{ \tau } ( \mathbf{X}_i ) $ given the individual treatment response estimator $ \hat{ \tau } $, i.e., $ \hat{ \pi }_n \in \mathrm{argmax}_{ \pi \in \Pi } \hat{A}_n^{ \tau } ( \pi ) $, where $\Pi $ indicates the policy function class. Notably, $ \hat{ \pi }_n $ tends to assign treatment to units with positive treatment effect and control to units with negative responses.

Now, consider the outcome variable $ Y_i $ under network interference. For notational simplicity and clarity of the later proof, we assume first-order interference from nearest neighboring units, hence the outcome variable can be written as $ Y_i ( T_i, \mathbf{X}_{ \mathcal{N}_i }, T_{ \mathcal{N}_i } ) $. Inspired by the definition of $ A ( \pi ) $, the utility function of a policy $ \pi $ under interference is defined as
\begin{align}
  S ( \pi ) & := \mathbb{E} [ ( 2 \pi ( \mathbf{X}_i ) - 1 ) ( Y_i ( T_i=1, \mathbf{X}_{ \mathcal{N}_i }, T_{ \mathcal{N}_i } = \pi ( \mathbf{X}_{ \mathcal{N}_i } ) ) \nonumber \\
  & \quad \quad \ - Y_i ( T_i=0, \mathcal{G} = \emptyset ) ) ],  
  \label{eq:utility_under_interference}
\end{align}
where $ Y_i ( T_i=0, \mathcal{G} = \emptyset ) $ with an empty graph represents the individual outcome under control without any network influence~\footnote{Hence $ \mathbf{X}_{ \mathcal{N}_i } $ and $ T_{ \mathcal{N}_i } $ are omitted in the expression.}. After some manipulations, $ S ( \pi ) $ equals the sum of individual treatment effect and spillover effect, i.e., $ S ( \pi ) = \mathbb{E} [ ( 2 \pi ( \mathbf{X}_i ) - 1 ) ( \tau_i + \delta_i ( \pi ) ) ] $, where $ \tau_i := \mathbb{E} [ Y_i (T_i = 1, \mathcal{G} = \emptyset) - Y_i ( T_i=0, \mathcal{G} = \emptyset ) | \mathbf{X}_i ] $ and $ \delta_i ( \pi ) := \mathbb{E} [ Y_i ( T_i=1, \mathbf{X}_{ \mathcal{N}_i }, T_{ \mathcal{N}_i } = \pi ( \mathbf{X}_{ \mathcal{N}_i } ) ) - Y_i ( T_i=1, \mathcal{G} = \emptyset ) | \mathbf{X}_i, \mathbf{X}_{ \mathcal{N}_i } ] $. To be more specific, $ \tau_i $ is the conventional individual treatment effect, while $ \delta_i ( \pi ) $ represents the spillover effect under the policy $ \pi $ and when $T_i = 1$. Due to the network-dependency in the spillover effect, an optimal policy will not merely treat units with positive individual treatment effect but also adjust its intervention on the entire graph to maximize the spillover effects.

Next, we introduce the regret of learned intervention policy. Let $ \hat{ \tau }_i  $ and $ \hat{ \delta }_i ( \pi ) $ denote the estimator of $ \tau_i $ and $ \delta_i ( \pi ) $, respectively. Given the true models $ \tau_i $ and $ \delta_i ( \pi ) $, let $ S_n^{ \pi, \delta } ( \pi ) := \frac{1}{n} \sum_{i=1}^n ( 2 \pi ( \mathbf{X}_i ) - 1 ) ( \tau_i + \delta_i ( \pi ) ) $ be the empirical analogue of $ S ( \pi ) $, and let
\begin{equation}
  \hat{S}_n^{ \pi, \delta } ( \pi ) := \frac{1}{n} \sum_{i=1}^n ( 2 \pi ( \mathbf{X}_i ) - 1 ) ( \hat{\tau}_i + \hat{\delta}_i ( \pi ) ) 
\end{equation}
be the empirical utility with estimators plugged in. Using learned causal estimators, an optimal intervention policy from the empirical utility perspective can be obtained from $ \hat{ \pi }_n \in \mathrm{argmax}_{ \pi \in \Pi } \hat{S}_n^{ \pi, \delta } ( \pi ) $. Moreover, the best possible intervention policy from the functional class $ \Pi $ with respect to the utility $ S ( \pi ) $ is written as $ \pi^{ \star } := \mathrm{argmax}_{ \pi \in \Pi } S ( \pi ) $, and the policy regret between $ \pi^{ \star } $ and $ \hat{ \pi }_n $ is defined as $ \mathcal{R} ( \hat{ \pi }_n ) := S ( \pi^{ \star } ) - S (  \hat{ \pi }_n ) $.


We briefly mention the dependency of the policy regret bound on the network structure. Throughout the estimation of policy regret, we maintain the following assumptions.  
\begin{my_assumption}{\ } \\ 
  (BO) Bounded treatment and spillover effects: There exist $ 0 < M_1, M_2 < \infty $ such that the individual treatment effect satisfies $ | \tau_i | \leq M_1 $ and the spillover effect satisfies $ \forall \pi \in \Pi, | \delta_i ( \pi ) | \leq M_2 $.  \\
  (WI) Weak independence assumption: For any node indices $i$ and $j$, the weak independence assumption assumes that $ \mathbf{X}_i \bot \mathbf{X}_j \ \text{if} \ A_{ij} = 0 \text{, or} \ \nexists  k \ \text{with} \ A_{ik} = A_{kj} = 1 $. \\
  (LIP) Lipschitz continuity of the spillover effect w.r.t. policy: Given two treatment policies $ \pi_1 $ and $ \pi_2 $, for any node $i$ the spillover effect satisfies $ | \delta_i ( \pi_1 ) - \delta_i ( \pi_2 ) | \leq L || \pi_1 - \pi_2 ||_{ \infty } $, where the Lipschitz constant satisfies $ L > 0$ and $ || \pi_1 - \pi_2 ||_{ \infty } := \sup_{ \mathbf{X} \in \boldsymbol{ \chi } } | \pi_1 ( \mathbf{X} ) - \pi_2 ( \mathbf{X} ) |  $.  \\
  (ES) Uniformly consistency: after fitting experimental or observational data on $ \mathcal{G} $, individual treatment effect estimator satisfies
\begin{equation*}
   \frac{1}{n} \sum_{i=1}^n | \tau_i - \hat{ \tau }_i | < \frac{ \alpha_{\tau} }{ n^{\zeta_{\tau}} }, 
\end{equation*}  
and spillover estimator satisfies
\begin{equation}
  \forall \pi \in \Pi, \ \frac{1}{n} \sum_{i=1}^n | \delta_i ( \pi ) - \hat{ \delta }_i ( \pi ) | < \frac{ \alpha_{ \delta } }{ n^{ \zeta_{\delta} } }
\end{equation}
where $\alpha_{\tau} > 0 $ and $ \alpha_{ \delta } > 0 $ are scaling factors that characterize the errors of estimators. $ \zeta_{\tau} $ and $ \zeta_{ \delta }$ control the convergence rate of estimators for individual treatment effect and spillover effect, respectively, which satisfy $ 0 < \zeta_{\tau}, \zeta_{\delta} < 1 $.  
  \label{ass:assumption}
\end{my_assumption}

Note that the (ES) assumption corresponds to Claim~\ref{claim:estimator}. It assumes a more general convergence rate than  $ \frac{1}{ \sqrt{n} } $ when $ d_{ \max } (n) $ depends on the number of units. Therefore, we use the coefficients $\zeta_{\tau}$ and $\zeta_{\delta}$ to characterize the convergence rates, which is in line with the assumption made in~\cite{athey2017efficient} (see Assumption 2 of~\cite{athey2017efficient}).

Besides, (LIP) assumes that the change of received spillover effect is bounded after modifying the treatment assignments of one unit's neighbors. We will use hypergraph techniques, instead of chromatic number arguments, to give a tighter bound of policy regrets. Another advantage is that the weak independence (WI) assumption can be relaxed to support longer dependencies on the network. However, by relaxing (WI), the power of $ d_{ \max } $ in Theorem~\ref{theorem:regret_bound} and the following Theorem~\ref{theorem:regret_bound_constrained} needs to be modified correspondingly. For example, if we assume a next-nearest neighbors dependency of covariates, i.e., $ \mathbf{X}_i \perp \mathbf{X}_j $ for $ j \not\in {i} \cup \mathcal{N}_i \cup \mathcal{N}_i^{ (2) } $, then the term $ d_{ \max }^2 $ in Theorem~\ref{theorem:regret_bound} and~\ref{theorem:regret_bound_constrained} needs to be modified to $ d_{ \max }^4 $.

Under Assumption~\ref{ass:assumption}, we can derive the following bound. 
\begin{my_theorem}
  By Assumption~\ref{ass:assumption}, for any small $\epsilon > 0$, the policy regret is bounded by $ \mathcal{R} ( \hat{ \pi }_n ) \leq 2 \left( \frac{\alpha_{\tau}}{n^{\zeta_{\tau}}} +  \frac{\alpha_{\delta}}{n^{\zeta_{\delta}}} \right) + 2 \epsilon $ with probability at least $   1 - \mathcal{N} \left( \Pi, \frac{ \epsilon }{ 4 ( 2M_1 + 2M_2 + L ) } \right) \exp \left( - \frac{ n \epsilon^2 }{ 32 ( d_{ \max }^2 + 1 ) ( M_1 + M_2 )^2 } \right) $, where $ \mathcal{N} \left( \Pi, \frac{ \epsilon }{ 4 ( 2M_1 + 2M_2 + L ) } \right) $ indicates the covering number on the functional class $ \Pi $ with radius $ \frac{ \epsilon }{ 4 ( 2M_1 + 2M_2 + L ) } $, and $ d_{ \max } $ is the maximal node degree in the graph $ \mathcal{G} $. 
  \label{theorem:regret_bound}
\end{my_theorem}
\begin{proof}
  Under (WI) and (BO), we can use concentration inequalities of networked random variables defined on a hypergraph, which is derived from graph $ \mathcal{G} $ to bound the convergence rate. Moreover, the Lipschitz assumption (LIP) allows an estimation of the covering number of the policy functional class $ \Pi $. Detailed derivations of policy regret bound is relegated to Appendix I.   
\end{proof}

Suppose that the policy functional class $ \Pi $ is finite and its capacity is bounded by $ | \Pi | $.  According to Theorem~\ref{theorem:regret_bound}, with probability at least $ 1 - \delta $, the policy regret is bounded by $   \mathcal{R} ( \hat{ \pi }_n ) \leq 2 \left( \frac{\alpha_{\tau}}{n^{\zeta_{\tau}}} +  \frac{\alpha_{\delta}}{n^{\zeta_{\delta}}} \right) + 8 ( M_1 + M_2 ) \sqrt{ \frac{2 ( d_{ \max }^2 + 1 ) }{n} \log \frac{ | \Pi | }{ \delta } }  \approx 2 \left( \frac{\alpha_{\tau}}{n^{\zeta_{\tau}}} +  \frac{\alpha_{\delta}}{n^{\zeta_{\delta}}} \right) + 8 d_{ \max } ( M_1 + M_2 ) \sqrt{ \frac{ 2 }{n} \log \frac{ | \Pi | }{ \delta } } $. It indicates that optimal policies are more difficult to find in a dense graph even under weak interactions between neighboring nodes.

In a real-world setting, treatments could be expensive. So the policymaker usually encounters a budget or capacity constraints, e.g., the proportion of patients receiving treatment is limited, and to decide who should be treated under constraints is a challenging problem~\citep{kitagawa2017should}. Through the interference-free welfare function $ A ( \pi ) $,  a policy is trained to make treatment choices using only each individual's features. In contrast, under interference, a smart policy should maximize the utility function Eq.~\eqref{eq:utility_under_interference} by deciding whether to treat an individual or expose it under neighboring treatment effects such that a required constraint can be satisfied. Therefore, in the second part of the experiments, after fitting causal estimators, we investigate policy networks that maximize the utility function $ S(\pi) $ on the graph and satisfy a treatment proportion constraint.

To be more specific, we consider the constraint where only $p_t$ percentage of the population can be assigned to treatment~\footnote{Note that here $p_t$ differs from the treatment probability $p$ from causal structural equations in the randomized experiment setting.}. The corresponding sample-averaged loss function for a policy network $ \pi $ under capacity constraint  is defined as $\mathcal{L}_{ \mathrm{pol} } ( \pi ) := - \hat{S}_n^{ \tau, \delta } ( \pi ) + \gamma ( \frac{ 1 }{ n } \sum_{i=1}^n \pi ( \mathbf{X}_i ) - p_t )$, where $ \gamma $ is a hyperparameter for the constraint. The optimal policy under capacity constraint is obtained by $ \hat{ \pi }_n^{ p_t } \in \min_{ \pi \in \Pi } \mathcal{ L }_{ \mathrm{ pol } } ( \pi ) $.


Let $ \mathcal{R} ( \hat{ \pi }_n^{ p_t } ) $ denote the capacity-constrained policy regret with pre-assigned treatment percentage $p_t$. The upper bound of $ \mathcal{R} ( \hat{ \pi }_n^{ p_t } ) $ is provided in Theorem~\ref{theorem:regret_bound_constrained} and proved in Appendix J. It indicates that if, in the constraint, $p_t$ is small, then the optimal capacity-constrained policy will be challenging to find. Increasing the treatment probability cannot guarantee the improvement of the group's interest due to the non-linear network effect. Therefore, finding the balance between optimal treatment probability, treatment assignment, and group's welfare is a provocative question in social science.

\begin{my_theorem}
  By Assumption~\ref{ass:assumption}, for any small $\epsilon > 0$, the policy regret under the capacity constraint $ p_t $ is bounded by $ \mathcal{R} ( \hat{ \pi }_n^{ p_t } ) \leq 2 \left( \frac{\alpha_{\tau}}{n^{\zeta_{\tau}}} +  \frac{\alpha_{\delta}}{n^{\zeta_{\delta}}} \right) + 2 \epsilon $ with probability at least $ 1 - \mathcal{N} \exp \left( - \frac{ n \epsilon^2 }{ 32 ( d_{ \max }^2 + 1 ) ( M_1 + M_2 )^2 } \right) $, where $ \mathcal{N} := \mathcal{N} \left( \Pi, \frac{ \epsilon }{ 8 [ ( M_1 + M_2 + L ) + \frac{1}{p_t} ( M_1 + M_2 ) ] } \right)$ indicates the covering number on the functional class $ \Pi $ with radius $ \frac{ \epsilon }{ 8 [ ( M_1 + M_2 + L ) + \frac{1}{p_t} ( M_1 + M_2 ) ] } $, and $ d_{ \max } $ is the maximal node degree in the graph $ \mathcal{G} $. 
  \label{theorem:regret_bound_constrained}
\end{my_theorem}

\section{Experiments}
\label{sec:exp}

\begin{table*}[thp]
\centering
\scriptsize
\begin{tabular}{c | c c | c c }
  & \multicolumn{2}{c}{Wave1} & \multicolumn{2}{c}{Pokec} \\
  & $ \sqrt{ \mathcal{ MSE } } $ & $ \epsilon_{ PEHE } $ & $ \sqrt{ \mathcal{ MSE } } $ & $ \epsilon_{ PEHE } $ \\
  \hline
DA GB                     & $0.721 \pm 0.054$ & $0.289 \pm 0.061$ & $0.713 \pm 0.016$ & $0.321 \pm 0.057$ \\
DA RF                     & $1.037 \pm 0.122$ & $0.790 \pm 0.215$ & $0.749 \pm 0.023$ & $0.840 \pm 0.087$ \\
DR GB                     & $0.831 \pm 0.109$ & $0.499 \pm 0.185$ & $0.686 \pm 0.020$ & $0.275 \pm 0.051$ \\
DR EN                     & $0.929 \pm 0.091$ & $0.733 \pm 0.135$ & $0.695 \pm 0.019$ & $0.247 \pm 0.060$ \\
GPS                       & $0.238 \pm 0.012$ & $0.150 \pm 0.047$ & $0.329 \pm 0.010$ & $0.147 \pm 0.010$ \\
  \hline
GCN + $ \hat{HSIC}^{\Phi / \mathit{GNN}}$       & $0.192 \pm 0.019$ & $0.047 \pm 0.018$ & $0.305 \pm 0.011$ & $0.136 \pm 0.009$ \\
GraphSAGE + $ \hat{HSIC}^{\Phi / \mathit{GNN}}$ & $0.181 \pm 0.016$ & $0.042 \pm 0.020$ & $0.303 \pm 0.008$ & $\mathbf{0.123 \pm 0.003}$ \\
$1$-GNN + $ \hat{HSIC}^{\Phi / \mathit{GNN}}$   & $\mathbf{0.176 \pm 0.011}$ & $\mathbf{0.035 \pm 0.011}$ & $\mathbf{0.302 \pm 0.004}$ & $0.130 \pm 0.006$ \\
  \hline
  \textbf{Improve}       &  $ 26.1 \% $  &  $ 76.7 \% $  &  $ 8.2 \% $  &  $ 16.3 \% $  \\  
\hline
\end{tabular}
  \caption{Experimental results of randomized experiments on the Wave1 and Pokec datasets using linear response generation function $G_0$. For Wave1, we set (node degree) $ k=10 $, (decay parameter)$ \alpha = 0.5 $, and (treatment probability) $ p = 0.1 $, and for Pokec $ p = 0.1 $. Improvements are obtained by comparing with the best baselines.}
  \label{tab:exp_wave_pokec}
\end{table*}
\normalsize

\subsection{Datasets}
\label{sec:datasets}

The difficulties of evaluating the performance of the proposed estimators lie in the broad set of missing outcomes under counterfactual inference. Therefore, we conduct randomized experiments on two semi-synthetic datasets with \emph{ground-truth} response generation functions, and observational studies on one real dataset with \emph{unknown} treatment assignment and response generation functions. Notably, in the randomized experiment setting, we consider a linear response generation function inspired by Eq. 5 of~\citep{toulis2013estimation}, $ G_0 :  Y_i = Y_i (T_i=0, \mathcal{G} = \emptyset ) + T_i \tau ( \mathbf{X}_i ) + \delta_i ( \mathbf{X}, \mathbf{T}, \mathcal{G} ) + \epsilon_{Y_i} $, where $ Y_i (T_i=0, \mathcal{G} = \emptyset ) $ is the outcome under control and without network interference, and $ \epsilon_{Y_i} $ represents Gaussian noise. $ \tau ( \mathbf{X}_i ) $ and $ \delta_i ( \mathbf{X}, \mathbf{T}, \mathcal{G} ) $ represent individual treatment effect and spillover effect, respectively, whose forms are dataset-dependent and discussed below.


To further investigate the superiority of the GNN-based causal estimators on nonlinear causal responses, we also consider nonlinear data generation functions inspired by Section 4.2 of~\citep{toulis2013estimation}. For instance, a generation approach $ G_1 $ contains the quadratic spillover effect $  \delta^2_i ( \mathbf{X}, \mathbf{T}, \mathcal{G} ) $, or a more complicated approach $ G_2 $ with an additional interaction term between individual treatment and spillover effect, namely $ \tau ( \mathbf{X}_i ) \delta_i ( \mathbf{X}, \mathbf{T}, \mathcal{G} ) $. Detailed expressions of approaches $G_1$ and $G_2$ can be fund in Appendix E.

\textbf{Wave1 } Wave1 is an in-school questionnaire data collected through the National Longitudinal Study of Adolescent Health project~\citep{chantala1999national}. The questionnaire contains questions such as age, grade, health insurance, etc. Due to the anonymity of Wave1, we use the symmetrized $k$-NN graph derived from the questionnaire data as the friendship network. In our experiments, we choose $k=10$, and the resulting friendship network has $5,578$ nodes and $100,158$ links. We assume a randomized experiment conducted on the friendship network which describes students' improvements of performance through assigning to a tutoring program or through the peer effect. Hence $ Y_i (T_i=0, \mathcal{G} = \emptyset ) $ represents the overall performance of student $i$ before assignment to a tutoring program and before being exposed to peer influences, $\tau( \mathbf{X}_i )$ the simulated performance difference after an assignment, and $ \delta_i ( \mathbf{X}, \mathbf{T}, \mathcal{G} ) $ the synthetic peer effect. Exact forms of $ Y_i (T_i=0, \mathcal{G} = \emptyset ) $ and $ \tau( \mathbf{X}_i ) $ depend nonlinearly on the features of each student. Moreover, the first-order peer effect is simulated as $ \delta_i ( \mathbf{X}, \mathbf{T}, \mathcal{G} ) := \alpha \frac{ 1 }{ | \mathcal{N}_i | } \sum\limits_{ j \in \mathcal{N}_i } T_j \tau ( \mathbf{X}_j ) $, where the decay parameter $ \alpha $ characterizes the decay of influence. For randomized experiments presented here, we randomly assign $ 10 \% $ of the population to the treatment, creating an under-treated population. Details of the generating process and more experiment results with different settings are relegated to Appendix C and G.

\textbf{Pokec } The friendship network derived from the Wave1 questionnaire data may violate the power-law degree distribution of real networks. Hence, we further conduct experiments on the real social network Pokec~\citep{takac2012data} with generated responses. Pokec is an online social network in Slovakia with profile data, including age, gender, education, etc. We consider randomized experiments on the Pokec social network, in which personalized advertisements of a new health medicine are pushed to some users. We assume that the response of exposed users to the advertisement only depends on a few properties, such as age, weight, smoking status, etc. We keep profiles with complete information on these properties, and the resulting Pokec social network contains $11,623$ nodes and $76,752$ links. Let $ Y_i (T_i=0, \mathcal{G} = \emptyset ) $ represent the purchase of this new health medicine without external influence on the decision, $ \tau ( \mathbf{X}_i ) $ the purchase difference after seeing the advertisement, $ \delta_i ( \mathbf{X}, \mathbf{T}, \mathcal{G} ) $ the purchase difference due to social influences. For randomized experiments on the Pokec social network, we also consider peer effects from next-nearest neighbors by defining $ \delta_i ( \mathbf{X}, \mathbf{T}, \mathcal{G}) := \alpha \frac{1}{ | \mathcal{N}_i | } \sum\limits_{ j \in \mathcal{N}_i }  T_j \tau ( \mathbf{X}_j ) + \alpha^2  \frac{1}{ | \mathcal{N}_i^{ (2) } | } \sum\limits_{ k \in \mathcal{N}_i^{ (2) } }  T_k \tau ( \mathbf{X}_k ) $, where the decay parameter $\alpha$ characterizes the decay of influence. Details and more experimental results with different hyperparameter settings are given in Appendix D and G.

\textbf{Amazon } The co-purchase dataset from Amazon contains product details, review information, and a list of similar products. Therefore, there is a directed network of products that describes whether a substitutable or complementary product is getting co-purchased with another product~\citep{leskovec2007dynamics}. To study the causal effect of reviews on the sales of products, \citep{rakesh2018linked} generates a dataset containing products with only positive reviews from the Amazon co-purchase dataset, named as pos Amazon, and Amazon for short. In this dataset, all items have positive reviews, i.e., the average rating is larger than $3$, and one item is considered to be treated if there are more than three reviews under this item; otherwise, an item is in the control group. In this setting, pos Amazon is an over-treated dataset with more than $70 \%$ of products being in the treatment group. Word2vec embedding of an item's review serves as the feature vector of this item. Moreover, the individual treatment effect of an item is approximated by matching it to other items having similar features and under minimal exposure to neighboring nodes' treatments.

\subsection{Results of Causal Estimators}

\paragraph{Evaluation Metrics} 

One evaluation metric is the square root of MSE for the prediction of the observed outcomes on the test dataset $ \mathcal{U}_T $, which is defined as $ \sqrt{ \mathcal{ MSE } } := \sqrt{ \frac{1}{ | \mathcal{U}_T | } \sum_{i \in \mathcal{U}_T } ( Y_i - h_{ T_i } )^2 } $, where $ h_{ T_i } $ denotes the output of the outcome prediction network (see $h_0$ and $h_1$ in Fig.~\ref{fig:model}). This metric reflects how well an estimator can predict the superimposed individual treatment and spillover effects on a network. Another evaluation metric that quantifies the quality of extracted individual treatment effect is the Precision in Estimation of Heterogeneous Effect studied in~\citep{hill2011bayesian}, which is defined as $ \epsilon_{ PEHE } := \frac{1}{ | \mathcal{U}_T | } \sum_{ i \in \mathcal{U}_T } ( \tau ( \mathbf{X}_i ) - \hat{ \tau } ( \mathbf{X}_i ) )^2 $, where $ \hat{ \tau } ( \mathbf{X}_i ) $ is defined in Eq.~\eqref{eq:tau_estimation}.

\paragraph{Baselines}

Baseline models are domain adaption method~\citep{kunzel2019metalearners} with gradient boosting regression (\textbf{DA GB}), with random forest regression (\textbf{DA RF}), doubly-robust estimator~\citep{funk2011doubly} with gradient boosting regression (\textbf{DR GB}), and elastic net regression (\textbf{DR EN}). They are implemented via EconML~\citep{econml} with grid-searched hyperparameters. These baselines incorporate the feature vectors as inputs and exposure as the control variable into the model. For randomized experiments on Wave1 and Pokec, the predefined treatment probability $p$ is provided, while for the observational studies on the Amazon dataset, the covariate-dependent treatment probability is estimated. Moreover, the generalized propensity score (GPS) method is reproduced and enhanced for a fair comparison, equipped with the same feature map $ \Phi $ function. More details of baselines, the sketch of the training procedure, and hyperparameters are relegated to Appendix G.

\begin{table}[thp]
\centering
\scriptsize
\begin{tabular}{c | c c }
  &  $ \sqrt{ \mathcal{ MSE } } $  &  $ \epsilon_{ PEHE } $  \\
  \hline
  DA GB  &  $0.601 \pm 0.007$  &  $1.370 \pm 0.016$  \\
  DA RF  &  $0.604 \pm 0.019$  &  $1.398 \pm 0.013$  \\
  DR GB  &  $0.615 \pm 0.022$  &  $1.222 \pm 0.020$  \\
  DR EN  &  $1.104 \pm 0.001$  &  $1.929 \pm 0.003$  \\
  GPS    &  $0.399 \pm 0.003$  &  $1.968 \pm 0.025$  \\
  \hline 
  GCN                                 &  $0.312 \pm 0.002$  &  $2.400 \pm 0.201$  \\
  GCN + $\hat{HSIC}^{\textit{GNN}}$            &  $0.303 \pm 0.006$  &  $1.881 \pm 0.076$  \\
  GCN + $\hat{HSIC}^{\Phi}$           &  $0.301 \pm 0.002$  &  $1.531 \pm 0.024$  \\
  GraphSAGE                           &  $0.305 \pm 0.001$  &  $1.984 \pm 0.026$  \\
  GraphSAGE + $\hat{HSIC}^{\textit{GNN}}$      &  $0.296 \pm 0.002$  &  $1.567 \pm 0.051$  \\
  GraphSAGE + $\hat{HSIC}^{\Phi}$     &  $0.300 \pm 0.002$  &  $1.358 \pm 0.025$  \\
  $1$-GNN                             &  $0.279 \pm 0.000$  &  $1.512 \pm 0.111$  \\
  $1$-GNN + $\hat{HSIC}^{\textit{GNN}}$        &  $\mathbf{0.276 \pm 0.002}$  &  $1.434 \pm 0.030$  \\
  $1$-GNN + $\hat{HSIC}^{\Phi}$       &  $\mathbf{0.277 \pm 0.002}$  &  $\mathbf{1.098 \pm 0.031}$  \\
  \hline
  \textbf{Improve}    &  $ 30.8 \% $  &  $ 10.1 \% $  \\ 
  \hline
\end{tabular}
  \caption{Experimental result on the pos Amazon dataset without representation balancing and under different imbalance penalties.}
  \label{tab:exp_amazon}
\end{table} 
\normalsize

\paragraph{Experiments}

We use partial outcomes, both in the randomized experiments and observational settings, to train the GNN-based causal estimators. We investigate the effect of penalizing representation imbalance in the observational studies on the Amazon dataset. The entire data points $( \mathbf{X}_i, T_i, G_i, Y_i )$ are randomly divided into training ($ 80 \% $), validation ($ 5 \% $), and test ($ 15 \% $) sets. Note that the entire network $ \mathcal{G} $ and the covariates of all units $ \mathbf{X} $ are given during the training and test, while only the causal responses of units in the training set are provided in the training phase. For the randomized experiments using the Wave1 and Pokec datasets, we repeat the experiments $3$ times and use different random parameters in the response generation process each time.

Experimental results on the Wave1 and Pokec data generated via linear model $G_0$ are presented in Table~\ref{tab:exp_wave_pokec}. Both representation balancing $ \hat{HSIC}^{ \Phi} $ and $ \hat{HSIC}^{ \mathit{GNN} } $ are deployed in the GNN-based estimators for searching for the best performance. GNN-based estimators, especially the $1$-GNN estimator, are superior for superimposed causal effects prediction. One can observe a $ 26.1 \% $ improvement of the $ \sqrt{ \mathcal{MSE} } $ metric on the Wave1 dataset when comparing the $1$-GNN estimator with the enhanced GPS method and a $ 8.2 \% $ improvement on the Pokec dataset. The covariates of neighboring units in the Pokec dataset actually have strong cosine similarity, hence the improvement on the Pokec dataset is not significant, and the network effect can be approximately captured from the exposure variable. Table~\ref{tab:exp_amazon} shows the experimental results on the pos Amazon dataset in the observational study. In particular, we demonstrate the effects of without representation penalty, and with different penalties. It shows that representation penalties can significantly improve the individual treatment effect recovery, serving as a regularization to avoid over-fitting the network interference. Furthermore, GNN-based estimators using $ \hat{HSIC}^{\textit{GNN}} $ penalty are slightly better than those using $ \hat{HSIC}^{\Phi} $ penalty; however, by sacrificing the metric $ \epsilon_{PEHE} $.

\begin{table*}[thp]
\centering
\scriptsize
\begin{tabular}{c | c c | c c }
  &  \multicolumn{2}{c}{Wave1}  &  \multicolumn{2}{c}{Pokec}  \\
  &  $ \Delta \hat{ S } ( \hat{ \pi }_n^{ p_t } ) $  &  $ \Delta  S ( \hat{ \pi }_n^{ p_t } ) $  &  $ \Delta \hat{ S } ( \hat{ \pi }_n^{ p_t } ) $  &  $ \Delta  S ( \hat{ \pi }_n^{ p_t } ) $  \\
  \hline
DA GB         &  $0.276 \pm 0.033$  &  $0.002 \pm 0.025$  &  $0.231 \pm 0.051$  &  $0.001 \pm 0.036$  \\
DA RF         &  $0.302 \pm 0.029$  &  $0.003 \pm 0.021$  &  $0.198 \pm 0.080$  &  $0.001 \pm 0.057$  \\
DR GB         &  $0.322 \pm 0.023$  &  $0.002 \pm 0.019$  &  $0.338 \pm 0.060$  &  $0.002 \pm 0.046$  \\
DR EN         &  $0.311 \pm 0.019$  &  $0.001 \pm 0.018$  &  $0.329 \pm 0.028$  &  $0.001 \pm 0.026$  \\
GPS           &  $0.235 \pm 0.042$  &  $0.004 \pm 0.032$  &  $0.362 \pm 0.069$  &  $0.001 \pm 0.053$  \\
\hline
GCN           &  $0.260 \pm 0.024$  &  $0.163 \pm 0.020$  &  $0.270 \pm 0.007$  &  $0.190 \pm 0.012$  \\
GraphSAGE     &  $0.283 \pm 0.031$  &  $0.176 \pm 0.025$  &  $0.376 \pm 0.049$  &  $0.211 \pm 0.034$  \\
$1$-GNN       &  $0.327 \pm 0.038$  &  $\mathbf{0.208 \pm 0.026}$  &  $0.377 \pm 0.041$  &  $\mathbf{0.225 \pm 0.031}$  \\
\hline
\end{tabular}
  \caption{Intervention policy improvements on the Wave1 and Pokec semi-synthetic datasets under treatment capacity constraint with $p_t=0.3$. $ \Delta \hat{ S } ( \hat{ \pi }_n^{ p_t } ) $ and $  \Delta S ( \hat{ \pi }_n^{ p_t } ) $ represent utility differences evaluated from learned estimators and ground truth, respectively. Note that \emph{only} $ \Delta S ( \hat{ \pi }_n^{ p_t } ) $  reflects the real policy improvement. }
  \label{tab:poly_diff_wave_pokec_tp0.3}
\end{table*}  
\normalsize

\begin{table*}[thp]
\centering 
\scriptsize
\begin{tabular}{c | c c c | c c c c}
  &  DA GB &  DA RF  &  GPS  &  GCN  & GraphSAGE  &  $1$-GNN  \\
  \hline
  $ \Delta \hat{S} ( \hat{ \pi }_n^{ p_t } ) $  &  $38.9 \pm 1.1$  &  $84.1 \pm 2.3$  &  $ \underline{98.6 \pm 10.8} $  &  $80.7 \pm 0.9$  &  $ \mathbf{ 86.0 \pm 0.9 } $  &  $84.1 \pm 1.3$ \\
  \hline
\end{tabular}
  \caption{Intervention policy improvements on the pos Amazon dataset under treatment capacity constraint with $p_t=0.5$.   Only domain adaption methods and GPS are compared since they are the best baseline estimators according to Table~\ref{tab:exp_amazon}. }
  \label{tab:poly_diff_amazon}
\end{table*}

The performance of GNN-based causal estimators on nonlinear response models generated by $G_1$ and $G_2$ with different hyperparameters are reported in Appendix E. In general, for the $ \sqrt{ \mathcal{MSE} } $ metric, GNN-based estimators outperform the best baseline GPS dramatically, showing the effectiveness of predicting the superimposed causal responses even under nonlinear generation mechanisms. Moreover, a significant performance improvement on the $ \epsilon_{PEHE}$ metric with the Wave1 and Pokec datasets shows that setting an empty graph, i.e., $ \mathcal{G} = \emptyset $, in the GNN-based estimators according to Eq.~\ref{eq:tau_estimation} is an appropriate approach for disentangling and extracting individual causal effects from interfered causal responses (see Table 8 and 9 in Appendix E.).

\subsection{Results on Improved Intervention Policy}

\paragraph{Experiment Settings}

After obtaining the optimal causal effect estimators and feature map $ \Phi $ (see Fig.~\ref{fig:model}), we subsequently optimize intervention policy on the same graph. A neural network having two hidden layers, with ReLU activation between hidden layers and sigmoid activation at the end, is employed as the policy network. The output of the policy network lies in $[0, 1]$, and it is interpreted as the probability of treating a node. The real intervention choice is then sampled from this probability via the Gumbel-softmax trick~\citep{jang2016categorical} such that gradients can be back-propagated. Sampled treatment choices along with corresponding node features are then fed into the feature map $ \Phi $ and subsequent causal estimators to evaluate the utility function under network interference defined in Eq.~\eqref{eq:utility_under_interference}. Each experiment setting is repeated $5$ times until convergence. The hyperparameter $ \gamma $ in $ \mathcal{ L }_{ \mathrm{pol} } $ is tuned such that the constraint for the percentage $ p_t $ is satisfied within the tolerance $ \pm 0.01 $. More details of experiment settings and hyperparameters are relegated to Appendix G.3 and F.

To quantify the optimized policy $ \hat{ \pi }_n^{ p_t } $, we evaluate the difference
\begin{equation*}
    \Delta \hat{ S } ( \hat{ \pi }_n^{ p_t }  ) := \hat{ S }_n^{ \tau, \delta } ( \hat{ \pi }_n^{ p_t } ) - \hat{ S }_n^{ \tau, \delta } ( \pi_{ R }^{ p_t } ), 
\end{equation*}
where $ \pi_{ R }^{ p_t } $ represents a randomized intervention underlying the same capacity constraint. The difference $ \Delta \hat{ S } ( \hat{ \pi }_n^{ p_t }  ) $ indicates how a learned policy can outperform a randomized policy with the same constraint evaluated via learned causal effect estimators. However, from its definition, it is concerned that the policy improvement $ \hat{ \pi }_n^{ p_t } $ may be very biased, such that any ``expected improvement'' may come from the inaccurate causal estimators. Hence, for the Wave1 and Pokec datasets, knowing the generating process of treatment and spillover effects, we also compare the actual utility difference
\begin{equation*}
     \Delta  S ( \hat{ \pi }_n^{ p_t }  ) := S_n^{ \tau, \delta } ( \hat{ \pi }_n^{ p_t } ) - S_n^{ \tau, \delta } ( \pi_{ R }^{ p_t } ). 
\end{equation*}

Table~\ref{tab:poly_diff_wave_pokec_tp0.3} displays policy optimization results on the under-treated Wave1 and Pokec simulation datasets, where initially only $10 \%$ of nodes are randomly assigned to treatment. It shows that an optimized policy network cannot even outperform a randomized policy in ground truth when the causal estimators perform poorly. Hence, policy networks learned from the utility function with plugged in doubly-robust or domain adaption estimators are not reliable. By contrast, the small difference between genuine utility improvement $ \Delta  S ( \hat{ \pi }_n^{ p_t } ) $ and estimated improvement $ \Delta \hat{S} ( \hat{ \pi }_n^{ p_t } ) $ for the GNN-based causal estimators indicates the reliability of the optimized policy. Moreover, comparing the ground-truth utility improvement on GPS and GCN-based estimator shows that the policy network sensitively relies on the accuracy of the employed causal estimator. Furthermore, one might argue that through baseline estimators, a simple policy network cannot adjust its treatment choice according to neighboring nodes' features and responses, unlike through GNN-based estimators. For a fair comparison, in Appendix F, we also provide experimental results using a GNN-based policy network. However, we still cannot observe genuine utility improvements on $ \Delta  S ( \hat{ \pi }_n^{ p_t } ) $ when using baseline models as causal estimators.

Next, we conduct experiments for intervention policy learning on the over-treated pos Amazon dataset under treatment capacity constraint. Since we do not have access to the ground truth of the pos Amazon dataset, Table~\ref{tab:poly_diff_amazon} shows the utility difference under treatment capacity constraint with $p_t=0.5$ evaluated only from learned causal estimators. Although the optimized utility improvement $ \Delta \hat{S} ( \hat{ \pi }_n^{ p_t } ) $ achieves the best result via the GPS causal estimator, it might be unreliable compared to the ground truth. A reliable policy improvement having comparable utility improvement via a GNN-based causal estimator is expected.

\section{Conclusion}

In this work, we first introduced the task of causal inference under general network interference and proposed causal effect estimators using GNNs of various types. We also defined a novel utility function for policy optimization on interconnected nodes, of which a graph-dependent policy regret bound can be derived. We conduct experiments on semi-synthetic simulation and real datasets. Experiment results show that GNN-based causal effect estimators with an HSIC discrepancy penalty, are superior in superimposed causal effect prediction, and the individual treatment effect can be recovered reasonably well. Subsequent experiments of intervention policy optimization under capacity constraint further confirm the importance of employing an optimal and reliable causal estimator for policy improvement. In future works, we will consider causal effects on partially observable and dynamic networks.


\newpage

\textbf{Acknowledgement} \\
We appreciate the fruitful discussion with Dr. Yuyi Wang. This project is supported by the Cognitive Deep Learning project funded by Siemens CT. This work has also been funded by the German Federal Ministry of Education and Research (BMBF) under Grant No. 01IS18036A. The authors of this work take full responsibilities for its content.
\begin{figure}[h]
    \includegraphics[width=0.2\linewidth]{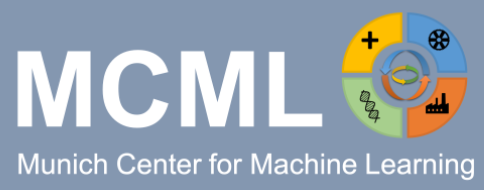}
\end{figure}
\bibliography{main}

\begin{thebibliography}{45}
\providecommand{\natexlab}[1]{#1}
\providecommand{\url}[1]{\texttt{#1}}
\expandafter\ifx\csname urlstyle\endcsname\relax
  \providecommand{\doi}[1]{doi: #1}\else
  \providecommand{\doi}{doi: \begingroup \urlstyle{rm}\Url}\fi

\bibitem[Arbour et~al.(2016)Arbour, Garant, and Jensen]{arbour2016inferring}
D.~Arbour, D.~Garant, and D.~Jensen.
\newblock Inferring network effects from observational data.
\newblock In \emph{Proceedings of the 22nd ACM SIGKDD International Conference
  on Knowledge Discovery and Data Mining}, pages 715--724. ACM, 2016.

\bibitem[Aronow et~al.(2017)Aronow, Samii, et~al.]{aronow2017estimating}
P.~M. Aronow, C.~Samii, et~al.
\newblock Estimating average causal effects under general interference, with
  application to a social network experiment.
\newblock \emph{The Annals of Applied Statistics}, 11\penalty0 (4):\penalty0
  1912--1947, 2017.

\bibitem[Athey and Wager(2017)]{athey2017efficient}
S.~Athey and S.~Wager.
\newblock Efficient policy learning.
\newblock \emph{arXiv preprint arXiv:1702.02896}, 2017.

\bibitem[Bartlett et~al.(2005)Bartlett, Bousquet, Mendelson,
  et~al.]{bartlett2005local}
P.~L. Bartlett, O.~Bousquet, S.~Mendelson, et~al.
\newblock Local rademacher complexities.
\newblock \emph{The Annals of Statistics}, 33\penalty0 (4):\penalty0
  1497--1537, 2005.

\bibitem[Bowers et~al.(2013)Bowers, Fredrickson, and
  Panagopoulos]{bowersFP2013}
J.~Bowers, M.~M. Fredrickson, and C.~Panagopoulos.
\newblock Reasoning about interference between units: {A} general framework.
\newblock \emph{Political Analysis}, 21:\penalty0 97--124, 2013.
\newblock \doi{10.1093/pan/mps038}.

\bibitem[Chantala and Tabor(1999)]{chantala1999national}
K.~Chantala and J.~Tabor.
\newblock National longitudinal study of adolescent health: Strategies to
  perform a design-based analysis using the add health data.
\newblock 1999.

\bibitem[Cucker and Zhou(2007)]{cucker2007learning}
F.~Cucker and D.~X. Zhou.
\newblock \emph{Learning theory: an approximation theory viewpoint}, volume~24.
\newblock Cambridge University Press, 2007.

\bibitem[Devroye et~al.(2013)Devroye, Gy{\"o}rfi, and
  Lugosi]{devroye2013probabilistic}
L.~Devroye, L.~Gy{\"o}rfi, and G.~Lugosi.
\newblock \emph{A probabilistic theory of pattern recognition}, volume~31.
\newblock Springer Science \& Business Media, 2013.

\bibitem[Forastiere et~al.(2016)Forastiere, Airoldi, and
  Mealli]{forastiere2016identification}
L.~Forastiere, E.~M. Airoldi, and F.~Mealli.
\newblock Identification and estimation of treatment and interference effects
  in observational studies on networks.
\newblock \emph{arXiv preprint arXiv:1609.06245}, 2016.

\bibitem[Funk et~al.(2011)Funk, Westreich, Wiesen, St{\"u}rmer, Brookhart, and
  Davidian]{funk2011doubly}
M.~J. Funk, D.~Westreich, C.~Wiesen, T.~St{\"u}rmer, M.~A. Brookhart, and
  M.~Davidian.
\newblock Doubly robust estimation of causal effects.
\newblock \emph{American journal of epidemiology}, 173\penalty0 (7):\penalty0
  761--767, 2011.

\bibitem[Gretton et~al.(2005)Gretton, Bousquet, Smola, and
  Sch{\"o}lkopf]{gretton2005measuring}
A.~Gretton, O.~Bousquet, A.~Smola, and B.~Sch{\"o}lkopf.
\newblock Measuring statistical dependence with hilbert-schmidt norms.
\newblock In \emph{International conference on algorithmic learning theory},
  pages 63--77. Springer, 2005.

\bibitem[Hamilton et~al.(2017)Hamilton, Ying, and
  Leskovec]{hamilton2017inductive}
W.~Hamilton, Z.~Ying, and J.~Leskovec.
\newblock Inductive representation learning on large graphs.
\newblock In \emph{Advances in Neural Information Processing Systems}, pages
  1024--1034, 2017.

\bibitem[Harris and Udry(2018)]{harris2018national}
K.~M. Harris and J.~R. Udry.
\newblock National longitudinal study of adolescent to adult health (add
  health), 1994-2008 [public use].
\newblock \emph{Ann Arbor, MI: Carolina Population Center, University of North
  Carolina-Chapel Hill [distributor], Inter-university Consortium for Political
  and Social Research [distributor]}, pages 08--06, 2018.

\bibitem[Hill(2011)]{hill2011bayesian}
J.~L. Hill.
\newblock Bayesian nonparametric modeling for causal inference.
\newblock \emph{Journal of Computational and Graphical Statistics}, 20\penalty0
  (1):\penalty0 217--240, 2011.

\bibitem[Hudgens and Halloran(2008)]{hudgens2008toward}
M.~G. Hudgens and M.~E. Halloran.
\newblock Toward causal inference with interference.
\newblock 103\penalty0 (482), 2008.

\bibitem[Jang et~al.(2017)Jang, Gu, and Poole]{jang2016categorical}
E.~Jang, S.~Gu, and B.~Poole.
\newblock Categorical reparameterization with gumbel-softmax.
\newblock \emph{International Conference on Learning Representations (ICLR)},
  2017.

\bibitem[Janson(2004)]{janson2004large}
S.~Janson.
\newblock Large deviations for sums of partly dependent random variables.
\newblock \emph{Random Structures \& Algorithms}, 24\penalty0 (3):\penalty0
  234--248, 2004.

\bibitem[Johansson et~al.(2016)Johansson, Shalit, and
  Sontag]{johansson2016learning}
F.~Johansson, U.~Shalit, and D.~Sontag.
\newblock Learning representations for counterfactual inference.
\newblock In \emph{International conference on machine learning}, pages
  3020--3029, 2016.

\bibitem[Kallus(2018)]{kallus2018balanced}
N.~Kallus.
\newblock Balanced policy evaluation and learning.
\newblock In \emph{Advances in Neural Information Processing Systems}, pages
  8895--8906, 2018.

\bibitem[Kallus and Zhou(2018)]{kallus2018confounding}
N.~Kallus and A.~Zhou.
\newblock Confounding-robust policy improvement.
\newblock In \emph{Advances in Neural Information Processing Systems}, pages
  9269--9279, 2018.

\bibitem[Kipf and Welling(2017)]{kipf2016semi}
T.~N. Kipf and M.~Welling.
\newblock Semi-supervised classification with graph convolutional networks.
\newblock \emph{International Conference on Learning Representations (ICLR)},
  2017.

\bibitem[Kitagawa and Tetenov(2017)]{kitagawa2017should}
T.~Kitagawa and A.~Tetenov.
\newblock Who should be treated? empirical welfare maximization methods for
  treatment choice.
\newblock Technical report, Cemmap working paper, 2017.

\bibitem[Kitagawa and Tetenov(2018)]{kitagawa2018should}
T.~Kitagawa and A.~Tetenov.
\newblock Who should be treated? empirical welfare maximization methods for
  treatment choice.
\newblock \emph{Econometrica}, 86\penalty0 (2):\penalty0 591--616, 2018.

\bibitem[K{\"u}nzel et~al.(2019)K{\"u}nzel, Sekhon, Bickel, and
  Yu]{kunzel2019metalearners}
S.~R. K{\"u}nzel, J.~S. Sekhon, P.~J. Bickel, and B.~Yu.
\newblock Metalearners for estimating heterogeneous treatment effects using
  machine learning.
\newblock \emph{Proceedings of the National Academy of Sciences}, 116\penalty0
  (10):\penalty0 4156--4165, 2019.

\bibitem[Leskovec et~al.(2007)Leskovec, Adamic, and
  Huberman]{leskovec2007dynamics}
J.~Leskovec, L.~A. Adamic, and B.~A. Huberman.
\newblock The dynamics of viral marketing.
\newblock \emph{ACM Transactions on the Web (TWEB)}, 1\penalty0 (1):\penalty0
  5, 2007.

\bibitem[Liu and Hudgens(2014)]{liu2014large}
L.~Liu and M.~G. Hudgens.
\newblock Large sample randomization inference of causal effects in the
  presence of interference.
\newblock \emph{Journal of the american statistical association}, 109\penalty0
  (505):\penalty0 288--301, 2014.

\bibitem[Manski(2009)]{manski2009identification}
C.~F. Manski.
\newblock \emph{Identification for prediction and decision}.
\newblock Harvard University Press, 2009.

\bibitem[Morris et~al.(2019)Morris, Ritzert, Fey, Hamilton, Lenssen, Rattan,
  and Grohe]{morris2018weisfeiler}
C.~Morris, M.~Ritzert, M.~Fey, W.~L. Hamilton, J.~E. Lenssen, G.~Rattan, and
  M.~Grohe.
\newblock Weisfeiler and leman go neural: Higher-order graph neural networks.
\newblock In \emph{Proceedings of the AAAI Conference on Artificial
  Intelligence}, volume~33, pages 4602--4609, 2019.

\bibitem[Ogburn et~al.(2017)Ogburn, Sofrygin, Diaz, and van~der
  Laan]{ogburn2017causal}
E.~L. Ogburn, O.~Sofrygin, I.~Diaz, and M.~J. van~der Laan.
\newblock Causal inference for social network data.
\newblock \emph{arXiv preprint arXiv:1705.08527}, 2017.

\bibitem[Rakesh et~al.(2018)Rakesh, Guo, Moraffah, Agarwal, and
  Liu]{rakesh2018linked}
V.~Rakesh, R.~Guo, R.~Moraffah, N.~Agarwal, and H.~Liu.
\newblock Linked causal variational autoencoder for inferring paired spillover
  effects.
\newblock In \emph{Proceedings of the 27th ACM International Conference on
  Information and Knowledge Management}, pages 1679--1682. ACM, 2018.

\bibitem[Research(2019)]{econml}
M.~Research.
\newblock {EconML}: {A Python Package for ML-Based Heterogeneous Treatment
  Effects Estimation}.
\newblock https://github.com/microsoft/EconML, 2019.
\newblock Version 0.x.

\bibitem[Rubin(1974)]{rubin1974estimating}
D.~B. Rubin.
\newblock Estimating causal effects of treatments in randomized and
  nonrandomized studies.
\newblock \emph{Journal of educational Psychology}, 66\penalty0 (5):\penalty0
  688, 1974.

\bibitem[Rubin(1980)]{rubin1980randomization}
D.~B. Rubin.
\newblock Randomization analysis of experimental data: The fisher randomization
  test comment.
\newblock \emph{Journal of the American Statistical Association}, 75\penalty0
  (371):\penalty0 591--593, 1980.

\bibitem[Scarselli et~al.(2008)Scarselli, Gori, Tsoi, Hagenbuchner, and
  Monfardini]{scarselli2008computational}
F.~Scarselli, M.~Gori, A.~C. Tsoi, M.~Hagenbuchner, and G.~Monfardini.
\newblock Computational capabilities of graph neural networks.
\newblock \emph{IEEE Transactions on Neural Networks}, 20\penalty0
  (1):\penalty0 81--102, 2008.

\bibitem[Shalit et~al.(2017)Shalit, Johansson, and
  Sontag]{shalit2017estimating}
U.~Shalit, F.~D. Johansson, and D.~Sontag.
\newblock Estimating individual treatment effect: generalization bounds and
  algorithms.
\newblock In \emph{Proceedings of the 34th International Conference on Machine
  Learning-Volume 70}, pages 3076--3085. JMLR. org, 2017.

\bibitem[Splawa-Neyman et~al.(1990)Splawa-Neyman, Dabrowska, and
  Speed]{splawa1990application}
J.~Splawa-Neyman, D.~M. Dabrowska, and T.~Speed.
\newblock On the application of probability theory to agricultural experiments.
  essay on principles. section 9.
\newblock \emph{Statistical Science}, pages 465--472, 1990.

\bibitem[Takac and Zabovsky(2012)]{takac2012data}
L.~Takac and M.~Zabovsky.
\newblock Data analysis in public social networks.
\newblock In \emph{International Scientific Conference and International
  Workshop Present Day Trends of Innovations}, volume~1, 2012.

\bibitem[Tchetgen and VanderWeele(2012)]{tchetgen2012causal}
E.~J.~T. Tchetgen and T.~J. VanderWeele.
\newblock On causal inference in the presence of interference.
\newblock \emph{Statistical methods in medical research}, 21\penalty0
  (1):\penalty0 55--75, 2012.

\bibitem[Tchetgen et~al.(2017)Tchetgen, Fulcher, and
  Shpitser]{tchetgen2017auto}
E.~J.~T. Tchetgen, I.~Fulcher, and I.~Shpitser.
\newblock Auto-g-computation of causal effects on a network.
\newblock \emph{arXiv preprint arXiv:1709.01577}, 2017.

\bibitem[Toulis and Kao(2013)]{toulis2013estimation}
P.~Toulis and E.~Kao.
\newblock Estimation of causal peer influence effects.
\newblock In \emph{International conference on machine learning}, pages
  1489--1497, 2013.

\bibitem[Van Der~Vaart and Wellner(1996)]{van1996weak}
A.~W. Van Der~Vaart and J.~A. Wellner.
\newblock Weak convergence.
\newblock In \emph{Weak convergence and empirical processes}, pages 16--28.
  Springer, 1996.

\bibitem[Viviano(2019)]{viviano2019policy}
D.~Viviano.
\newblock Policy targeting under network interference.
\newblock \emph{arXiv preprint arXiv:1906.10258}, 2019.

\bibitem[Wainwright(2019)]{wainwright2019high}
M.~J. Wainwright.
\newblock \emph{High-dimensional statistics: A non-asymptotic viewpoint},
  volume~48.
\newblock Cambridge University Press, 2019.

\bibitem[Wang et~al.(2017)Wang, Guo, and Ramon]{Wang2017learning}
Y.~Wang, Z.-C. Guo, and J.~Ramon.
\newblock {Learning from Networked Examples}.
\newblock In \emph{{28th International Conference on Algorithmic Learning
  Theory (ALT), Kyoto, Japan}}, October 2017.

\bibitem[Z{\"u}gner et~al.(2018)Z{\"u}gner, Akbarnejad, and
  G{\"u}nnemann]{zugner2018adversarial}
D.~Z{\"u}gner, A.~Akbarnejad, and S.~G{\"u}nnemann.
\newblock Adversarial attacks on neural networks for graph data.
\newblock In \emph{Proceedings of the 24th ACM SIGKDD International Conference
  on Knowledge Discovery \& Data Mining}, pages 2847--2856. ACM, 2018.

\end{thebibliography}

\onecolumn
\appendix

\aistatstitle{Causal Inference under Networked Interference and Intervention Policy Enhancement: \\ Supplementary Materials}

\section{HSIC}

The empirical HSIC using a Gaussian RBF kernel is written as $ \hat{HSIC}_{ \mathcal{K}_{ \sigma } }$. According to~\cite{gretton2005measuring}, given samples $ \{ \Phi( \mathbf{X}_i ), T_i \}_{ i=1 }^n $, the empirical estimation of HSIC in Gaussian kernel $ \mathcal{K}_{ \sigma } $ reads
\begin{align*}
  & \hat{HSIC}_{ \mathcal{K}_{\sigma} } = \frac{1}{n^2} \sum\limits_{i, j=1}^n \mathcal{K}_{ \sigma } ( \Phi( \mathbf{X}_i ), \Phi( \mathbf{X}_j ) ) \mathcal{K}_{ \sigma } ( T_i, T_j ) \\
  & \ + \frac{1}{n^4} \sum\limits_{i,j,k,l=1}^n \mathcal{K}_{\sigma} ( \Phi( \mathbf{X}_i ), \Phi( \mathbf{X}_j ) ) \mathcal{K}_{ \sigma } ( T_k, T_l ) - \frac{2}{ n^3 } \sum\limits_{i, j, k=1}^n \mathcal{K}_{\sigma} ( \Phi( \mathbf{X}_i ), \Phi( \mathbf{X}_j ) ) \mathcal{K}_{\sigma} ( T_i, T_k ). 
\end{align*}

\section{Nonparametric Identifiability of Causal Effect}
\label{seq:identifiability}

The nonparametric identifiability of expected causal response is guaranteed following~\cite{ogburn2017causal,forastiere2016identification}. For the sake of simplicity, we assume that influences are only from the first-order neighbors. To prove the identifiability, we introduce a variable $ V_i := S_{ V, i } ( \mathbf{X}_{ \mathcal{N}_i }, \mathbf{T}_{ \mathcal{N}_i } ) $, where 
\begin{equation*}
  S_{ V, i } : \{ 0, 1 \}^{ | \mathcal{N}_i | } \otimes \boldsymbol{ \chi }^{ \otimes | \mathcal{N}_i | } \rightarrow \mathcal{V}_i, 
\end{equation*}
for $i = 1, \dots, n$, represents the aggregation of neighboring covariates and treatment assignments, e.g., the average of neighboring treatments and the output of a GNN. Following reasonable assumptions are necessary for the nonparametric identifiability. 

\begin{my_assumption}{\ } \\
  (1) Given summary function $S_{V, i}$, for $i=1, \dots, n$, $ \forall \mathbf{T}_{ \mathcal{N}_i }, \mathbf{T}_{ \mathcal{N}_i }' $,  $ \forall \mathbf{X}_{ \mathcal{N}_i }, \mathbf{X}_{ \mathcal{N}_i }' $, $ \forall \mathbf{T}_{ \mathcal{N}_{-i} }, \mathbf{T}_{ \mathcal{N}_{-i} }' $, and $ \forall \mathbf{X}_{ \mathcal{N}_{-i} }, \mathbf{X}_{ \mathcal{N}_{-i} }' $, with $ S_{V, i} ( \mathbf{T}_{ \mathcal{N}_i}, \mathbf{X}_{ \mathcal{N}_i } ) = S_{V, i} ( \mathbf{T}_{ \mathcal{N}_i}', \mathbf{X}_{ \mathcal{N}_i }' ) $, then it holds
  \begin{equation*}
    Y_i ( T_i, \mathbf{T}_{ \mathcal{N}_i}, \mathbf{X}_{ \mathcal{N}_i}, \mathbf{T}_{ \mathcal{N}_{-i} }, \mathbf{X}_{ \mathcal{N}_{-i} } ) = Y_i ( T_i, \mathbf{T}_{ \mathcal{N}_i}', \mathbf{X}_{ \mathcal{N}_i}', \mathbf{T}_{ \mathcal{N}_{-i} }', \mathbf{X}_{ \mathcal{N}_{-i} }' ).   \\
  \end{equation*}
  (2) Unconfoundedness assumption: $ Y_i ( t_i, v_i ) \perp T_i, V_i | \mathbf{X}_i $, $ \forall t_i \in \{0, 1\} $ and $ v_i \in \mathcal{V}_i $, for $i = 1, \dots, n$.
\end{my_assumption}

Hence, the expected response of one unit under network inference can be identified as $ \mathbb{E} [  Y_i ( t_i, v_i ) ] = \mathbb{E} [ Y_i | T_i = t_i, V_i = v_i, \mathbf{X}_i ] $, $ \forall t_i \in \{0, 1\} $, and $ v_i \in \mathcal{V}_i $, for $i=1, \dots, n$. It is derived by 
\begin{align*}
  \mathbb{E} [ Y_i | T_i = t_i, V_i = v_i, \mathbf{X}_i ] & \overset{Asm. (1)}{=} \mathbb{E} [ Y_i ( t_i, v_i ) | T_i = t_i, V_i = v_i, \mathbf{X}_i ]  \\
  & \overset{Asm. (2)}{=}  \mathbb{E} [ Y_i ( t_i, v_i ) | \mathbf{X}_i ]. 
\end{align*}

\begin{table}[thp]
\centering
\small
\begin{tabular}{| c | c | c | c |}
   \hline
   H1GH52  &  Do you get enough sleep?  &  H1ED3   &  Have you skipped a grade?  \\
   \hline
   H1ED5   &  Have you repeated a grade?  &  H1ED7   &  Have you received an suspension?  \\
   \hline
   H1HS1   &  Have you had a routine physical examination?  &  H1HS3  &  Have you received psychological counseling? \\
   \hline
   H1WP17B &  Played a sport in the past 4 weeks?  &  H1TO51  &  Is alcohol easily available in your home?  \\
   \hline
   H1TO53  &  Is a gun easily available in your home?  &  H1NB5   &  Do you feel safe in your neighborhood?  \\
   \hline
   H1EE3   &  Did you work for pay in the last 4 weeks?  &  PA57D   &  Food stamps? \\
   \hline
   H1DA5   &  How often do you play sport?  &  H1DA7   &  How do you hang out with friends?  \\
   \hline
   H1ED11  &  Your grade in English or language arts?  &  H1ED12  &  Your grade in mathematics?  \\
   \hline
   H1ED13  &  Your grade in history or social studies?  &  H1ED14  &  Your grade in science?  \\
   \hline
   H1DS12  &  How often did you sell marijuana or other drugs  &  $-$  &  $-$  \\
   \hline
\end{tabular}
  \caption{Selected questions from the Wave1 data~\cite{harris2018national} that are used as feature vectors.}
  \label{tab:selected_features}
\end{table}
\normalsize

\section{Synthetic Randomized Experiments on Wave1}
\label{sec:wave_exp_app}

On the in-school friendship network derived from the Wave1 questionnaire data, we conduct randomized intervention experiments that simulate the improvement of performance after assigning a student to a tutoring or support program. Recall that $ Y_i (T_i=0, \mathcal{G} = \emptyset ) $ indicates the overall performance of student $i$ before assigning it to a tutoring program or being influenced by peers. We select specific questions from the questionnaire and regard the corresponding answers as the features of corresponding students. These feature vectors are further used to construct a symmetrized $k$-NN similarity graph as the in-school friendship network. Questions related to the potential performance of students are list in Table~\ref{tab:selected_features}.

\begin{table}[thp]
\centering
\small
\begin{tabular}{c | c c | c c }
  &  \multicolumn{2}{c}{$k=5$}  &  \multicolumn{2}{c}{$k=10$}   \\
  & $ \sqrt{ \mathcal{ MSE } } $ & $ \epsilon_{ PEHE } $ & $ \sqrt{ \mathcal{ MSE } } $ & $ \epsilon_{ PEHE } $  \\
  \hline
  GPS             &  $0.279 \pm 0.071$  &  $0.210 \pm 0.043$  &  $0.281 \pm 0.049$  &  $0.139 \pm 0.052$   \\
  \hline
  GCN             &  $0.212 \pm 0.035$  &  $0.095 \pm 0.055$  &  $0.211 \pm 0.013$  &  $0.058 \pm 0.036$   \\
  
  GraphSAGE       &  $\mathbf{0.200 \pm 0.032}$  &  $\mathbf{0.088 \pm 0.054}$  &  $\mathbf{0.199 \pm 0.030}$  &  $\mathbf{0.057 \pm 0.039}$   \\
  
  $1$-GNN         &  $0.214 \pm 0.039$  &  $0.096 \pm 0.062$  &  $0.203 \pm 0.033$  &  $ \mathbf{0.057 \pm 0.040}$  \\
  \hline
  \textbf{Improve} & $ 28.3 \% $  &  $ 58.1 \% $  &  $ 29.2 \% $  &  $ 59.0 \% $  \\ 
  \hline
\end{tabular}
  \caption{Evaluation metrics on under-treated synthetic data with $ \mathbf{p=0.1} $, $\alpha=0.5$, and $k=5, 10$. Improvements are obtained by comparing with the GPS baseline. Both representation balancing $ \hat{HSIC}^{ \Phi} $ and $ \hat{HSIC}^{ \mathit{GNN} } $ are deployed in the GNN-based estimators for searching for the best performance.}
  \label{tab:exp_under_treated}
\end{table}  
\normalsize

Using the answers of selected questions and their abbreviations, $ Y_i (T_i=0, \mathcal{G} = \emptyset ) $ is generated as follows
\begin{align*}
   Y_i (T_i=0, \mathcal{G} = \emptyset ) & := - X_{i, H1GH52} + 2 X_{i, H1ED3} - X_{i, H1ED5} - 2 X_{i, H1ED7}   \\ 
   & - 0.5 ( X_{i, H1ED11 } + X_{i, H1ED12 } + X_{i, H1ED13 } + X_{i, H1ED14 }  )  \\
   & + 0.5 ( X_{i, H1DA5} + X_{i, H1DA7 } ) - 3 X_{i, H1DS12} + f_{ \mathcal{N} } (X_{i, H1HS1} \\ & + X_{i, H1HS3 } + X_{i, H1WP17B} + X_{i, H1TO51} + X_{i, H1TO53} \\
   & + X_{i, H1NB5} + X_{i, H1EE3} + X_{i, PA57D}), 
\end{align*}
where $ f_{ \mathcal{N} } ( \cdot ) $ represents a 1-layer neural network with random coefficients.

The generating process of the individual treatment response also depends on the selected properties. For example, by assigning a student who has repeated grade will probably improve this student's performance. The treatment effect is simulated as follows: 
\begin{align*}
  \tau ( \mathbf{X}_i ) & := X_{i, H1ED3} + 0.5 ( X_{i, H1GH52} + X_{i, H1ED5} + X_{i, H1ED7} ) \\ 
  & + 0.5 ( X_{i, H1ED11}+ X_{i, H1ED12} + X_{i, H1ED13} + X_{i, H1ED14} ) \\
  & + X_{i, H1DS12} + f_{ \mathcal{N} }, 
\end{align*}
where $ f_{ \mathcal{N} } $ represents a nonlinear random function depending on the rest of variables. Furthermore, peer effect in this synthetic experiment is generated by 
\begin{equation}
  \delta_i ( \mathbf{X}, \mathbf{T}, \mathcal{G} ) := \alpha \frac{ 1 }{ | \mathcal{N}_i | } \sum\limits_{ j \in \mathcal{N}_i } T_j \tau ( \mathbf{X}_j ), 
  \label{eq:simulated_peer_effect}
\end{equation}
where the decay parameter $ \alpha $ characterizes the decay of influence. Eq.~\ref{eq:simulated_peer_effect} means that the peer effect applied to the node $i$ is determined by individual treatment responses of its neighbors who are under treatment. Finally, the outcome, e.g., the linear response $G_0$, is simulated by 
\begin{equation}
  Y_i = Y_i ( T_i = 0, \mathcal{G} = \emptyset ) + T_i \tau ( \mathbf{X}_i ) + \delta_i ( \mathbf{X}, \mathbf{T}, \mathcal{G} ) + \epsilon_{Y_i}. 
  \label{eq:simulate_observational_data}
\end{equation}

\begin{table}[thp]
\centering
\small
\begin{tabular}{c | c c | c c }
  &  \multicolumn{2}{c}{$k=5$}  &  \multicolumn{2}{c}{$k=10$}   \\
  & $ \sqrt{ \mathcal{ MSE } } $ & $ \epsilon_{ PEHE } $ & $ \sqrt{ \mathcal{ MSE } } $ & $ \epsilon_{ PEHE } $  \\
  \hline
  GPS             &  $0.318 \pm 0.010$  &  $0.409 \pm 0.008$  &  $0.363 \pm 0.087$  &  $0.491 \pm 0.200$  \\
  \hline
  GCN             &  $0.277 \pm 0.007$  &  $0.051 \pm 0.007$  &  $0.288 \pm 0.063$  &  $0.087 \pm 0.053$  \\
  
  GraphSAGE       &  $0.276 \pm 0.024$  &  $\mathbf{0.050 \pm 0.007}$  &  $0.301 \pm 0.054$  &  $0.083 \pm 0.033$   \\
  
  $1$-GNN         &  $\mathbf{0.249 \pm 0.006}$  &  $0.054 \pm 0.015$  &  $\mathbf{0.278 \pm 0.056}$  &  $\mathbf{0.076 \pm 0.034}$   \\
  \hline
  \textbf{Improve}  &  $ 21.7 \% $  &  $ 87.8 \% $  & $ 23.4 \% $  &  $ 84.5 \% $  \\
  \hline
\end{tabular}
  \caption{Evaluation metrics on over-treated synthetic data with $ \mathbf{p=0.7} $, $\alpha=0.5$, and $k=5, 10$. Improvements are obtained by comparing with the GPS baseline. Both representation balancing $ \hat{HSIC}^{ \Phi} $ and $ \hat{HSIC}^{ \mathit{GNN} } $ are deployed in the GNN-based estimators for searching for the best performance.}
  \label{tab:exp_over_treated}
\end{table}  
\normalsize

\begin{table}[htp]
\centering
\small
\begin{tabular}{c | c c | c c }
  &  \multicolumn{2}{c}{$k=5$}  &  \multicolumn{2}{c}{$k=10$}    \\
  & $ \sqrt{ \mathcal{ MSE } } $ & $ \epsilon_{ PEHE } $ & $ \sqrt{ \mathcal{ MSE } } $ & $ \epsilon_{ PEHE } $  \\
  \hline
  GPS             &  $0.329 \pm 0.005$  &  $0.207 \pm 0.015$  &  $0.294 \pm 0.008$  &  $0.224 \pm 0.071$   \\
  \hline
  GCN             &  $0.269 \pm 0.011$  &  $0.047 \pm 0.006$  &  $0.215 \pm 0.020$  &  $0.050 \pm 0.012$   \\
  
  GraphSAGE       &  $0.279 \pm 0.015$  &  $0.044 \pm 0.003$  &  $0.223 \pm 0.018$  &  $0.037 \pm 0.011$   \\
  
  $1$-GNN         &  $\mathbf{0.268 \pm 0.015}$  &  $\mathbf{0.042 \pm 0.005}$  &  $\mathbf{0.214 \pm 0.015}$  &  $\mathbf{0.032 \pm 0.007}$  \\
  \hline
  \textbf{Improve} & $ 18.5 \% $ & $ 79.7 \% $  & $ 27.2 \% $  & $ 85.7 \% $  \\
  \hline
\end{tabular}
  \caption{Evaluation metrics on balanced synthetic data with $ \mathbf{p=0.5} $, $\alpha=0.5$, and $k=5, 10$. Improvements are obtained by comparing with the GPS baseline. Both representation balancing $ \hat{HSIC}^{ \Phi} $ and $ \hat{HSIC}^{ \mathit{GNN} } $ are deployed in the GNN-based estimators for searching for the best performance.}
  \label{tab:exp_balanced}
\end{table}  
\normalsize

The benefit of using synthetic data is that we can modify the experiment settings. Three parameters control the experimental settings: number of neighbors $k$, which determines the graph structure and density; the probability $p$ of assigning a node to treatment which controls the population imbalance between treatment and control groups; the decay parameter $\alpha$, which determines the intensity of peer effect. For the evaluation results reported in the main text we generate the simulation data with parameters $ k=10 $, $ p=0.1 $, and $ \alpha = 0.5 $. We report more evaluations in Table~\ref{tab:exp_under_treated}, Table~\ref{tab:exp_over_treated}, and Table~\ref{tab:exp_balanced}. One observation is that in the randomized experiment setting with linear response, the GraphSAGE-based estimator is a good candidate for causal inference in an under-treated population, while $1$-GNN-based estimator is superior in a balanced- or over-treated population.

\section{Synthetic Randomized Experiments on Pokec}
\label{sec:pokec_exp_app}

\begin{figure}[thp]
\begin{center}
  \includegraphics[width=0.4\linewidth]{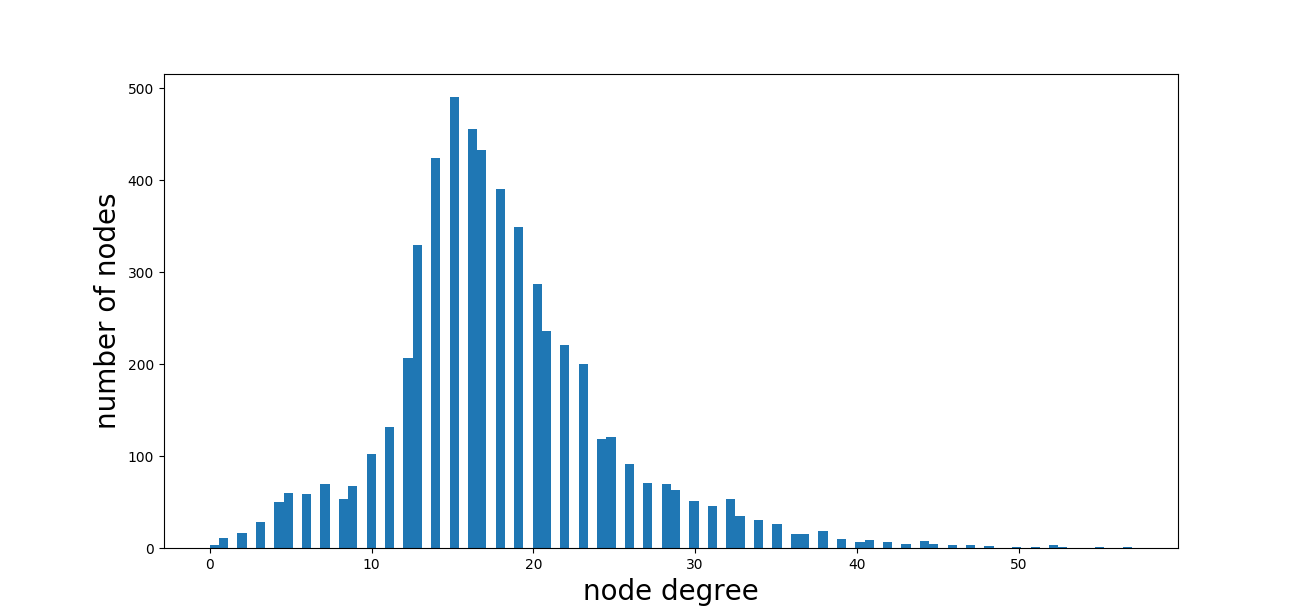} 
  \includegraphics[width=0.4\linewidth]{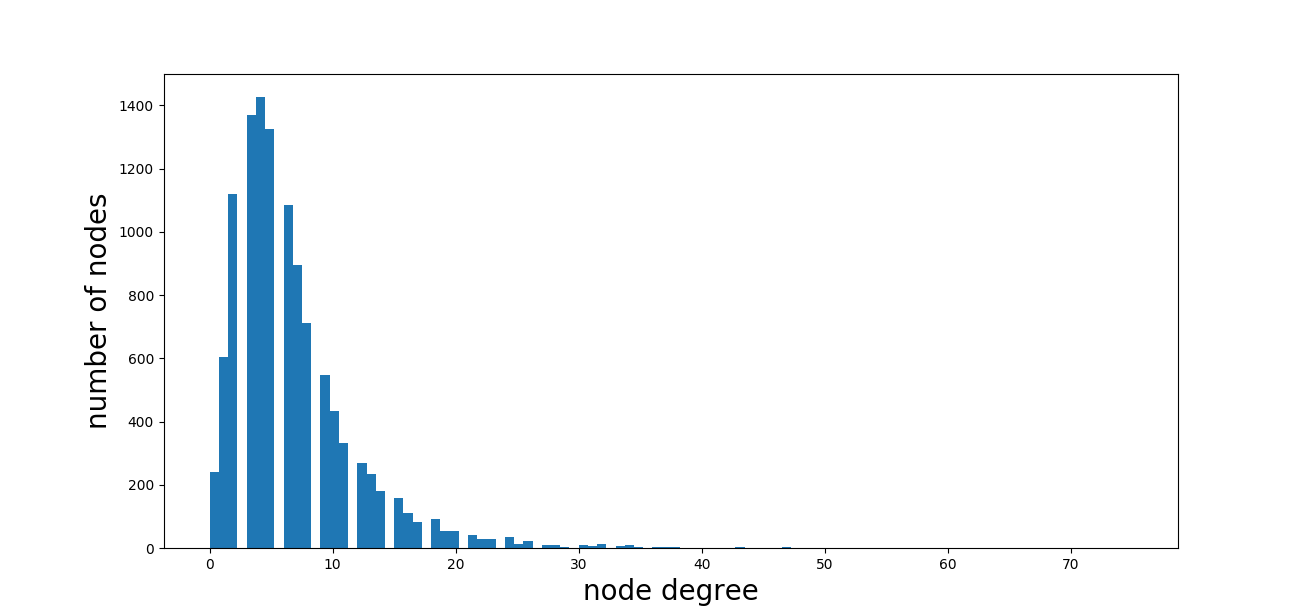}  \\
\end{center}
  \caption{Number of nodes vs. node degree from the $k$-NN similarity graph of Wave1 with $k=10$ (left), and from the Pokec social network (right).}
  \label{fig:node_degree}
\end{figure}

The motivation for using a real social network dataset is that the $k$-NN similarity graph can violate the power-law degree distribution, as shown in Fig.~\ref{fig:node_degree}.  Consider hypothetical intervention experiments to the users of the Pokec social network. After reading a personalized advertisement or getting influenced by social contacts, a user is encouraged to purchase a new medicine. To simulate the individual buying behavior, we use profile features that are related to the health condition of a user. Table~\ref{tab:pokec_data} lists the related features used in semi-synthetic experiments.

\begin{table}[thp]
\centering
\small
\begin{tabular}{| c | c | c | c |}
\hline
features  &  values  &  features  &  values  \\
\hline
gender  &  $ [0, 1] $  &  age  &  $ [15, 16, \cdots, 60] $  \\
\hline
height  &  $ [140, 141, \cdots, 200] $  &  weight  &  $ [30, 31, \cdots, 200] $  \\
\hline
completed level of education  &  $ [0, 1, 2, 3] $  &  eyesight  &  $ [0, 1] $  \\
\hline
relation to smoking  &  $ [0, 1, 2, 3] $  &  relation to alcohol  &  $ [0, 1, 2, 3] $  \\
\hline
relation to casual sex  &  $ [0, 1, 2] $  &  $-$  &  $-$  \\
\hline
\end{tabular}
  \caption{Characteristics of users and corresponding ranges of values selected from the Pokec social network data.}
  \label{tab:pokec_data}
\end{table}

We assume that a healthy person with good habits is self-motivated to purchase health medicine even without external influences. Hence, $ Y_i ( T_i = 0, \mathcal{G} = \emptyset ) $ is simulated as follows:
\small
\begin{align*}
   Y_i ( T_i = 0, \mathcal{G} = \emptyset ) & := 0.2 ( 1 - X_{i, gender} ) + 0.5 X_{i, age} - 0.2 X_{i, weight} + 0.5 X_{i, education} \\
   & - 0.6 (3 - X_{i, smoke} ) + 0.2 X_{i, sex} - 0.6 ( 3 - X_{i, alcohol} ) + \epsilon,  
\end{align*} 
\normalsize
where $\epsilon$ is a Gaussian random variable with mean $0.1$. Suppose that new health medicine is advertised to offer miraculous effects on weight loss, quit smoking, abstinence, etc. Then the individual treatment response can be generated by
\small
\begin{align*}
  \tau ( \mathbf{X}_i ) & := 0.8 ( 1 - X_{i, gender} ) + X_{i, age} + 0.3 X_{i, weight} + 0.5 (1 - X_{i, eyesight})  \\
  &  0.5 ( X_{i, education} + 0.5 ) + 0.6 X_{i, smoke} + 0.5 X_{i, alcohol} + \epsilon. 
\end{align*} 
\normalsize

\begin{table}[thp]
\centering
\small
\begin{tabular}{c | c c | c c }
  &  \multicolumn{2}{c}{$\alpha=0.1$}  &  \multicolumn{2}{c}{$\alpha=0.9$}    \\
  & $ \sqrt{ \mathcal{ MSE } } $ & $ \epsilon_{ PEHE } $ & $ \sqrt{ \mathcal{ MSE } } $ & $ \epsilon_{ PEHE } $ \\
  \hline
  GPS            &  $0.263 \pm 0.001$  &  $0.156 \pm 0.017$  &  $0.595 \pm 0.005$  &  $0.185 \pm 0.005$  \\
  \hline
  GCN            &  $0.230 \pm 0.017$  &  $0.147 \pm 0.031$  &  $0.573 \pm 0.033$  &  $0.163 \pm 0.005$  \\
  GraphSAGE      &  $\mathbf{0.227 \pm 0.005}$  &  $\mathbf{0.128 \pm 0.015}$  &  $\mathbf{0.569 \pm 0.032}$  &  $\mathbf{0.151 \pm 0.011}$  \\
  $1$-GNN        &  $0.231 \pm 0.006$  &  $0.132 \pm 0.014$  &  $0.571 \pm 0.033$  &  $0.197 \pm 0.020$  \\
  \hline
  \textbf{Improve} &  $ 13.5 \% $  &  $ 17.9 \% $  &  $ 4.4 \% $  &  $ 18.4 \% $  \\
  \hline 
\end{tabular}
  \caption{Evaluation metrics on under-treated Pokec social network with $ \mathbf{p=0.1} $, $\alpha=0.1, 0.9 $. Improvements are obtained by comparing with the GPS baseline. Both representation balancing $ \hat{HSIC}^{ \Phi} $ and $ \hat{HSIC}^{ \mathit{GNN} } $ are deployed in the GNN-based estimators for searching for the best performance.}
  \label{tab:pokec_exp_under_treated_0.1}
\end{table}

\begin{table}[thp]
\centering
\small
\begin{tabular}{c | c c | c c }
  &  \multicolumn{2}{c}{$\alpha=0.1$}  &  \multicolumn{2}{c}{$\alpha=0.9$}    \\
  & $ \sqrt{ \mathcal{ MSE } } $ & $ \epsilon_{ PEHE } $ & $ \sqrt{ \mathcal{ MSE } } $ & $ \epsilon_{ PEHE } $ \\
  \hline
  GPS            &  $0.404 \pm 0.007$  &  $0.126 \pm 0.004$  &  $1.438 \pm 0.000$  &  $\mathbf{0.533 \pm 0.015}$  \\
  \hline
  GCN            &  $0.247 \pm 0.008$  &  $0.044 \pm 0.003$  &  $1.426 \pm 0.030$  &  $0.594 \pm 0.039$  \\
  GraphSAGE      &  $0.240 \pm 0.006$  &  $0.041 \pm 0.001$  &  $1.417 \pm 0.021$  &  $0.662 \pm 0.061$  \\
  $1$-GNN        &  $\mathbf{0.233 \pm 0.001}$  &  $\mathbf{0.039 \pm 0.002}$  &  $\mathbf{1.390 \pm 0.033}$  &  $1.076 \pm 0.094$  \\
  \hline
  \textbf{Improve}  &  $ 42.3 \% $  &  $ 69.0 \% $  &  $ 3.3 \% $  &  $ -11.4 \% $  \\
  \hline 
\end{tabular}
  \caption{Evaluation metrics on over-treated Pokec social network with $ \mathbf{p=0.7} $, $\alpha=0.1, 0.9 $. Improvements are obtained by comparing with the GPS baseline. Both representation balancing $ \hat{HSIC}^{ \Phi} $ and $ \hat{HSIC}^{ \mathit{GNN} } $ are deployed in the GNN-based estimators for searching for the best performance.}
  \label{tab:pokec_exp_over_treated_0.7}
\end{table}

Since Pokec is a social network, in the semi-synthetic experiments, we also take into account long-range influences to simulate opinion propagation in the social network. To be more specific, the spillover effect on one node not only depends on the nearest neighboring nodes but also next-nearest neighboring nodes. Formally, it is defined as  
\begin{equation}
  \delta_i ( \mathbf{X}, \mathbf{T}, \mathcal{G}) := \alpha \frac{1}{ | \mathcal{N}_i | } \sum\limits_{ j \in \mathcal{N}_i }  T_j \tau ( \mathbf{X}_j ) + \alpha^2  \frac{1}{ | \mathcal{N}_i^{ (2) } | } \sum\limits_{ k \in \mathcal{N}_i^{ (2) } }  T_k \tau ( \mathbf{X}_k ), 
  \label{eq:pokec_spillover}
\end{equation}
where $\alpha$ is the decay factor and $ \mathcal{N}_i^{ (2) } $ represents the next-nearest neighbors of $i$. Finally, the observed data in the randomized experiments can be derived from $Y_i ( T_i = 0, \mathcal{G} = \emptyset )$, $ \tau ( \mathbf{X}_i ) $, and social network structure $ \mathcal{G}_{Pokec} $ using Eq.~\ref{eq:pokec_spillover} and Eq.~\ref{eq:simulate_observational_data} for the linear response or Eq.~\ref{eq:G_1} and Eq.~\ref{eq:G_2} for nonlinear responses. The experiments reported in the main text use the setting $ \alpha=0.5 $ and $ p=0.1 $.

Since the network structure $ \mathcal{G}_{Pokec} $ is given, we provide more experiment results in Table~\ref{tab:pokec_exp_under_treated_0.1} and Table~\ref{tab:pokec_exp_over_treated_0.7} to understand the effect of decay parameter $ \alpha $. In particular, we consider regimes from negligible peer effects with $\alpha=0.1$ to significant peer effects with $ \alpha = 0.9 $. Since the covariates of neighboring units in the Pokec dataset have strong cosine similarity, and the simulation generation process is relatively simple, GNN-based causal estimators might overfit the superimposed causal effects and poorly recover the individual treatment effect. It is becoming more evident if the peer effects are strong and the population is over-treated, where the GPS baseline can achieve comparable results as other GNN-based estimators using only the information of exposure level (see Table~\ref{tab:pokec_exp_over_treated_0.7}).

\begin{table*}[thp]
\begin{center}
\small
\begin{tabular}{c | c | c | c | c  }
	\hline
    & \multicolumn{4}{c}{ Wave1 }  \\
    \hline
    & \multicolumn{2}{c}{ $ G_1 $ }  &  \multicolumn{2}{|c}{ $ G_2 $ } \\
    & $ \sqrt{ \mathcal{MSE} }$ &  $ \epsilon_{PEHE} $  & $ \sqrt{ \mathcal{MSE} }$ &  $ \epsilon_{PEHE} $  \\
    \hline
 DA GB            &  $0.770 \pm .017$  &  $0.379 \pm .126$  &  $0.763 \pm .047$   &  $0.248 \pm .121$  \\
 
 DA RF            &  $1.047 \pm .046$  &  $0.701 \pm .029$  &  $0.977 \pm .021$   &  $0.599 \pm .193$  \\
 
 DR GB            &  $0.814 \pm .058$  &  $0.392 \pm .029$  &  $0.771 \pm .014$   &  $0.401 \pm .028$  \\
 
 DR EN            &  $1.063 \pm .037$  &  $0.843 \pm .005$  &  $0.886 \pm .010$   &  $0.636 \pm .173$  \\
 
 GPS              &  $0.236 \pm .001$  &  $0.158 \pm .031$  &  $0.262 \pm .071$   &  $0.163 \pm .063$  \\
 \hline
 GCN              &  $0.192 \pm .003$  &  $0.050 \pm .007$  &  $0.201 \pm .034$   &  $0.044 \pm .026$  \\
 
 GraphSAGE        &  $\mathbf{0.191 \pm .004}$  &  $\mathbf{0.049 \pm .003}$  &  $0.198 \pm .022$   &  $\mathbf{0.039 \pm .018}$  \\
 
 $1$-GNN          &  $0.207 \pm .003$  &  $0.058 \pm .006$  &  $\mathbf{0.188 \pm .020}$   &  $0.043 \pm .024$  \\
 \hline
 \textbf{Improve} & $ 19.1 \% $  &  $19.0 \% $  &  $ 28.2 \% $  &  $ 76.1 \% $  \\
 \hline 
  & \multicolumn{4}{c}{ Pokec }  \\
  \hline
 DA GB            &  $0.988 \pm .005$   &  $0.419 \pm .046$   &  $1.189 \pm .017$   &  $0.376 \pm .033$  \\
 DA RF            &  $1.016 \pm .024$   &  $1.075 \pm .031$   &  $1.225 \pm .009$   &  $1.016 \pm .037$  \\
 DR GB            &  $0.943 \pm .024$   &  $0.297 \pm .057$   &  $1.173 \pm .012$   &  $0.314 \pm .020$  \\
 DR EN            &  $0.947 \pm .023$   &  $0.181 \pm .031$   &  $1.172 \pm .013$   &  $0.282 \pm .041$  \\
 GPS              &  $0.420 \pm .006$   &  $0.212 \pm .070$   &  $0.475 \pm .004$   &  $0.220 \pm .013$  \\
 \hline
 GCN              &  $0.367 \pm .005$   &  $0.162 \pm .004$   &  $0.423 \pm .017$   &  $0.183 \pm .010$  \\
 GraphSAGE        &  $\mathbf{0.360 \pm .000}$   &  $\mathbf{0.146 \pm .001}$   &  $0.425 \pm .018$   &  $0.167 \pm .005$  \\
 $1$-GNN          &  $0.366 \pm .013$   &  $0.151 \pm .006$   &  $\mathbf{0.408 \pm .009}$   &  $\mathbf{0.158 \pm .004}$  \\
     \hline
\textbf{Improve}  &  $ 14.3 \% $  &  $ 19.3 \% $  &  $ 14.1 \% $  &  $ 28.2 \% $  \\
    \hline
    
\end{tabular}
\caption{Experimental results of randomized experiments on the Wave1 and Pokec dataset using nonlinear response generation functions $G_1$ and $G_2$ with $ \kappa = \mathbf{0.2} $. For Wave1, we set (node degree) $ k=10 $, (decay parameter) $ \alpha = 0.5 $, and (treatment probability) $ p = 0.1 $. For Pokec, we set $ p = 0.1 $. $ \hat{HSIC}^{ \Phi} $ and $ \hat{HSIC}^{ \mathit{GNN} } $ are deployed in the GNN-based estimators. $ \hat{HSIC}^{ \Phi} $ and $ \hat{HSIC}^{ \mathit{GNN} } $ are deployed in the GNN-based estimators. Improvements are obtained by comparing with the best baselines.}
\label{tab:wave_pokec_nonlinear_0.2}
\end{center}
\end{table*}
\normalsize

\section{Experimental Results of Nonlinear Causal Responses}

To further investigate the superiority of the GNN-based causal estimators on nonlinear causal responses, we consider the following nonlinear data generation function inspired by Section 4.2 of~\cite{toulis2013estimation}, 
\begin{equation}
G_1: Y_i = Y_i (T_i=0, \mathcal{G} = \emptyset ) + T_i \tau ( \mathbf{X}_i ) + \delta_i ( \mathbf{X}, \mathbf{T}, \mathcal{G} )  +  \kappa \delta^2_i ( \mathbf{X}, \mathbf{T}, \mathcal{G} ) + \epsilon_{Y_i},  
  \label{eq:G_1}
\end{equation}
where $ \kappa $ characterizes the strength of nonlinear effects. In addition, a more complicated nonlinear response generation function 
\begin{equation}
    G_2:  Y_i = Y_i (T_i=0, \mathcal{G} = \emptyset ) + T_i \tau ( \mathbf{X}_i ) + \delta_i ( \mathbf{X}, \mathbf{T}, \mathcal{G} )  +  \frac{ \kappa }{2} \delta^2_i ( \mathbf{X}, \mathbf{T}, \mathcal{G} )  + \frac{\kappa}{2}  \tau ( \mathbf{X}_i ) \delta_i ( \mathbf{X}, \mathbf{T}, \mathcal{G} ) + \epsilon_{Y_i}
    \label{eq:G_2}
\end{equation}
is considered, where the quadratic terms signify the spillover effect depending on the individual treatment effect.

Table~\ref{tab:wave_pokec_nonlinear_0.2} reports the performance of GNN-based causal estimators on nonlinear causal effects prediction tasks. Nonlinear responses are generated via $ G_1 $ and $ G_2 $ with $ \kappa = 0.2 $. For the $ \sqrt{ \mathcal{MSE} } $ metric, GNN-based estimators outperform the best baseline GPS dramatically, showing the effectiveness of predicting nonlinear causal responses. Moreover, a $ 19.0 \% (G_1) $ and $ 76.1 \% (G_2) $ performance improvement on the $ \epsilon_{PEHE}$ metric with the Wave1 dataset shows that setting an empty graph, i.e., $ \mathcal{G} = \emptyset $, in the GNN-based estimators is an appropriate approach for extracting individual causal effect.

\textbf{\begin{table*}[thp]
\begin{center}
\small
\begin{tabular}{c | c | c | c | c }
	\hline
    & \multicolumn{4}{c}{ Wave1 }  \\
    \hline
    & \multicolumn{2}{c}{ $ G_1 $ }  &  \multicolumn{2}{|c}{ $ G_2 $ }  \\
    & $ \sqrt{ \mathcal{MSE} }$ &  $ \epsilon_{PEHE} $  & $ \sqrt{ \mathcal{MSE} }$ &  $ \epsilon_{PEHE} $  \\
    \hline
 DA GB            &  $0.742 \pm .083$  &  $0.210 \pm .008$  &  $1.060 \pm .047$   &  $0.400 \pm .054$  \\
 DA RF            &  $1.007 \pm .027$  &  $0.527 \pm .141$  &  $1.243 \pm .089$   &  $1.056 \pm .222$  \\
 DR GB            &  $0.784 \pm .019$  &  $0.352 \pm .074$  &  $1.116 \pm .106$   &  $0.633 \pm .195$  \\
 DR EN            &  $0.882 \pm .053$  &  $0.575 \pm .015$  &  $1.258 \pm .176$   &  $0.841 \pm .293$  \\
 GPS              &  $0.280 \pm .017$  &  $0.142 \pm .032$  &  $0.289 \pm .012$   &  $0.244 \pm .066$  \\
 \hline
 GCN              &  $0.224 \pm .008$  &  $\mathbf{0.038 \pm .003}$  &  $0.237 \pm .020$   &  $0.095 \pm .010$  \\
 GraphSAGE        &  $\mathbf{0.214 \pm .007}$  &  $0.045 \pm .002$  &  $\mathbf{0.231 \pm .014}$   &  $\mathbf{0.072 \pm .003}$  \\
 $1$-GNN          &  $0.216 \pm .003$  &  $0.040 \pm .001$  &  $0.250 \pm .020$   &  $0.103 \pm .015$  \\
 \hline
 \textbf{Improve} &  $ 23.6 \% $  &  $ 73.2 \% $  &  $ 20.1 \% $  &  $ 70.5 \% $  \\ 
 \hline
 & \multicolumn{4}{c}{ Pokec }  \\
 \hline
 DA GB            &  $1.342 \pm .070$  &  $0.551 \pm .026$  &  $2.095 \pm .070$   &  $0.828 \pm .282$  \\
 DA RF            &  $1.369 \pm .060$  &  $1.015 \pm .074$  &  $2.125 \pm .080$   &  $1.389 \pm .109$  \\
 DR GB            &  $1.324 \pm .081$  &  $0.306 \pm .011$  &  $2.038 \pm .090$   &  $0.438 \pm .005$  \\
 DR EN            &  $1.325 \pm .078$  &  $0.336 \pm .032$  &  $2.043 \pm .089$   &  $0.338 \pm .040$  \\
 GPS              &  $0.693 \pm .058$  &  $0.450 \pm .042$  &  $0.813 \pm .068$   &  $0.375 \pm .089$  \\
 \hline
 GCN              &  $0.483 \pm .010$  &  $0.193 \pm .001$  &  $0.729 \pm .007$   &  $0.242 \pm .032$  \\
 GraphSAGE        &  $0.480 \pm .009$  &  $0.198 \pm .004$  &  $\mathbf{0.713 \pm .017}$   &  $\mathbf{0.217 \pm .025}$  \\
 $1$-GNN          &  $\mathbf{0.454 \pm .003}$  &  $\mathbf{0.159 \pm .005}$  &  $0.767 \pm .023$   &  $\mathbf{0.218 \pm .002}$  \\
 \hline
 \textbf{Improve} &  $ 34.5 \% $  &  $ 48.0 \% $  &  $ 12.3 \% $  &  $ 35.8 \% $ \\ 
 \hline
\end{tabular}
\end{center}
\caption{Experimental results of randomized experiments on the Wave1 and Pokec datasets using nonlinear response generation functions $G_1$ and $G_2$ with $ \mathbf{\kappa = 0.5} $. For Wave1, other parameters are set as (node degree) $ k=10 $, (decay parameter) $ \alpha = 0.5$, and (treatment probability) $ p = 0.1 $. For Pokec, we set the treatment probability as $ p = 0.1 $ and the decay parameter as $ \alpha = 0.5$. Both representation balancing $ \hat{HSIC}^{ \Phi} $ and $ \hat{HSIC}^{ \mathit{GNN} } $ are deployed in the GNN-based estimators for searching for the best performance. Improvements are obtained by comparing with the best baseline.}
\label{tab:wave_pokec_nonlinear_0.5}
\end{table*}
\normalsize}

Table~\ref{tab:wave_pokec_nonlinear_0.5} reports the performance of GNN-based causal estimators on the Wave1 and Pokec datasets using nonlinear response models. Nonlinear responses are generated via $ G_1 $ and $ G_2 $ under $ \kappa = 0.5 $. For the $ \sqrt{ \mathcal{MSE} } $ metric, GNN-based estimators outperform the best baseline by $ 23.6 \% (G_1) $ and $ 20.1 \% (G_2) $ on Wave1, and by $ 34.5 \% (G_1) $ and $ 12.3 \% (G_2) $ on the Pokec dataset. Moreover, GNN-based causal estimators significantly outperform the best baseline in the individual treatment effect recovery task. Especially, a $ 73.2 \% ( G_1 ) $ and a $ 70.5 \% (G_2) $ improvement on Wave1 are observed, and a $ 48.0 \% ( G_1 ) $ and a $ 35.8 \% (G_2) $ improvement on Pokec. The significantly improved metric $ \epsilon_{PEHE}$ indicates that even in the regime with higher nonlinear causal effects, GNN-based causal estimators can disentangle and extract individual treatment effects from strong interference.

\section{Additional Experiments for Intervention Policy Optimization}
\label{sec:policy_exp_add}

In addition to the policy optimization experiments on the Wave1 and Pokec simulation data under the treatment capacity constraint $p_t=0.3$, in Table~\ref{tab:poly_diff_wave_pokec_tp0.5} we also report the intervention policy improvement under the treatment capacity constraint with $p_t=0.5$.  

\begin{table}[thp]
\centering
\small
\begin{tabular}{c | c c c c }
  &  \multicolumn{2}{c}{Wave1}  &  \multicolumn{2}{c}{Pokec}  \\
  &  $ \Delta \hat{ S } ( \hat{ \pi }_n^{ p_t } ) $  &  $ \Delta  S ( \hat{ \pi }_n^{ p_t } ) $  &  $ \Delta \hat{ S } ( \hat{ \pi }_n^{ p_t } ) $  &  $ \Delta  S ( \hat{ \pi }_n^{ p_t } ) $  \\
  \hline
DA GB         &  $0.636 \pm 0.028$  &  $0.012 \pm 0.025$  &  $0.479 \pm 0.066$  &  $0.002 \pm 0.055$  \\
DA RF         &  $0.644 \pm 0.027$  &  $0.016 \pm 0.023$  &  $0.477 \pm 0.049$  &  $0.008 \pm 0.045$  \\
DR GB         &  $0.761 \pm 0.037$  &  $0.003 \pm 0.031$  &  $0.712 \pm 0.133$  &  $0.001 \pm 0.089$  \\
DR EN         &  $0.901 \pm 0.150$  &  $0.006 \pm 0.100$  &  $0.708 \pm 0.093$  &  $0.001 \pm 0.078$  \\
GPS           &  $0.964 \pm 0.091$  &  $0.018 \pm 0.076$  &  $0.841 \pm 0.072$  &  $0.007 \pm 0.060$  \\
\hline
GCN           &  $0.725 \pm 0.015$  &  $0.544 \pm 0.012$  &  $0.747 \pm 0.041$  &  $0.566 \pm 0.035$  \\
GraphSAGE     &  $0.712 \pm 0.031$  &  $0.532 \pm 0.024$  &  $0.754 \pm 0.099$  &  $0.559 \pm 0.079$  \\
$1$-GNN       &  $0.722 \pm 0.052$  &  $\mathbf{0.546 \pm 0.041}$  &  $0.806 \pm 0.031$  &  $\mathbf{0.586 \pm 0.023}$  \\
\hline
\end{tabular}
  \caption{Intervention policy improvements on the Wave1 and Pokec semi-synthetic datasets under treatment capacity constraint with $p_t=0.5$. Note that \emph{only} $ \Delta S ( \hat{ \pi }_n^{ p_t } ) $  reflects the genuine policy improvement.}
  \label{tab:poly_diff_wave_pokec_tp0.5}
\end{table}  
\normalsize

\begin{table}[htp]
\centering
\small
\begin{tabular}{c | c c  }
  &  \multicolumn{2}{c}{Wave1} \\
  &  $ \Delta \hat{ S } ( \hat{ \pi }_n^{ p_t } ) $  &  $ \Delta  S ( \hat{ \pi }_n^{ p_t } ) $  \\
  \hline
DA GB         &  $0.291 \pm 0.031$  &  $0.004 \pm 0.026$   \\
DA RF         &  $0.310 \pm 0.041$  &  $0.003 \pm 0.032$   \\
DR GB         &  $0.102 \pm 0.057$  &  $0.002 \pm 0.048$   \\
DR EN         &  $0.360 \pm 0.044$  &  $0.002 \pm 0.037$   \\
GPS           &  $0.278 \pm 0.061$  &  $0.006 \pm 0.051$   \\
\hline
GCN           &  $0.279 \pm 0.029$  &  $0.179 \pm 0.026$   \\
GraphSAGE     &  $0.268 \pm 0.023$  &  $0.169 \pm 0.019$   \\
$1$-GNN       &  $0.310 \pm 0.022$  &  $\mathbf{0.201 \pm 0.016}$   \\
\hline
\end{tabular}
  \caption{Intervention policy improvements on the Wave1 semi-synthetic dataset under treatment capacity constraint with $p_t=0.3$. The policy network employed is another $1$-GNN. Note that \emph{only} $ \Delta S ( \hat{ \pi }_n^{ p_t } ) $  reflects the real policy improvement.}
  \label{tab:graph_poly_diff_wave_tp0.3}
\end{table}  
\normalsize

Until now, we have only employed a simple neural network as the policy network with feature vectors as input. For GNN-based methods, the policy learner can adjust its treatment rules according to the neighboring nodes' features and responses through the GNN-based causal estimators. However, through baseline estimators, e.g., doubly-robust estimators, a simple policy network cannot access the neighboring features of a node. Therefore, for a fair comparison, we employ another $1$-GNN as the policy network, and the evaluations on the Wave1 dataset are given in Table~\ref{tab:graph_poly_diff_wave_tp0.3}. The results further confirm that the accuracy of causal effect estimators is crucial for intervention policy optimization on interconnected units.

\section{Experiment Settings}
\label{sec:setting_app}

\subsection{GNN-based Estimators in Causal Inference Experiments}
For GNN-based estimators, we use Adam as a default optimizer with learning rate $0.001$ and weight decay $0.0001$. The number of total epochs is $20,000$; early stopping is employed by monitoring the loss on the validation set every $2000$ epochs. Hyperparameter $ \kappa $ in $ \mathcal{L}_{ \mathrm{est} } $ for penalizing the distribution discrepancy is searched from $ \{ 0.001, 0.005, 0.1, 0.2 \} $ for the Wave1 and Pokec datasets, and from $ \{ 0.1, 0.2, 0.5, 1.\} $ for the Amazon dataset. The feature map neural network $ \Phi $ has hidden dimensions $[64, 64]$ for the Wave1 and Pokec datasets, and $[256, 128, 128]$ for the Amazon dataset. GNNs have hidden dimensions $[128, 32]$ for the Wave1 and Pokec datasets, and $[256, 128, 64]$ for the Amazon dataset. Outcome prediction networks $h_0$ and $h_1$ have hidden dimensions $[64, 32]$ for the Wave1 and Pokec datasets, and $[256, 128, 64]$ for the Amazon dataset. ReLU is used as the activation function between hidden layers. Dropout is also employed between hidden layers with dropout rate a $0.5$.

\subsection{Baseline Estimators in Causal Inference Experiments}

For baseline models, learning rate of the DR EN model is searched from $ \{ 0.001, 0.01, 0.1 \}$ with maximal iteration $10000$. For the DA RF model, the number of estimators is searched from $ \{ 5, 10, 20 \} $, the maximal depth from $ \{ 5, 10, 20 \} $, and the minimum number of samples at a leaf node from $ \{ 5, 10, 20 \} $. For the DR GB and DA GB models, the number of estimators is searched from $ \{ 10, 50 \} $, and the maximal depth is searched from $ \{ 5, 10 \}$. In our experiments, the training procedure of Domain Adaption estimators for causal inference under interference is given as below
\begin{align*}
  \hat{\mu}_0 & = M_1 \left( Y_i^0 \sim [ \mathbf{X}_i^0 ; G_i ], \text{weights} = \frac{ g( \mathbf{X}_i^0 ) }{ 1 - g( \mathbf{X}_i^0 ) } \right), \\
  \hat{\mu}_1 & = M_2 \left( Y_i^1 \sim [ \mathbf{X}_i^1 ; G_i ], \text{weights} = \frac{ 1 - g( \mathbf{X}_i^1 ) }{ g( \mathbf{X}_i^1 ) } \right), \\
  \hat{D}_i^1 & = Y_i^1 - \hat{\mu}_0 ( [ \mathbf{X}_i^1, G_i ] ), \\
  \hat{D}_i^0 & = \hat{\mu}_1 ( [ \mathbf{X}_i^0; G_i] ) - Y_i^0,  \\
  \hat{ \tau } & = M_3 ( \hat{D}_i^0 | \hat{D}_i^1 \sim \mathbf{X}_i^0| \mathbf{X}_i^1 ), 
\end{align*} 
where $ M_1, M_2, M_3 $ are machine learning algorithms; $ Y_i^0, \mathbf{X}_i^0 $ represent the outputs and covariates of units under control in the training dataset, and $ Y_i^1, \mathbf{X}_i^1 $ under treatment. To capture the interference, the exposure variable $ G_i $ is concatenated to the covariates. $ g (  \mathbf{X}_i ) $ is an estimation of $ \Pr [ T_i = 1 | \mathbf{X}_i ] $ in the observational study using the Amazon dataset, while it is the predefined treatment probability $p$ in  randomized experiments using the Wave1 and Pokec datasets. Similarly, the training procedure of Doubly Robust estimators for causal inference under interference is given as 
\begin{align*}
  \hat{\mu}_0 & = M_1 ( Y_i^0 \sim [ \mathbf{X}_i^0 ; G_i ] ), \\
  \hat{\mu}_1 & = M_2 ( Y_i^1 \sim [ \mathbf{X}_i^1 ; G_i ] ), \\
  \hat{D}_i^1 & = \hat{\mu}_1 ( [ \mathbf{X}_i; G_i] ) + \frac{ Y_i - \hat{\mu}_1 ( [ \mathbf{X}_i; G_i] ) }{ g( \mathbf{X}_i ) } \mathds{ 1 } \{ T_i = 1 \}, \\
  \hat{D}_i^0 & = \hat{\mu}_0 ( [ \mathbf{X}_i; G_i] ) + \frac{ Y_i - \hat{\mu}_0 ( [ \mathbf{X}_i; G_i] ) }{ 1 - g( \mathbf{X}_i ) } \mathds{ 1 } \{ T_i = 0 \}, \\
  \hat{\tau} & = M_3 ( ( \hat{D}_i^1 - \hat{D}_i^0 ) \sim \mathbf{X}_i ), 
\end{align*}
where $ M_1, M_2, M_3 $ are machine learning algorithms; $ g (  \mathbf{X}_i ) $ is an estimation of $ \Pr [ T_i = 1 | \mathbf{X}_i ] $ in the observational study using the Amazon dataset, while it is the predefined treatment probability $p$ in  randomized experiments using the Wave1 and Pokec datasets.

\subsection{Intervention Policy Experiments}

Causal estimators with the best performance will be saved and fixed for the subsequent intervention policy improvement experiments on the same dataset. We use Adam as a default optimizer for the policy network with a learning rate of $0.001$. The policy network has hidden dimensions $[64, 32]$ for the Wave1 and Pokec datasets, and $[128, 64, 64]$ for the Amazon dataset. ReLU is employed as the activation function between hidden layers, and a sigmoid function is applied to the output. Treatment is then sampled from a Bernoulli distribution using the output of the policy network as the probability. The Gumbel-softmax trick~\cite{jang2016categorical} is employed such that errors can be back-propagated. Hyperparameter $ \gamma $ in $ \mathcal{L}_{ \mathrm{pol} } $ for enforcing the constraint is chosen from $\{ 5, 50, 100, 200, 500 \}$, such that the pre-defined constraint can be satisfied within the tolerance $\pm 0.01$. Besides, we also penalize the distribution discrepancy under the new intervention policy given by the policy network, and the hyperparameter for penalizing this term is chosen from $\{0.0, 0.0001, 0.001, 0.01, 0.1, 1\}$. The number of training epochs is $2000$, and each experiment is repeated $5$ times.

\section{Error Bound of Causal Estimators}
\label{sec:error_bound_proof}

In this section we will give a heuristic explanation why the causal estimators are difficult to obtain under interference. We briefly summarize the theoretical result of this section in the following claim.
\begin{my_claim}
GNN-based causal estimators restricted to a particular class for predicting the superimposed causal effects have an error bound $ \mathcal{O} ( \sqrt{ \frac{ D_{max}^3 \ln D_{ max } }{ n } } ) $, where $ D_{ max } := 1 + d_{ max } + d_{ max }^2 $ and $ d_{ max } $ is the maximal node degree in the graph.
  \label{claim:estimator}
\end{my_claim}
The above claim indicates that an accurate and consistent causal estimator is difficult with large network effects. Worse case is that the $ \frac{1}{ \sqrt{n} } $ convergence rate, or sample dependency, becomes unreachable when $ d_{ \max } (n) $ depends on the number of units, namely the maximal node degree increases with the number of nodes. The exact convergence rate of causal estimators is impossible to derive since it depends on the topology of the network, and it beyond the theoretical scope of this work. This claim will be used as one of the important assumptions for proving the policy regret bounds.

In the following, we will first motivate GNN-based causal estimators and then prove Claim~\ref{claim:estimator} step by step. First, with abuse of notation, we consider the following linear model with deterministic outcome 
\begin{equation}
  \mu_{ \star } ( \mathbf{X}_i, \mathbf{X}, \mathbf{T}, \mathcal{G} ) = T_i \tau_{ \star } ( \mathbf{X}_i ) + \alpha_1 \sum\limits_{ j \in \mathcal{N}_i }  T_j \tau_{ \star } ( \mathbf{X}_j ) + \alpha_2 \sum\limits_{ k \in \mathcal{N}_i^{ (2) } }  T_k \tau_{ \star } ( \mathbf{X}_k )
  \label{eq:deterministic_outcome}
\end{equation}
by setting $ Y_i ( T_i = 0 ) = 0 $, $ \alpha = 1 $ and letting $ \alpha_1 = \frac{1}{ | \mathcal{N}_i | } $, $ \alpha_2 = \frac{1}{ | \mathcal{N}_i^{ (2) } | } $, where $ \tau_{ \star } $ stands for the ground truth individual treatment response which is bounded by $ || \tau_{ \star } ||_{ \infty } \leq M $.

One motivation for employing localized graph convolution network, such as GraphSAGE, is that the surrogate model of a $2$-layer GraphSAGE can recover the linear model, especially, when $ \mathbf{T} = \mathbf{1} $. To be more specific, consider the following form of a $2$-layer GraphSAGE
\begin{align*}
  \mathbf{X}_i^{ (1) } & = \text{ ReLU } ( \mathbf{X}_i + \sum\limits_{ j \in \mathcal{N}_i } \mathbf{X}_j \mathbf{W}^{ (1) } )  \\
  \mathbf{X}_i^{ (2) } & = \text{ ReLU } ( \mathbf{X}_i^{ (1) } + \sum\limits_{ j \in \mathcal{N}_i } \mathbf{X}_j^{ (1) } \mathbf{W}^{ (2) } )  \\
  & = \text{ ReLU } [ \text{ ReLU }  ( \mathbf{X}_i + \sum\limits_{ j \in \mathcal{N}_i } \mathbf{X}_j \mathbf{W}^{ (1) } ) +  \sum\limits_{ j \in \mathcal{N}_i } \text{ ReLU } ( \mathbf{X}_j + \sum\limits_{ k \in \mathcal{N}_i^{ (2) } }  \mathbf{X}_k \mathbf{W}^{ (1) } ) \mathbf{W}^{ (2) }  ]. 
\end{align*}
A prediction from it reads $ o( \mathbf{X}_i ) =  \mathbf{X}_i^{ (2) \intercal  } \mathbf{v} $, where $ \mathbf{v} $ is a vector mapping the second hidden layer to the outcome prediction. In a \emph{surrogate model}~\footnote{ The \emph{surrogate models} of graph convolutional networks are first studied in~\cite{zugner2018adversarial} for designing adversarial attacks on GNNs and finding robust nodes.}, where an identity mapping replaces the ReLU activation function, the model returns the outcome prediction 
\begin{equation*}
  o_{surrogate} ( \mathbf{X}_i ) = \mathbf{X}_i^{ \intercal } \mathbf{v} + \sum\limits_{ j \in \mathcal{N}_i }  ( \mathbf{X}_j \mathbf{W}^{ (1) } + \mathbf{X}_j \mathbf{W}^{ (2) }  )^{ \intercal } \mathbf{v} + \sum\limits_{ k \in \mathcal{N}_i^{ (2) } }  ( \mathbf{X}_k \mathbf{W}^{ (1) } \mathbf{W}^{ (2) } )^{ \intercal } \mathbf{v}, 
\end{equation*} 
which correctly recovers the linear model and the simulation protocol of spillover effects when all units are assigned to treatment. Moreover, according to the universal approximation properties of GNNs \cite{scarselli2008computational}, $ \mu_{ \star } $ can be approximated. However, this claim cannot reflect an explicit dependence of estimation error on the graph structure. Hence, motivated by the \emph{surrogate model} and the universal approximation property, we study the following class of functions derived from the universal GNN. Let $ \mathcal{T} $ be a class of bounded functions with envelop $ M < \infty $ and finite VC-dimension $ VC ( \mathcal{T} ) < \infty $, and let 
\begin{equation}
  \mathcal{M}_{ GNN } := \{ \tau_1 + \cdots + \tau_{ D_{max} }, \ \tau_i \in \mathcal{T} \cup \{ 0 \} , \ i = 1, \dots, D_{max}, \ || \tau_1 + \cdots + \tau_{ D_{max} } ||_{ \infty } \leq 3M  \}, 
  \label{eq:m_gnn}
\end{equation}  
where $ D_{ max } $ is related to the maximal degree of the graph, for a $2$-layer GNN $ D_{ max } := 1 + d_{ max } + d_{ max }^2 $. Function from $ \mathcal{M}_{ GNN } $ takes $ ( \mathbf{ X }_i, \mathbf{X}_{ j \in \mathcal{N}_i } , \mathbf{X}_{ k \in \mathcal{N}_i^{ (2) } } )_{i=1}^n $ as input~\footnote{ Note that, treatment assignments can be combined with the covariates and fed into the function. In the experiments, we fed $ T_i \mathbf{X}_i $ into the GNNs, meaning that only covariates of treated units are non-zero. } and returns outcome prediction. The maximal subscript $ D_{ max } $ serves as a padding, to fit it, the function class $ \mathcal{T} $ is extended to $ \mathcal{T} \cup \{ 0 \}  $. As an example, one can find a function $ \mu_{ GNN } \in \mathcal{ M }_{ GNN } $ which approximates $ \mu_{ \star } ( \mathbf{X}_i, \mathbf{X}, \mathbf{T}, \mathcal{G} ) $ as 
\begin{equation*}
  \mu_{ GNN } ( \mathbf{X}_i, \mathbf{X}, \mathbf{T}, \mathcal{G} ) = \tau_{0} ( \mathbf{X}_i ) + \sum\limits_{ j \in \mathcal{N}_i } \tau_j ( \mathbf{X}_j ) + \sum\limits_{ k \in \mathcal{N}_i^{ (2) } } \tau_k ( \mathbf{X}_k ), 
\end{equation*}
where $ \tau_0, \tau_j, \tau_k \in \mathcal{T} $, for $ j \in \mathcal{N}_i $, $ k \in \mathcal{N}_i^{ (2) } $. In other words, there exists a function in the class $ \mathcal{M}_{ GNN } $ which, for every node in the network, only uses the representations of this node, this node's neighbors, and this node's 2-hop neighbors, similar to the surrogate model. Assumptions used in this section are summarized in Assumption~\ref{ass:assumption_estimator}.

\begin{my_assumption}{ \ }  \\
  (A1) Outcome simulation under interference follows the protocol given in Eq.~\ref{eq:deterministic_outcome} with $ || \mu_{ \star } ||_{ \infty } \leq 3 M $ due to the requirement $ || \tau_{ \star } ||_{ \infty } \leq M $.  \\
  (A2) Outcome prediction model is drawn from $ \mathcal{M}_{ GNN } $ defined in Eq.~\ref{eq:m_gnn}. \\
  (A3) There are no isolated nodes in the network~\footnote{ This assumption will be used later }. 
\label{ass:assumption_estimator}
\end{my_assumption}

Define the best approximation realized by the class $ \mathcal{M}_{ GNN } $ as
\begin{equation*}
  \tilde{ \mu }_{ GNN } := \mathrm{argmin}_{ \mu \in \mathcal{M}_{ GNN } } || \mu -  \mu_{ \star } ||_{ \infty }, 
\end{equation*}
and the approximation error
\begin{equation}
  \epsilon_{ GNN } := || \tilde{ \mu }_{ GNN } - \mu_{ \star } ||_{ \infty }. 
  \label{eq:epsilon_gnn}
\end{equation}
Moreover, define the optimal empirical estimator as
\begin{equation*}
  \hat{ \mu }_{ GNN } := \mathrm{argmin}_{  \mu \in \mathcal{M}_{ GNN } } \sum_{ i=1 }^n \ell ( \mu ( \mathbf{ X }_i, \mathbf{ X }, \mathbf{ T }, \mathcal{G} ), Y_i ).
\end{equation*}
Since both $ \tilde{ \mu }_{ GNN } $ and $ \hat{ \mu }_{ GNN } $ belong to the same class $ \mathcal{M}_{ GNN } $, it is easy to see
\begin{equation*}
  \mathbb{ E }_n [ \ell ( \tilde{ \mu }_{ GNN } ( \mathbf{ X }_i ) , Y_i ) ] \geq \mathbb{ E }_n [ \ell ( \hat{ \mu }_{ GNN } ( \mathbf{ X }_i  ), Y_i ) ], 
\end{equation*}
where we write $ \tilde{ \mu }_{ GNN } ( \mathbf{ X }_i )  $ and $ \hat{ \mu }_{ GNN } ( \mathbf{ X }_i ) $ for the sake of simplicity.

We can decompose the approximation error of the empirical causal estimator using the following fact
\begin{align*}
  & \mathbb{E} [ \ell ( \hat{ \mu }_{ GNN } ( \mathbf{X}_i ), Y_i ) - \ell ( \mu_{ \star } ( \mathbf{X}_i ), Y_i ) ] \\
  & \quad = \mathbb{E} \mathbb{E}_{ Y_i } [ \hat{ \mu }_{ GNN }^2 ( \mathbf{X}_i ) - 2 Y_i \hat{ \mu }_{ GNN } ( \mathbf{X}_i ) + 2 Y_i \mu_{ \star } ( \mathbf{X}_i ) - \mu_{ \star }^2 ( \mathbf{X}_i ) ]  \\
  & \quad  = \mathbb{E} [ \hat{ \mu }_{ GNN }^2 ( \mathbf{X}_i ) - 2 \mu_{ \star } ( \mathbf{X}_i ) \hat{ \mu }_{ GNN } ( \mathbf{X}_i ) + \mu_{ \star }^2 ( \mathbf{X}_i ) ] \\
  & \quad = \mathbb{E} [ ( \hat{ \mu }_{ GNN } ( \mathbf{X}_i ) - \mu_{ \star } ( \mathbf{X}_i ) )^2 ]. 
\end{align*}
It then yields 
\begin{align*}
  \mathbb{E} [ ( \hat{ \mu }_{ GNN } ( \mathbf{X}_i ) - \mu_{ \star } ( \mathbf{X}_i ) )^2 ] & = \mathbb{E} [ \ell ( \hat{ \mu }_{ GNN } ( \mathbf{X}_i ), Y_i ) - \ell ( \mu_{ \star } ( \mathbf{X}_i ), Y_i ) ]  \\
  & \leq \mathbb{E} [ \ell ( \hat{ \mu }_{ GNN } ( \mathbf{X}_i ), Y_i ) - \ell ( \mu_{ \star } ( \mathbf{X}_i ), Y_i ) ] \\
  & \quad - \mathbb{E}_n [ \ell ( \hat{ \mu }_{ GNN } ( \mathbf{X}_i ), Y_i ) ] + \mathbb{E}_n [ \ell ( \tilde{ \mu }_{ GNN } ( \mathbf{X}_i ), Y_i ) ]  \\
  & = \underbrace{ ( \mathbb{E} - \mathbb{E}_n ) [ \ell ( \hat{ \mu }_{ GNN } ( \mathbf{X}_i ), Y_i ) - \ell ( \mu_{ \star } ( \mathbf{X}_i ) , Y_i ) ] }_{ ( \mathrm{I} ) } \\
  & \quad + \underbrace{ \mathbb{E}_n [ \ell ( \tilde{ \mu }_{ GNN } ( \mathbf{X}_i ), Y_i ) - \ell ( \mu_{ \star } ( \mathbf{X}_i ), Y_i ) ] }_{ ( \mathrm{II} ) }. 
  \label{eq:decomp} 
\end{align*}

The second term $( \mathrm{II} )$ can be bounded by applying the Bernstein inequality. The following inequality holds with probability at least $ 1- \mathrm{e}^{ - \gamma } $
\begin{align}
  ( \mathrm{II} ) & \leq \mathbb{E} [ \ell ( \tilde{ \mu }_{ GNN } ( \mathbf{X}_i ), Y_i ) - \ell ( \mu_{ \star } ( \mathbf{X}_i ), Y_i ) ] + \sqrt{ \frac{ 2 C_{ \ell }^2 || \tilde{ \mu }_{ GNN } - \mu_{ \star } ||_{ \infty }^2 \gamma }{ n } } \nonumber \\
  & \quad + \frac{ 2 C_{ \ell } || \tilde{ \mu }_{ GNN } - \mu_{ \star } ||_{ \infty } \gamma }{ 3n }  \nonumber \\
  & = \mathbb{E} [ ( \tilde{ \mu }_{ GNN } ( \mathbf{X}_i ) - \mu_{ \star } ( \mathbf{X}_i ) )^2 ] + \sqrt{ \frac{ 2 C_{ \ell }^2 \epsilon_{ GNN }^2 \gamma }{ n } } + \frac{ 2 C_{ \ell } \epsilon_{ GNN } \gamma }{ 3n } \nonumber  \\
  & \leq \epsilon_{ GNN }^2 + \epsilon_{ GNN } \sqrt{ \frac{ 2 C_{ \ell }^2 \gamma }{ n } } + \frac{ 4 C_{ \ell } M \gamma }{ n }
\end{align}
using the facts $ || \ell ( \tilde{ \mu }_{ GNN } ( \mathbf{X}_i ), Y_i ) - \ell ( \mu_{ \star } ( \mathbf{X}_i ), Y_i ) ||_{ \infty } \leq C_{ \ell }  || \tilde{ \mu }_{ GNN } - \mu_{ \star } ||_{ \infty } $ and $ \epsilon_{ GNN } := || \tilde{ \mu }_{ GNN } - \mu_{ \star } ||_{ \infty } \leq 6M $ where $ C_{ \ell } $ represents the finite Lipschitz constant of loss function.

Furthermore, the first term $ ( \mathrm{I} ) $, the maximal deviation between empirical and true means, can be bounded using the standard symmetrization method (see Theorem 2.1 in~\cite{bartlett2005local}). Consider a class of functions $ \mathcal{F} $, for any $ f \in \mathcal{F} $, assume that $ ||f||_{ \infty } \leq F $ and $ \mathbb{V} [ f ] \leq V $. Then for every $ \gamma > 0 $, with probability at least $ 1 - \mathrm{e}^{ - \gamma } $ 
\begin{equation*}
  \sup_{ f \in \mathcal{F} } ( \mathbb{E} [f] - \mathbb{E}_n [f] ) \leq \inf_{ \alpha > 0 } \left( 2 (1 + \alpha) \mathcal{R}_n \mathcal{F} + \sqrt{ \frac{ 2 V \gamma }{ n } } + 2 F ( \frac{1}{3} + \frac{1}{ \alpha } ) \frac{ \gamma }{ n } \right), 
\end{equation*}
where $ \mathcal{R}_n \mathcal{F} $ indicates the Rademacher complexity of $ \mathcal{F} $. Hence, it gives 
\begin{equation}
  ( \mathrm{I} ) \leq 4 \mathcal{R}_n \{ \ell ( \mu ) - \ell ( \mu_{ \star } ): \mu \in \mathcal{M}_{ GNN }  \} + 6 \sqrt{ \frac{ 2 C_{ \ell }^2 M^2 \gamma }{ n } } + \frac{ 16 C_{\ell} M \gamma }{ n } 
\end{equation}
by setting $ \alpha = 1$ and using $ || \ell ( \mu ( \mathbf{X}_i ), Y_i ) - \ell ( \mu_{ \star } ( \mathbf{X}_i ), Y_i ) ||_{ \infty } \leq C_{ \ell }  || \mu - \mu_{ \star } ||_{ \infty } \leq 6 C_{ \ell } M $, $ \mathbb{V} [ \ell ( \mu ( \mathbf{X}_i ), Y_i ) - \ell ( \mu_{ \star } ( \mathbf{X}_i ), Y_i ) ] \leq \mathbb{E} [ ( \ell ( \mu ( \mathbf{X}_i ), Y_i ) - \ell ( \mu_{ \star } ( \mathbf{X}_i ), Y_i ) )^2 ] \leq 36 C_{ \ell }^2 M^2 $ for any $ \mu \in \mathcal{M}_{ GNN } $. Moreover, the Rademacher complexity term is defined as 
\begin{align}
  & \mathcal{R}_n \{ \ell ( \mu ) - \ell ( \mu_{ \star } ): \mu \in \mathcal{M}_{ GNN }  \} \nonumber \\
  & \quad := \mathbb{E}_{ \sigma } \left[ \left. \sup_{ \mu \in \mathcal{M}_{ GNN } } | \frac{1}{n} \sum\limits_{i=1}^n \sigma_i  ( \ell ( \mu ( \mathbf{X}_i ), Y_i ) - \ell ( \mu_{ \star } ( \mathbf{X}_i ), Y_i ) )  |  \right\vert \mathbf{X}, \mathbf{T}, \mathcal{G} \right]  \nonumber  \\
  & \quad \leq \underbrace{ C_{ \ell } \mathbb{E}_{ \sigma } \left[ \left. \sup_{ \mu \in \mathcal{M}_{ GNN } } | \frac{1}{n} \sum\limits_{i=1}^n \sigma_i  ( \mu ( \mathbf{X}_i ) - \mu_{ \star } ( \mathbf{X}_i ) )  |  \right\vert \mathbf{X}, \mathbf{T}, \mathcal{G} \right] }_{ ( \# ) }, 
  \label{eq:rad_1}
\end{align}
where $ \{ \sigma_i \}_{i=1}^n $ are Rademacher random variables. Before using the covering number arguments to further bound the Rademacher complexity term we introduce the following lemmas.

\begin{my_lemma} [Theorem 29.6 in \cite{devroye2013probabilistic}]
  Let $ \mathcal{F}_1, \dots, \mathcal{F}_k $ be classes of real functions on $ \mathbb{R}^d $. For $n$ arbitrary fixed points $ z_1^n = (z_1, \dots, z_n) $ in $ \mathbb{R}^d $, define the sets $ \mathcal{F}_1 ( z_1^n ), \dots, \mathcal{F}_k ( z_1^n ) $ by $ \mathcal{F}_j ( z_1^n ) = \{ f_j(z_1), \dots, f_j (z_n): f_j \in \mathcal{F}_j \} $, $ j = 1 \dots, k $. Also introduce $ \mathcal{F} = \{ f_1 + \cdots + f_k: f_j \in \mathcal{F}_j, \ j=1, \dots, k \} $. Then for every $ \epsilon > 0 $ and $ z_1^n $, 
  \begin{equation}
    \mathcal{N}_1 ( \epsilon, \mathcal{F} ( z_1^n ) ) \leq \prod\limits_{ j=1 }^k \mathcal{N}_1 ( \epsilon / k , \mathcal{F}_j ( z_1^n ) ). 
  \end{equation}
\label{lemma:cover_add_fun}
\end{my_lemma}

\begin{my_lemma}
  Let $ \mathcal{F}_1, \dots, \mathcal{F}_k $ be classes of bounded real functions on $ \mathbb{R}^d $ with envelop $F$ and finite VC-dimension $ v < \infty $, for $ 3 \leq k \leq K $. Also introduce $ \mathcal{F} = \{ f_1 + \cdots + f_k, f_j \in \mathcal{F}_j, \ j = 1, \dots, k \} $ and let $ \mathcal{F} ( z_1^n ) = \{ f(z_1), \dots, f(z_n), f \in \mathcal{F} \} $ for arbitrary fixed points $z_1^n $ in $ \mathbb{R}^d $. Then we have the following bound 
  \begin{equation}
    \mathbb{E}_{ \sigma } [ \sup_{ f \in \mathcal{F} } | \frac{1}{n} \sum\limits_{ i=1 }^n \sigma_i f ( z_i ) | ] \leq C_F \sqrt{ \frac{ k v \ln k }{ n } },
  \end{equation}
where $ C_F $ is a constant which depends only on the envelop. 
\label{lemma:k_rad_bound}
\end{my_lemma}
\begin{proof}
  According to the Theorem 5.22 in~\cite{wainwright2019high}, the Rademacher complexity term is bounded as 
\begin{equation*}
  \mathbb{E}_{ \sigma } [ \sup_{ f \in \mathcal{F} } | \frac{1}{n} \sum\limits_{ i=1 }^n \sigma_i f ( z_i ) | ] \leq \underbrace{ \frac{ 32 }{ \sqrt{n} } \int_0^{ 2F } \sqrt{ \ln \mathcal{N}_1 ( \epsilon, \mathcal{F} ( z_1^n ) ) } \ d \epsilon }_{ ( \star ) }. 
\end{equation*}
Using Lemma~\ref{lemma:cover_add_fun} and $ \mathcal{N}_1 ( \epsilon, \mathcal{F} ) \leq \mathcal{N}_2 ( \epsilon, \mathcal{F} ) $, it gives  
\begin{equation*}
  ( \star ) \leq \frac{ 32 }{ \sqrt{n} } \int_0^{ 2F } \sqrt{ \sum_{j=1}^k \ln \mathcal{N}_2 ( \epsilon / k , \mathcal{F}_j ( z_1^n ) ) }.
\end{equation*}
Moreover, a uniform entropy bound for the covering number is given by the Theorem 2.6.7 in~\cite{van1996weak}. A small modification gives
\begin{equation*}
  \mathcal{N}_2 ( \epsilon, \mathcal{F}_j ( z_1^n ) ) \leq C ( v + 1 ) (16 e)^{ ( v + 1 ) } ( k / \epsilon )^{ 2 v }, j = 1, \dots, k, 
\end{equation*}
where $C$ is a universal constant. Furthermore, following the same technique used by Eq. A.6 in~\cite{kitagawa2018should}, we obtain
\begin{align*}
  ( \star ) & \leq \frac{ 32 }{ \sqrt{n} } \sqrt{k} \int_0^{ 2F } \sqrt{ \ln C + \ln (v+1) + (v+1) \ln (16e) +2v \ln k - 2v \ln \epsilon } \ d \epsilon  \\
  & \stackrel{ (1) }{ \leq } \frac{ 32 }{ \sqrt{n} } \sqrt{k v } \int_0^{ 2F } \sqrt{ \ln C + \ln 2 + \ln (16e) + 2 \ln k - 2 \ln \epsilon } \ d \epsilon  \\
  & \stackrel{ (2) }{ \leq } \frac{ 32 }{ \sqrt{n} } \sqrt{k v \ln k } \int_0^{ 2F } \sqrt{ \ln C + \ln 2 + \ln (16e) + 2 - 2 \ln \epsilon / \ln K } \ d \epsilon := C_F \sqrt{ \frac{ k v \ln k }{ n } },  
\end{align*} 
where $(1)$ uses the fact that usually $v$ is large enough and $(2)$ is due to the condition $ 3 \leq k \leq K $. 
\end{proof}

Now, we can further bound the term $ \mathcal{R}_n \{ \ell ( \mu ) - \ell ( \mu_{ \star } ): \mu \in \mathcal{M}_{ GNN } \} $ after Eq.~\ref{eq:rad_1}. Note that
\begin{align}
  ( \# ) &  = C_{ \ell } \mathbb{E}_{ \sigma } \Bigg[ \sup_{ f \in \mathcal{M}_{ GNN } } \Bigg|   \frac{1}{n} \sum\limits_{i=1}^n  \sigma_i [ ( \tau_0 ( \mathbf{X}_i ) - T_i \tau_{ \star } ( \mathbf{X}_i ) ) + \sum\limits_{ j \in \mathcal{N}_i } ( \tau_{j} ( \mathbf{X}_j ) - T_j \tau_{ \star } ( \mathbf{X}_j ) ) \nonumber \\
  & \quad \quad  + \sum\limits_{ k \in \mathcal{N}_i^{ (2) } } ( \tau_k ( \mathbf{X}_k ) - T_k \tau_{ \star } ( \mathbf{X}_k ) )  ]    \Bigg| \Bigg]. 
  \label{eq:rad_2}
\end{align}
Define a new constant for each node $ D_i := 1 + | \mathcal{N}_i | + | \mathcal{N}_i^{ (2) } | $, $ i = 1, \dots, n $. According to (A3) in Assumption~\ref{ass:assumption_estimator}, we have $ D_i \geq 3 $. Also introduce a new class of function $ \Omega := \{ \mathcal{T} \pm \tau_{ \star } \} $. Note that class $ \Omega $ has the same VC-dimension as $ \mathcal{T} $, i.e., $ VC ( \Omega ) = VC ( \mathcal{T} ) $, and $ || \omega ||_{ \infty } \leq 2 M $ for any $ \omega \in \Omega $. Recall the definition of $ D_{max} := 1 + d_{max} + d_{max}^2 $. By decomposing the node subscript $i$ into groups with the same $ D_i $, Eq.~\ref{eq:rad_2} can be further written as 
\begin{align*}
  ( \# ) & = C_{ \ell } \mathbb{E}_{ \sigma } \left[ \sup_{ f \in \mathcal{M}_{ GNN } } \left\vert  \frac{1}{n} \sum\limits_{ i=1 }^n \sigma_i \sum\limits_{ l=1 }^{ D_i } \omega_l ( \mathbf{X}_i, \mathbf{X}, \mathbf{T}, \mathcal{G} )  \right\vert \right]  \quad \omega_l \in \Omega  \\
  & = C_{ \ell } \mathbb{E}_{ \sigma } \left[ \sup_{ f \in \mathcal{M}_{ GNN } } \left\vert  \sum\limits_{ k=3 }^{ D_{max} } \frac{1}{n} \sum\limits_{ i: D_i = k } \sigma_i \sum\limits_{ l=1 }^k \omega_l ( \mathbf{X}_i, \mathbf{X}, \mathbf{T}, \mathcal{G} ) \right\vert \right]  \\
  & \stackrel{ (1) }{ \leq } C_{ \ell } \sum\limits_{ k=3 }^{ D_{max} }  \mathbb{E}_{ \sigma } \left[ \sup_{ f \in \mathcal{M}_{ GNN } } \left\vert \frac{1}{n} \sum\limits_{ i: D_i = k } \sigma_i \sum\limits_{ l=1 }^k \omega_l ( \mathbf{X}_i, \mathbf{X}, \mathbf{T}, \mathcal{G} )  \right\vert \right]  \\
  & \stackrel{ (2) }{ \leq } C_{ \ell } C_F \sum\limits_{ k=3 }^{ D_{max} } \frac{1}{n} \sqrt{ | i: D_i = k | k VC ( \mathcal{T} ) \ln k } \leq C_{ \ell } C_F  \sum\limits_{ k=3 }^{ D_{max} }  \sqrt{ \frac{ k VC ( \mathcal{T} ) \ln k }{ n } } \\
  & \leq C_{ \ell } C_F \sqrt{ \frac{ D_{max}^3 VC ( \mathcal{T} ) \ln D_{ max } }{ n } }, 
\end{align*}
where $(1)$ uses the triangle inequality and $(2)$ uses Lemma~\ref{lemma:k_rad_bound}. Hence, the $ ( \mathrm{I } ) $ term is bounded by
\begin{equation*}
  ( \mathrm{I} ) \leq 4 C_{ \ell } C_F \sqrt{ \frac{ D_{max}^3 VC ( \mathcal{T} ) \ln D_{ max } }{ n } } +  6 \sqrt{ \frac{ 2 C_{ \ell }^2 M^2 \gamma }{ n } } + \frac{ 16 C_{\ell} M \gamma }{ n }. 
\end{equation*}
By combining $ ( \mathrm{I} ) $ and $ ( \mathrm{ II } ) $ we have the following theorem.

\begin{my_theorem}
  Suppose Assumption \ref{ass:assumption_estimator} holds. Let $ \hat{ \mu }_{ GNN } $ be the optimal causal estimator obtained by minimizing an empirical loss function using the data $ \{ \mathbf{X}_i, \mathbf{X}, \mathbf{T}, \mathcal{G} \}_{ i=1 }^n $. Suppose that the loss function has a finite Lipschitz constant $  C_{ \ell } $ and $ \hat{ \mu }_{ GNN } $ is restricted to $ \mathcal{M}_{ GNN } $, Then with probability at least $ 1 - 2 \mathrm{e}^{ - \gamma } $, the causal estimator under interference has an error bound
  \begin{align}
    \mathbb{E} [ ( \hat{ \mu }_{ GNN } ( \mathbf{X}_i ) - \mu_{ \star } ( \mathbf{X}_i ) )^2 ] & \leq  4 C_{ \ell } C_F \sqrt{ \frac{ D_{max}^3 VC ( \mathcal{T} ) \ln D_{ max } }{ n } }  + 6 \sqrt{ \frac{ 2 C_{ \ell }^2 M^2 \gamma }{ n } } \nonumber \\
    &  + \epsilon_{ GNN } \sqrt{ \frac{ 2 C_{ \ell }^2 \gamma }{ n } } + \frac{ 20 C_{ \ell } M \gamma }{ n } + \epsilon_{ GNN }^2, 
  \end{align}
where $ \epsilon_{ GNN } $ is defined in Eq.~\ref{eq:epsilon_gnn}. 
\label{theo:estimator_error_bound}
\end{my_theorem}
Keeping only the leading term with $ D_{ max } $, under network interference, the causal estimator has an error bound $ \mathcal{O} ( \sqrt{ \frac{ D_{max}^3 \ln D_{ max } }{ n } } ) $. It indicates that an accurate causal estimator is difficult to obtain under large network interference. Recall that the prediction outcome from the GNN causal estimator is actually the superposition of individual treatment effect and spillover effect. Hence, it is expected that, similarly, the individual treatment effect becomes more and more difficult to recover under more substantial network interference. This intuitive expectation can be observed in the following experimental results in Table~\ref{tab:increased_error}. We observe that the error of individual treatment effect estimator increases from $k=1$ to $k=4$.

\begin{table}[thp]
\centering
\begin{tabular}{ c | c c c }
 & $ k = 1$ & $k=2$ & $ k = 4 $  \\
 \hline
 GraphSAGE  & $0.048$ & $0.129$ & $0.152$  \\
\end{tabular}
\caption{$ \epsilon_{ PEHE } $ on the semi-synthetic Wave1 data with $p=0.1$, $\alpha=0.5$, and $k=1, 2, 4$. To fit the theoretical analysis, exposure level is not fed into the model.}
\label{tab:increased_error}
\end{table}

\section{Policy Regret Bound}
\label{subsec:policy_regret_proof}

In this section, we provide a regret bound for the intervention policy that employs GNN-based causal estimators. The policy regret bound is first summarized in the following theorem. 
\begin{my_theorem}
  By Assumption~\ref{ass:ass_policy_regret}, for any small $\epsilon > 0$, the policy regret is bounded by $ \mathcal{R} ( \hat{ \pi }_n ) \leq 2 \left( \frac{\alpha_{\tau}}{n^{\zeta_{\tau}}} +  \frac{\alpha_{\delta}}{n^{\zeta_{\delta}}} \right) + 2 \epsilon $ with probability at least
\begin{equation*}
  1 - \mathcal{N} \left( \Pi, \frac{ \epsilon }{ 4 ( 2M_1 + 2M_2 + L ) } \right) \exp \left( - \frac{ n \epsilon^2 }{ 32 ( d_{ \max }^2 + 1 ) ( M_1 + M_2 )^2 } \right)
\end{equation*}
where $ \mathcal{N} \left( \Pi, \frac{ \epsilon }{ 4 ( 2M_1 + 2M_2 + L ) } \right) $ indicates the covering number~\footnote{The covering number characterizes the capacity of a functional class. Definition is provided in the Appendix~\ref{subsec:policy_regret_proof}} on the functional class $ \Pi $ with radius $ \frac{ \epsilon }{ 4 ( 2M_1 + 2M_2 + L ) } $, and $ d_{ \max } $ is the maximal node degree in the graph $ \mathcal{G} $. 
  \label{theorem:regret_bound}
\end{my_theorem}

Suppose that the policy functional class $ \Pi $ is finite and its capacity is bounded by $ | \Pi | $.  According to Theorem~\ref{theorem:regret_bound}, with probability at least $ 1 - \delta $, the policy regret is bounded by
\begin{align*}
  \mathcal{R} ( \hat{ \pi }_n ) & \leq 2 \left( \frac{\alpha_{\tau}}{n^{\zeta_{\tau}}} +  \frac{\alpha_{\delta}}{n^{\zeta_{\delta}}} \right) + 8 ( M_1 + M_2 ) \sqrt{ \frac{2 ( d_{ \max }^2 + 1 ) }{n} \log \frac{ | \Pi | }{ \delta } } \\        & \approx 2 \left( \frac{\alpha_{\tau}}{n^{\zeta_{\tau}}} +  \frac{\alpha_{\delta}}{n^{\zeta_{\delta}}} \right) + 8 d_{ \max } ( M_1 + M_2 ) \sqrt{ \frac{ 2 }{n} \log \frac{ | \Pi | }{ \delta } }
\end{align*}
It indicates that optimal policies are more difficult to find in a dense graph even under weak interactions between neighboring nodes.

Throughout the estimation of policy regret, we maintain the following assumptions.  
\begin{my_assumption}{\ } \\ 
  (BO) Bounded treatment and spillover effects: There exist $ 0 < M_1, M_2 < \infty $ such that the individual treatment effect satisfies $ | \tau_i | \leq M_1 $ and the spillover effect satisfies $ \forall \pi \in \Pi, | \delta_i ( \pi ) | \leq M_2 $.  \\
  (WI) Weak independence assumption: For any node indices $i$ and $j$, the weak independence assumption assumes that $ \mathbf{X}_i \bot \mathbf{X}_j \ \text{if} \ A_{ij} = 0 \text{, or} \ \nexists  k \ \text{with} \ A_{ik} = A_{kj} = 1 $. \\
  (LIP) Lipschitz continuity of the spillover effect w.r.t. policy: Given two treatment policies $ \pi_1 $ and $ \pi_2 $, for any node $i$ the spillover effect satisfies $ | \delta_i ( \pi_1 ) - \delta_i ( \pi_2 ) | \leq L || \pi_1 - \pi_2 ||_{ \infty } $, where the Lipschitz constant satisfies $ L > 0$ and $ || \pi_1 - \pi_2 ||_{ \infty } := \sup_{ \mathbf{X} \in \boldsymbol{ \chi } } | \pi_1 ( \mathbf{X} ) - \pi_2 ( \mathbf{X} ) |  $.  \\
  (ES) Uniformly consistency: after fitting experimental or observational data on $ \mathcal{G} $, individual treatment effect estimator satisfies
\begin{equation*}
   \frac{1}{n} \sum_{i=1}^n | \tau_i - \hat{ \tau }_i | < \frac{ \alpha_{\tau} }{ n^{\zeta_{\tau}} }, 
\end{equation*}  
and spillover estimator satisfies
\begin{equation}
  \forall \pi \in \Pi, \ \frac{1}{n} \sum_{i=1}^n | \delta_i ( \pi ) - \hat{ \delta }_i ( \pi ) | < \frac{ \alpha_{ \delta } }{ n^{ \zeta_{\delta} } }
\end{equation}
where $\alpha_{\tau} > 0 $ and $ \alpha_{ \delta } > 0 $ are scaling factors that characterize the errors of estimators. $ \zeta_{\tau} $ and $ \zeta_{ \delta }$ control the convergence rate of estimators for individual treatment effect and spillover effect, respectively, which satisfy $ 0 < \zeta_{\tau}, \zeta_{\delta} < 1 $.  
  \label{ass:ass_policy_regret}
\end{my_assumption}

Before proving Theorem~\ref{theorem:regret_bound} step by step, we first discuss the plausibility of Assumption~\ref{ass:ass_policy_regret}. Notice that the (ES) assumption requires consistent estimators of the individual treatment effect and the spillover effect, which is the fundamental problem of causal inference with interference. In our GNN-based model, these empirical errors are particularly difficult to estimate due to the lack of proper theoretical tools for understanding GNNs. To grasp how these GNN-based causal estimators are influenced by the network structure and network effect, in Appendix~\ref{sec:error_bound_proof}, we have studied a particular class of GNNs, which is inspired by the \emph{surrogate model} of nonlinear graph neural networks and derived Claim~\ref{claim:estimator}. Claim~\ref{claim:estimator} indicates that the $ \frac{1}{ \sqrt{n} } $ error bound of GNN-based causal estimators might be unreachable when $ d_{ \max } (n) $ depends on the number of units. Therefore, in the (ES) assumption, we assume the coefficients $\zeta_{\tau}$ and $\zeta_{\delta}$ to characterize the convergence rates, which is line with the assumption made in~\cite{athey2017efficient} (see Assumption 2 of~\cite{athey2017efficient}).

Besides, (LIP) assumes that the change of received spillover effect is bounded after modifying the treatment assignments of one unit's neighbors. This assumption is plausible, at least, in the synthetic experiments. For instance, consider the spillover effect in the simulated experiments generated by $ \delta_i ( \pi ) = \alpha \frac{1}{ | \mathcal{N}_i | } \sum_{ j \in \mathcal{N}_i } \pi ( \mathbf{X}_j ) \tau ( \mathbf{X}_j ) $ (see Eq.~\ref{eq:simulated_peer_effect}), then we can see
\begin{equation*}
  | \delta_i ( \pi_1 ) - \delta_i ( \pi_2 ) | \leq \alpha \frac{1}{ | \mathcal{N}_i | } \sum_{j \in \mathcal{N}_i } M_1 | \pi_1 ( \mathbf{X}_j ) - \pi_2 ( \mathbf{X}_j ) | \leq \alpha M_1 || \pi_1 - \pi_2 ||_{\infty}. 
\end{equation*}
Hence, in this example $L = \alpha M_1 $.

The underlying difficulty of estimating the intervention policy regret is the networked setting. Weak independence assumption (WI) allows us to use hypergraph-based method and derive concentration inequalities for the networked random variables. We will use hypergraph techniques, instead of chromatic number arguments, to give a tighter bound of policy regrets. Another advantage is that the weak independence (WI) assumption can be relaxed to support longer dependencies on the network. However, by relaxing (WI), the power of $ d_{ \max } $ in the regret bound needs to be modified correspondingly. For example, if we assume a next-nearest neighbors dependency of covariates, i.e., $ \mathbf{X}_i \perp \mathbf{X}_j $ for $ j \not\in {i} \cup \mathcal{N}_i \cup \mathcal{N}_i^{ (2) } $, then the term $ d_{ \max }^2 $ in Theorem~\ref{theorem:regret_bound} needs to be modified to $ d_{ \max }^4 $. This change remains the same for the policy regret bound under capacity constraint, which will be provided in Theorem~\ref{theorem:regret_bound_constrained}.

The flow of the proof for Theorem~\ref{theorem:regret_bound} can be summarized as: Under (WI) and (BO), we use concentration inequalities of networked random variables defined on a hypergraph, which is derived from graph $ \mathcal{G} $ to bound the convergence rate. Besides, using the Lipschitz assumption (LIP) enables us to estimate the covering number of the policy functional class $ \Pi $.

Concentration inequalities on partly dependent random variables are first given in~\cite{janson2004large}. Later,  \cite{Wang2017learning} provides tighter concentration inequalities using hypergraph and weak dependence assumption. A hypergraph is a generalization of graph in which a hyperedge groups a number of vertices in the graph. For instance, consider a graph with $n$ vertices, and let $ \mathcal{N} = \{ v_1, v_2, \dots, v_n \} $ represent the set of vertices. Hyperedges set $ \mathcal{E}_h = \{ e_{h, 1}, e_{h, 2} \cdots, e_{h, m} \} $ represents instances joining a number of vertices. In the following, let $ \mathcal{G}_h = ( \mathcal{N}, \mathcal{E}_h ) $ denote a hypergraph.

\begin{my_definition}[Definition 1 in~\cite{Wang2017learning}]
  Given a hypergraph $ \mathcal{G}_h $, we call $ \{ \xi_i \}_{ i=1 }^n $ $\mathcal{G}_h$-networked random variables if there exist functions $f_i : \boldsymbol{ \chi }^{ \otimes | e_{h, i} | } \rightarrow \mathbb{R} $ such that $ \xi_i = f_i ( \{ \mathbf{X}_v | v \in e_{h, i} \} ) $, where $  \{ \mathbf{X}_v | v \in e_{h, i} \}  $ represents the set of covariates of the vertices in the hyperedge $ e_{h, i} $.   
\end{my_definition}

Furthermore, we have the following concentration inequality.

\begin{my_theorem}[Corollary 7 in~\cite{Wang2017learning}]
  Let $ \{ \xi_i \}_{i=1}^n $ be $\mathcal{G}_h$-networked random variables with mean $ \mathbb{E} [ \xi_i ] = \mu $, and satisfying $ a < \xi_i < b $, $ \forall i \in \{ 1, 2, \dots, n \} $. Then for all $ \epsilon > 0 $, 
  \begin{equation}
    \Pr \left( \left| \frac{1}{n} \sum_{i=1}^n \xi_i - \mu\right| \geq \epsilon \right) \leq \exp \left( - \frac{ n \epsilon^2 }{ 2 \omega_{ \mathcal{G}_h } ( b - a )^2 }  \right), 
    \label{eq:concentration} 
  \end{equation}
where $ \omega_{ \mathcal{G}_h } := \max_{ v \in \mathcal{N} } | \{ e_h : v \in e_h \} | $ represents the maximal degree of $ \mathcal{G}_h $. 
  \label{theo:concentration} 
\end{my_theorem}

Recall the following definitions of utility functions $S_n^{\tau, \delta} ( \pi )$, $ \hat{S}_n^{ \tau, \delta } ( \pi )$, and $ S ( \pi ) $ 
\begin{align*}
  & S ( \pi ) := \mathbb{E} [ (2 \pi ( \mathbf{X}_i ) - 1 ) ( \tau_i + \delta_i ( \pi ) ) ]  \\
  & S_n^{ \tau, \delta } ( \pi ) := \frac{1}{n} \sum_{i=1}^n (2 \pi ( \mathbf{X}_i ) - 1 ) ( \tau_i + \delta_i ( \pi ) )  \\
  & \hat{S}_n^{ \tau, \delta } ( \pi ) := \frac{1}{n} \sum_{i=1}^n (2 \pi ( \mathbf{X}_i ) - 1 ) ( \hat{\tau}_i + \hat{\delta}_i ( \pi ) ), 
\end{align*}
where the policy $ \pi $ function has output in $ [0, 1] $. An optimal empirical policy is obtained via $ \hat{ \pi }_n \in \mathrm{argmax}_{ \pi \in \Pi } \hat{S}_n^{ \tau, \delta } ( \pi ) $. Note that in the definition of $ S ( \pi ) $ we still keep the subindex $i$ to emphasize the dependence of spillover effect on neighboring nodes. Next we provide several lemmas related to the utility functions.

\begin{my_lemma}  
  Let $ \mathcal{S} (\pi) := S_n^{\tau, \delta} ( \pi ) - S ( \pi ) $, for any $ \pi_1, \pi_2 \in \Pi $, where the policy class in contained in $[0, 1]$, according to the assumptions (BO) and (LIP) we have
  \begin{equation*}
    | \mathcal{S} ( \pi_1 ) - \mathcal{S} ( \pi_2 ) | \leq 2 ( 2 M_1 + 2 M_2 + L) || \pi_1 - \pi_2 ||_{\infty}
  \end{equation*}
  \label{lemma:s1_s2_lemma}
\end{my_lemma} 
\begin{proof}
  First note that  $ | \mathcal{S} ( \pi_1 ) - \mathcal{S} ( \pi_2 ) | \leq | S ( \pi_1 ) - S ( \pi_2 ) | + | S_n^{\tau, \delta} ( \pi_1 ) - S_n^{\tau, \delta} ( \pi_2 ) | $, and we have
  \begin{align*}
    & | S ( \pi_1 ) - S( \pi_2 ) | = |  \int_{ \boldsymbol{ \chi } } ( 2 \pi_1 ( \mathbf{X}_i ) - 1 ) ( \tau_i + \delta_i ( \pi_1 ) ) - ( 2 \pi_2 ( \mathbf{X}_i ) - 1 ) ( \tau_i + \delta_i ( \pi_2 ) ) \ d \mathbf{X}_i |  \\
    & \leq \int_{ \boldsymbol{ \chi } } 2 | \tau_i | || \pi_1 - \pi_2 ||_{ \infty } + | ( 2 \pi_1 ( \mathbf{X}_i ) - 1 ) ( \delta_i ( \pi_2 ) + L || \pi_1 - \pi_2 ||_{ \infty } ) - ( 2 \pi_2 ( \mathbf{X}_i ) -1 ) \delta_i ( \pi_2 ) | \ d \mathbf{X}_i  \\
    & = \int_{ \boldsymbol{ \chi } } 2 | \tau_i | || \pi_1 - \pi_2 ||_{ \infty } + | 2 ( \pi_1 ( \mathbf{X}_i ) - \pi_2 ( \mathbf{X}_i ) ) \delta_i ( \pi_2 ) + L ( 2 \pi_1 ( \mathbf{X}_i ) - 1 ) || \pi_1 - \pi_2 ||_{ \infty }  | \ d \mathbf{X}_i \\
    & \leq ( 2 | \tau_i | + 2 | \delta_i ( \pi_2 ) | + L ) || \pi_1 - \pi_2 ||_{\infty} \\
    & \leq ( 2 M_1 + 2 M_2 + L) || \pi_1 - \pi_2 ||_{\infty}. 
  \end{align*}  
Similarly, we have $ | S_n^{\tau, \delta} ( \pi_1 ) - S_n^{\tau, \delta} ( \pi_2 ) | \leq ( 2 M_1 + 2 M_2 + L) || \pi_1 - \pi_2 ||_{\infty} $. 
\end{proof}

Using the concentration inequality in Theorem~\ref{theo:concentration} we can obtain the convergence rate of the worst-case utility regret. We also use a capacity measure of the policy functional class $ \Pi $, namely the covering number, to prove the convergence rate, which is defined in the following. 
\begin{my_definition}[Definition 3.1 in \cite{cucker2007learning}]
  Let $ \Pi $ be a metric space and $ \epsilon > 0 $, the covering number $ \mathcal{N} ( \Pi, \epsilon ) $ is defined as the minimal $ l \in \mathbb{N} $ such that there exist $ l $ disks in $ \Pi $ with radius $ \epsilon $ covering $ \Pi $.  
\end{my_definition}

\begin{my_lemma}
  Under Assumption~\ref{ass:ass_policy_regret}, for any $ \{ \mathbf{X}_i \}_{ i=1 }^n \in \boldsymbol{ \chi }^{ \otimes n } $ and $\epsilon > 0$, it satisfies 
  \begin{align}
    & \Pr \left( \sup\limits_{ \pi \in \Pi } | S_n^{ \tau, \delta } ( \pi ) - S ( \pi ) | \leq \epsilon \right)  \nonumber \\
    & \quad \geq 1 - \mathcal{N} \left( \Pi, \frac{ \epsilon }{ 4 ( 2M_1 + 2M_2 + L ) } \right) \exp \left( - \frac{ n \epsilon^2 }{ 32 ( d_{ \max }^2 + 1 ) ( M_1 + M_2 )^2 } \right), 
  \end{align}
where $ \mathcal{N} \left( \Pi, \frac{ \epsilon }{ 4 ( 2M_1 + 2M_2 + L ) } \right) $ represents the covering number on the policy functional class $ \Pi $ with radius $ \frac{ \epsilon }{ 4 ( 2M_1 + 2M_2 + L ) } $. 
  \label{lemma:sup_bound}
\end{my_lemma}
\begin{proof}
  According to the assumption (BO), the summands are bounded as $ | ( 2 \pi ( \mathbf{X}_i ) - 1 ) ( \tau_i + \delta_i ( \pi ) ) | \leq M_1 + M_2 $, $ \forall i \in \{1, \dots, n \}$. Given the graph $ \mathcal{G} = ( \mathcal{N}, \mathcal{E} ) $ and its corresponding adjacency matrix $A$, using the weak independence assumption (WI) a dependence hypergraph can be defined as $ \mathcal{G}_h = ( \mathcal{N}, \mathcal{E}_h ) $, where a hyperedge $ e_{h, i} \in \mathcal{E}_h $ is defined as $ e_{h, i} := \{ v_i \} \cup \{ v_j| j \in \mathcal{N}_i \} \cup \{ v_k | \exists j :  A_{ij} = 1 \land A_{jk} = 1 \} $. Therefore, the maximal degree of the hypergraph $ \mathcal{G}_h $ satisfies $ \omega_{ \mathcal{G}_h } \leq d_{ \max }^2 + 1 $, where $ d_{ \max } $ indicates the maximal vertex degree of the graph $ \mathcal{G} $. Via Theorem~\ref{theo:concentration}, we have 
  \begin{equation}
    \Pr \left( | S_n^{\tau, \delta} ( \pi ) - S ( \pi ) | \geq \epsilon \right) \leq \exp \left( - \frac{ n \epsilon^2 }{ 8 ( d_{ \max }^2 + 1 ) ( M_1 + M_2 )^2 }  \right), \ \forall \pi \in \Pi. 
  \end{equation}
Let $ l = \mathcal{N} \left( \Pi, \frac{ \epsilon }{ 2 (2 M_1 + 2 M_2 + L) } \right) $ denote the covering number. Consider policies $ \pi_j$, with $j \in \{1, \dots, l \}$ located in the center of disks $D_j$ with radius $ \frac{ \epsilon }{ 2 (2 M_1 + 2 M_2 + L) } $ which cover the policy functional class $\Pi $. Recall the definition $ \mathcal{S} ( \pi ) := S_n^{ \tau, \delta } ( \pi ) - S ( \pi ) $, by Lemma~\ref{lemma:s1_s2_lemma}, for any $ \pi_j$ and $ \pi \in D_j $, we have 
\begin{equation*}
  | \mathcal{S} ( \pi ) - \mathcal{S} ( \pi_j ) | \leq 2 ( 2 M_1 + 2 M_2 + L) \frac{ \epsilon }{ 2 (2 M_1 + 2 M_2 + L) } = \epsilon.  
\end{equation*}
Then $ \forall \pi \in D_j $, $ \sup_{ \pi \in D_j }  \mathcal{S} ( \pi ) \geq 2 \epsilon  \Rightarrow \mathcal{S} ( \pi_j ) \geq \epsilon $, which indicates
\begin{equation*}
  \Pr ( \sup_{ \pi \in D_j }  \mathcal{S} ( \pi ) \geq 2 \epsilon ) \leq \Pr ( \mathcal{S} ( \pi_j ) \geq \epsilon ) \leq \exp \left( - \frac{ n \epsilon^2 }{ 8 ( d_{ \max }^2 + 1 ) ( M_1 + M_2 )^2 }  \right). 
\end{equation*}
Since $ \Pi = D_1 \cup \cdots \cup D_l $, it is easy to see
\begin{align*}
  \Pr \left( \sup_{ \pi \in \Pi } \mathcal{S} ( \pi ) \geq 2 \epsilon \right) &  \leq \sum\limits_{j=1}^l \Pr \left( \sup_{ \pi \in D_j } \mathcal{ S } ( \pi ) \geq 2 \epsilon \right) \\
  & \leq \mathcal{N} \left( \Pi, \frac{ \epsilon }{ 2 (2 M_1 + 2 M_2 + L) } \right) \exp \left( - \frac{ n \epsilon^2 }{ 8 ( d_{ \max }^2 + 1 ) ( M_1 + M_2 )^2 }  \right). 
\end{align*}
Upper bound for the probability $  \Pr \left( \sup_{ \pi \in \Pi } \mathcal{S} ( \pi ) \leq - 2 \epsilon \right) $ can be derived in the same way. The statement becomes valid by replacing $ \epsilon $ by $ \frac{\epsilon}{ 2 }$.
\end{proof}

Finally, we prove the policy regret bound in Theorem~\ref{theorem:regret_bound} as follows. 
\begin{proof}
Consider an arbitrary policy $ \tilde{ \pi } \in \Pi $, we have the following utility difference 
\begin{align*}
    S ( \tilde{\pi} ) - S ( \hat{ \pi }_n ) & = S_n^{ \tau, \delta } ( \tilde{ \pi } ) - S_n^{ \tau, \delta } ( \tilde{ \pi } ) + S_n^{ \tau, \delta } ( \hat{ \pi }_n ) - S_n^{ \tau, \delta } ( \hat{ \pi }_n ) \\
    & \quad + S ( \tilde{ \pi } ) -S ( \hat{ \pi }_n ) +  \hat{S}_n^{ \tau, \delta } ( \hat{ \pi }_n ) - \hat{S}_n^{ \tau, \delta } ( \hat{ \pi }_n )  \\
    & \leq \underbrace{ S_n^{ \tau, \delta } ( \tilde{ \pi } ) - \hat{S}_n^{ \tau, \delta } ( \tilde{ \pi } ) - S_n^{ \tau, \delta } ( \hat{ \pi }_n ) + \hat{S}_n^{ \tau, \delta } ( \hat{ \pi }_n ) }_{ (1) } \\
    & \quad + \underbrace{ S ( \tilde{ \pi } ) - S_n^{ \tau, \delta } ( \tilde{ \pi } ) + S_n^{ \tau, \delta } ( \hat{ \pi }_n ) - S ( \hat{ \pi }_n ) }_{ (2) }.
\end{align*}
Using $ \forall \pi \in \Pi $, $ \pi \in [0, 1] $ and assumption (ES) the term $(\star)$ can be bounded as 
\begin{align*}
  ( 1 ) & = \frac{1}{n} \sum_{i=1}^n 2 ( \tau_i - \hat{ \tau }_i ) ( \tilde{ \pi } ( \mathbf{X}_i ) - \hat{ \pi }_n ( \mathbf{X}_i ) ) \\
  & + \frac{1}{n} \sum_{i=1}^n  ( 2 \tilde{ \pi } ( \mathbf{X}_i ) - 1 ) ( \delta_i ( \tilde{ \pi } ) - \hat{ \delta }_i ( \tilde{ \pi } ) ) - \frac{1}{n}  \sum_{i=1}^n ( 2 \hat{ \pi }_n ( \mathbf{X}_i ) - 1 ) ( \delta_i ( \hat{ \pi }_n ) - \hat{ \delta }_i ( \hat{ \pi }_n ) ) \\
  & \leq \frac{1}{n} \sum_{i=1}^n 2 | \tau_i - \hat{ \tau }_i | + \frac{1}{n} \sum_{i=1}^n | \delta_i ( \tilde{ \pi } ) - \hat{ \delta }_i ( \tilde{ \pi } ) | + \frac{1}{n} \sum_{i=1}^n | \delta_i ( \hat{ \pi }_n ) - \hat{ \delta }_i ( \hat{ \pi }_n ) | \\
  &  \leq 2 \left( \frac{\alpha_{\tau}}{n^{\zeta_{\tau}}} +  \frac{\alpha_{\delta}}{n^{\zeta_{\delta}}} \right). 
\end{align*}
Furthermore, $ (2) \leq | S_n^{ \tau, \delta } ( \tilde{ \pi } ) - S ( \tilde{ \pi } ) | +  | S_n^{ \tau, \delta } ( \hat{ \pi }_n ) - S ( \hat{ \pi }_n ) | \leq 2 \sup_{ \pi \in \Pi } | S_n^{ \tau, \delta } ( \pi ) - S ( \pi ) | $. In summary, 
\begin{equation*}
  \mathcal{R} ( \hat{ \pi }_n ) := \sup_{ \tilde{\pi} \in \Pi } ( S ( \tilde{ \pi } ) - S ( \hat{ \pi }_n ) ) \leq 2 \left( \frac{\alpha_{\tau}}{n^{\zeta_{\tau}}} +  \frac{\alpha_{\delta}}{n^{\zeta_{\delta}}} \right) + 2 \epsilon, 
\end{equation*}
with probability at least $ 1 - \mathcal{N} \left( \Pi, \frac{ \epsilon }{ 4 ( 2M_1 + 2M_2 + L ) } \right) \exp \left( - \frac{ n \epsilon^2 }{ 32 ( d_{ \max }^2 + 1 ) ( M_1 + M_2 )^2 } \right) $ via Lemma~\ref{lemma:sup_bound}. 
\end{proof}

\section{Capacity-constrained Policy Regret}
\label{subsec:const_policy_regret_proof}

In this section, we provide an additional policy regret bound under capacity constraint. 
\begin{my_theorem}
  By Assumption~\ref{ass:ass_policy_regret}, for any small $\epsilon > 0$, the policy regret under the capacity constraint $ p_t $ is bounded by $ \mathcal{R} ( \hat{ \pi }_n^{ p_t } ) \leq 2 \left( \frac{\alpha_{\tau}}{n^{\zeta_{\tau}}} +  \frac{\alpha_{\delta}}{n^{\zeta_{\delta}}} \right) + 2 \epsilon $ with probability at least $ 1 - \mathcal{N} \exp \left( - \frac{ n \epsilon^2 }{ 32 ( d_{ \max }^2 + 1 ) ( M_1 + M_2 )^2 } \right) $, where $ \mathcal{N} := \mathcal{N} \left( \Pi, \frac{ \epsilon }{ 8 [ ( M_1 + M_2 + L ) + \frac{1}{p_t} ( M_1 + M_2 ) ] } \right)$ indicates the covering number on the functional class $ \Pi $ with radius $ \frac{ \epsilon }{ 8 [ ( M_1 + M_2 + L ) + \frac{1}{p_t} ( M_1 + M_2 ) ] } $, and $ d_{ \max } $ is the maximal node degree in the graph $ \mathcal{G} $. 
  \label{theorem:regret_bound_constrained}
\end{my_theorem}
This capacity-constrained policy regret bound indicates that if, in the constraint, $p_t$ is small, then the optimal capacity-constrained policy will be challenging to find. Increasing the treatment probability can not guarantee the improvement of the group's interest due to the non-linear network effect. Therefore, finding the balance between optimal treatment probability, treatment assignment, and group's welfare is a provocative question in social science.

Before proving Theorem~\ref{theorem:regret_bound_constrained}, let us first review the definition of utility function $ A ( \pi ) $ following Section 2 of~\cite{athey2017efficient}. The benefit of deploying the intervention policy $ \pi $ compared to assigning everyone in control group is defined as 
\begin{equation*}
  V( \pi ) := \mathbb{E} [ Y_i ( T_i = 1 ) \pi ( \mathbf{X}_i ) + Y_i ( T_i = 0 ) ( 1 - \pi ( \mathbf{X}_i ) ) ] - \mathbb{E} [ Y_i ( T_i = 0 ) ]  =  \mathbb{ E } [ \pi ( \mathbf{X}_i ) \tau ( \mathbf{X}_i ) ], 
\end{equation*} 
and the utility function equals 
\begin{equation*}
  A ( \pi ) := 2 V ( \pi ) - \mathbb{E} [ \tau ( \mathbf{X}_i ) ] = \mathbb{E} [ ( 2 \pi ( \mathbf{X}_i ) - 1 ) \tau ( \mathbf{X}_i )  ]. 
\end{equation*}
In the following, let us consider policy learning under treatment constraint $p_t$, and we will introduce a capacity-constrained utility function under network interference. If the distribution of covariates $ \mathcal{P}_{ \boldsymbol{\chi} } $ is known, and let $ \mathcal{P}_{ \boldsymbol{ \chi } } ( \pi ) $ denote the treatment rule on the covariates space, then a capacity-constrained welfare gain relative to treating no one is defined as (see also Section 4.1 of~\cite{kitagawa2017should})
\begin{align*}
  V_{ p_t } ( \pi ) & := \mathbb{E} [ [ Y_i (T_i=1) \min \{ 1, \frac{ p_t }{ \mathcal{P}_{ \boldsymbol{ \chi } } ( \pi ) } \} + Y_i ( T_i=0 ) ( 1- \min \{ 1, \frac{ p_t }{ \mathcal{P}_{ \boldsymbol{ \chi } } ( \pi ) } \} ) ] \pi ( \mathbf{X}_i ) \\
  & + Y_i ( T_i = 0 ) ( 1 - \pi ( \mathbf{X}_i ) ) ] - \mathbb{E} [ Y_i ( T_i = 0 ) ] \\
  & = \min \{ 1, \frac{ p_t }{ \mathcal{P}_{ \boldsymbol{ \chi } } ( \pi ) } \} \mathbb{E} [ \pi ( \mathbf{X}_i ) \tau ( \mathbf{X}_i ) ) ], 
\end{align*}
and the corresponding capacity-constrained utility function equals 
\begin{equation*}
  A_{ p_t } ( \pi ) := 2 V_{ p_t } ( \pi ) - \mathbb{E} [ \tau ( \mathbf{X}_i ) ] = \mathbb{E} [ ( 2 \min \{ 1, \frac{ p_t }{ \mathcal{P}_{ \boldsymbol{ \chi } } ( \pi ) } \} \pi ( \mathbf{X}_i ) - 1 )  \tau ( \mathbf{X}_i ) ) ]. 
\end{equation*}

Similarly, the capacity-constrained utility function under interference for interconnected units reads 
\begin{equation*}
  S_{ p_t } ( \pi ) := \mathbb{E} [ ( 2 \min \{ 1, \frac{ p_t }{ \mathcal{P}_{ \boldsymbol{ \chi } } ( \pi ) } \} \pi ( \mathbf{X}_i ) - 1 ) ( \tau_i + \delta_i ( \pi ) )]. 
\end{equation*}  
Moreover, the empirical version of $ S_{ p_t } ( \pi ) $ reads 
\begin{equation*}
  S_{n, p_t }^{ \tau, \delta } ( \pi ) := \frac{1}{n} \sum_{i=1}^n ( 2 \min \{ 1, \frac{ p_t }{ \mathcal{P}_{ \boldsymbol{ \chi } } ( \pi ) } \} \pi ( \mathbf{X}_i ) - 1 ) ( \tau_i + \delta_i ( \pi ) ). 
\end{equation*}
The empirical estimation of $ S_{ p_t } ( \pi ) $  with causal estimators being plugged in reads 
\begin{equation*}
  \hat{S}_{n, p_t}^{ \tau, \delta } ( \pi ) := \frac{1}{n} \sum_{i=1}^n ( 2 \min \{ 1, \frac{ p_t }{ \mathcal{P}_{ \boldsymbol{ \chi } } ( \pi ) } \} \pi ( \mathbf{X}_i ) - 1 ) ( \hat{\tau}_i + \hat{\delta}_i ( \pi ) ), 
\end{equation*}
an corresponding optimal capacity-constrained policy is obtained via~\footnote{This optimal capacity-constrained policy is, in principle, equivalent to the one obtained by minimizing the loss function $ \mathcal{L}_{ \mathrm{pol} } ( \pi ) := - \hat{S}_n^{ \tau, \delta } ( \pi ) + \gamma ( \frac{ 1 }{ n } \sum_{i=1}^n \pi ( \mathbf{X}_i ) - p_t ) $, since, in practice, treatment capacity constraint can be satisfied via Lagrangian multiplier.}
\begin{equation*}
  \hat{ \pi }_n^{p_t} \in \mathrm{argmax}_{ \pi\in \Pi } \hat{S}_{n, p_t}^{ \tau, \delta } ( \pi ).
\end{equation*}
Moreover, let $ \pi^{p_t \star} $ denote the best possible intervention policy from the functional class $ \Pi $ with respect to the utility $ S_{p_t} ( \pi ) $, namely $ \pi^{ p_t \star } \in \mathrm{argmax}_{ \pi \in \Pi } S_{ p_t } ( \pi ) $. The capacity-constrained policy regret is defined as $ \mathcal{R} ( \hat{ \pi }_n^{ p_t } ) := S_{ p_t } ( \pi^{ p_t \star } ) - S_{ p_t } ( \hat{ \pi }_n^{ p_t } ) $. Before estimating the capacity-constrained intervention policy regret we derive the following inequality similar to Lemma~\ref{lemma:s1_s2_lemma}.

\begin{my_lemma}
  Let $ \mathcal{S}_{ p_t } (\pi) := S_{n, p_t}^{\tau, \delta} ( \pi ) - S_{ p_t } ( \pi ) $, for any $ \pi_1, \pi_2 \in \Pi $, where the policy class in contained in $[0, 1]$, according to the assumptions (BO) and (LIP) we have 
  \begin{equation}
    | \mathcal{S}_{ p_t } ( \pi_1 ) - \mathcal{S}_{ p_t } ( \pi_2 ) | \leq 4 [ (M_1 + M_2 + L)  + \frac{ 1 }{ p_t } ( M_1 + M_2 )]  || \pi_1 - \pi_2 ||_{ \infty }. 
  \end{equation}
  \label{lemma:s1_s2_constrained_lemma}
\end{my_lemma}
\begin{proof}
  Note that $ | \mathcal{S}_{ p_t } ( \pi_1 ) - \mathcal{S}_{ p_t } ( \pi_2 ) | \leq | S_{ p_t } ( \pi_1 ) - S_{ p_t } ( \pi_2 )| + | S_{n, p_t}^{\tau, \delta} ( \pi_1 ) - S_{n, p_t}^{\tau, \delta} ( \pi_2 ) | $. We first rewrite $ S_{ p_t } ( \pi ) $ as
\begin{equation*}
   S_{ p_t } ( \pi ) = \min \{ 1, \frac{ p_t }{ \mathcal{P}_{ \boldsymbol{ \chi } } ( \pi ) } \}  \mathbb{E} [ (2 \pi ( \mathbf{X}_i ) - 1 ) ( \tau_i + \delta_i ( \pi ) ) ] + ( \min \{ 1, \frac{ p_t }{ \mathcal{P}_{ \boldsymbol{ \chi } } ( \pi ) } \} - 1 ) \mathbb{E} [ \tau_i + \delta_i ( \pi ) ]. 
\end{equation*}
Recall the definition of $S(\pi)$, and define $ T ( \pi ) := \mathbb{E} [ \tau_i + \delta_i ( \pi ) ] $, we have
\begin{align*}
  | S_{ p_t } ( \pi_1 ) - S_{ p_t } ( \pi_2 )| & = | \min \{ 1, \frac{ p_t }{ \mathcal{P}_{ \boldsymbol{ \chi } } ( \pi_1 ) } \} S ( \pi_1 ) - \min \{ 1, \frac{ p_t }{ \mathcal{P}_{ \boldsymbol{ \chi } } ( \pi_2 ) } \} S ( \pi_2 ) \\
  & \quad + ( \min \{ 1, \frac{ p_t }{ \mathcal{P}_{ \boldsymbol{ \chi } } ( \pi_1 ) } \} -1 ) T ( \pi_1 ) - ( \min \{ 1, \frac{ p_t }{ \mathcal{P}_{ \boldsymbol{ \chi } } ( \pi_2 ) } \} - 1 ) T ( \pi_2 ) | \\
  & \leq | \min \{ 1, \frac{ p_t }{ \mathcal{P}_{ \boldsymbol{ \chi } } ( \pi_1 ) } \} | | S ( \pi_1 ) - S ( \pi_2 ) | 
  \\
  & \quad + | S ( \pi_2 ) | | \min \{ 1, \frac{ p_t }{ \mathcal{P}_{ \boldsymbol{ \chi } } ( \pi_1 ) } \} - \min \{ 1, \frac{ p_t }{ \mathcal{P}_{ \boldsymbol{ \chi } } ( \pi_2 ) } \} | \\
  & \quad + | \min \{ 1, \frac{ p_t }{ \mathcal{P}_{ \boldsymbol{ \chi } } ( \pi_1 ) } \} - 1 | | T (\pi_1) - T (\pi_2) | \\
  & \quad + | T (\pi_2) | | \min \{ 1, \frac{ p_t }{ \mathcal{P}_{ \boldsymbol{ \chi } } ( \pi_1 ) } \} - \min \{ 1, \frac{ p_t }{ \mathcal{P}_{ \boldsymbol{ \chi } } ( \pi_2 ) } \} | \\
  & \leq | S (\pi_1) - S(\pi_2) | + | T(\pi_1) -T (\pi_2) | \\ 
  & \quad + ( | S(\pi_2) | + | T(\pi_2) | ) | \min \{ 1, \frac{ p_t }{ \mathcal{P}_{ \boldsymbol{ \chi } } ( \pi_1 ) } \} - \min \{ 1, \frac{ p_t }{ \mathcal{P}_{ \boldsymbol{ \chi } } ( \pi_2 ) } \} |. 
\end{align*}
Using the following bounds
\begin{align*}
  & | S ( \pi_1 ) - S ( \pi_2 ) | \leq ( 2 M_1 + 2 M_2 + L ) || \pi_1 - \pi_2 ||_{ \infty }, \\
  & | T ( \pi_1 ) - T ( \pi_2 ) | \leq L || \pi_1 - \pi_2 ||_{ \infty },  \\
  & | S ( \pi_2 ) | \leq M_1 + M_2,  \\
  & | T ( \pi_2 ) | \leq M_1 + M_2,  \\
  & | \min \{ 1, \frac{ p_t }{ \mathcal{P}_{ \boldsymbol{ \chi } } ( \pi_1 ) } \} - \min \{ 1, \frac{ p_t }{ \mathcal{P}_{ \boldsymbol{ \chi } } ( \pi_2 ) } \} | = | \frac{ p_t }{ \max \{ p_t, \mathcal{P}_{ \boldsymbol{\chi} } ( \pi_1 ) \} } - \frac{ p_t }{ \max \{ p_t, \mathcal{P}_{ \boldsymbol{\chi} } ( \pi_2 ) \} } | \\
  & \hspace{5.6cm} \leq \frac{1}{p_t} | \mathcal{P}_{ \boldsymbol{\chi} } ( \pi_1 ) - \mathcal{P}_{ \boldsymbol{\chi} } ( \pi_2 ) | \leq \frac{1}{p_t} || \pi_1 - \pi_2 ||_{ \infty }, 
\end{align*}
yields $ | S_{ p_t } ( \pi_1 ) - S_{ p_t } ( \pi_2 ) | \leq 2 [ ( M_1 + M_2 + L ) + \frac{1}{p_t} ( M_1 + M_2 ) ] || \pi_1 - \pi_2 ||_{ \infty } $. Similarly, we also have $ | S_{n, p_t}^{\tau, \delta} ( \pi_1 ) - S_{n, p_t}^{\tau, \delta} ( \pi_2 ) | \leq 2 [ ( M_1 + M_2 + L ) + \frac{1}{p_t} ( M_1 + M_2 ) ] || \pi_1 - \pi_2 ||_{ \infty } $. 
\end{proof}

In the same sense as Lemma~\ref{lemma:sup_bound}, using Lemma~\ref{lemma:s1_s2_constrained_lemma} we obtain the following bound for the policy functional class under a capacity constraint $p_t$. 
\begin{my_lemma}
    Under Assumption~\ref{ass:ass_policy_regret}, for any $ \{ \mathbf{X}_i \}_{ i=1 }^n \in \boldsymbol{ \chi }^{ \otimes n } $ and $\epsilon > 0$, it satisfies 
  \begin{equation*}
    \Pr \left( \sup\limits_{ \pi \in \Pi } | S_{n, p_t}^{ \tau, \delta } ( \pi ) - S_{ p_t } ( \pi ) | \leq \epsilon \right) \geq 1 - \mathcal{N} \exp \left( - \frac{ n \epsilon^2 }{ 32 ( d_{ \max }^2 + 1 ) ( M_1 + M_2 )^2 } \right), 
  \end{equation*}
where $ \mathcal{N} := \mathcal{N} \left( \Pi, \frac{ \epsilon }{ 8 [ ( M_1 + M_2 + L ) + \frac{1}{p_t} ( M_1 + M_2 ) ] } \right) $ represents the covering number on the policy functional class $ \Pi $ with radius $ \frac{ \epsilon }{ 8 [ ( M_1 + M_2 + L ) + \frac{1}{p_t} ( M_1 + M_2 ) ] } $. 
  \label{lemma:sup_constrained}
\end{my_lemma}

Finally, we can derive the capacity-constrained policy regret bound as follows. 
\begin{proof}
  Consider an arbitrary policy $ \tilde{ \pi } \in \Pi $, we have the following utility difference 
\begin{align*}
  S_{p_t} ( \tilde{\pi} ) - S_{p_t} ( \hat{\pi}_n^{ p_t } ) & \leq \underbrace{ S_{n, p_t}^{ \tau, \delta } ( \tilde{\pi} ) - \hat{S}_{n, p_t}^{ \tau, \delta } ( \tilde{\pi} ) - S_{n, p_t}^{ \tau, \delta } ( \hat{\pi}_n^{ p_t } ) + \hat{S}_{n, p_t}^{ \tau, \delta } ( \hat{\pi}_n^{ p_t } ) }_{ (1) } \\
  & \quad \underbrace{ S_{p_t} ( \tilde{ \pi } ) - S_{n, p_t}^{ \tau, \delta } ( \tilde{ \pi } ) + S_{n, p_t}^{ \tau, \delta } ( \hat{ \pi }_n^{ p_t } ) - S_{p_t} ( \hat{ \pi }_n^{ p_t } ) }_{ (2) }. 
\end{align*}
Using the fact that $ \forall \pi \in \Pi $,  $ | 2 \pi ( \mathbf{ X }_i ) \min \{ 1, \frac{ p_t }{ \mathcal{P}_{ \boldsymbol{  \chi } } ( \pi ) } \} - 1 | \leq 1 $, it is easy to see $ (1) \leq 2 \left( \frac{\alpha_{\tau}}{n^{\zeta_{\tau}}} +  \frac{\alpha_{\delta}}{n^{\zeta_{\delta}}} \right) $. Furthermore, 
\begin{equation*}
  (2) \leq | S_{n, p_t}^{ \tau, \delta } ( \tilde{ \pi } ) - S_{p_t} ( \tilde{ \pi } ) | +  | S_{n, p_t}^{ \tau, \delta } ( \hat{ \pi }_n^{p_t} ) - S_{p_t} ( \hat{ \pi }_n^{p_t} ) | \leq 2 \sup_{ \pi \in \Pi } | S_{n, p_t}^{ \tau, \delta } ( \pi ) - S_{p_t} ( \pi ) |. 
\end{equation*}
In summary, via Lemma~\ref{lemma:sup_constrained} it yields the statement. 
\end{proof}


\end{document}